\RequirePackage{silence}
\documentclass[final,onefignum,onetabnum]{siamart190516}

\usepackage{amssymb,framed,xcolor,bold-extra,microtype}
\usepackage[shortlabels]{enumitem}

\usepackage{thmtools,thm-restate}
\declaretheorem[name=Lemma,sibling=theorem]{restatable-lemma}
\declaretheorem[name=Theorem,sibling=theorem]{restatable-theorem}
\declaretheorem[name=Corollary,sibling=theorem]{restatable-corollary}

\crefname{restatable-lemma}{Lemma}{Lemmas}
\crefname{restatable-theorem}{Theorem}{Theorems}
\crefname{restatable-corollary}{Corollary}{Corollaries}

\crefname{subsection}{Section}{Sections}
\crefname{section}{Section}{Sections}

\usepackage{xspace}

\newcommand{\floor}[1]{\lfloor #1 \rfloor}
\newcommand{\ceil}[1]{\lceil #1 \rceil}
\newcommand{\OO}{\mathcal{O}}
\newcommand{\tildeOO}{\widetilde{\mathcal{O}}}

\newcommand{\pp}[1]{\textnormal{\textsc{#1}}\xspace}

\newcommand{\poly}{\operatorname{poly}}
\newcommand{\polylog}{\operatorname{polylog}}
\newcommand{\E}{\mathbb{E}}
\renewcommand{\Pr}{\mathbb{P}}
\newcommand{\pr}{\mathbb{P}}
\newcommand{\R}{\mathbb{R}}
\newcommand{\Z}{\mathbb{Z}}

\newcommand{\N}{\mathbb{N}}

\newcommand{\eps}{\varepsilon}
\renewcommand{\epsilon}{\varepsilon}
\newcommand{\pout}{p_\mathsf{out}}
\newcommand{\hmu}{\hat{\mu}}

\newcommand{\calC}{\mathcal{C}}
\newcommand{\calD}{\mathcal{D}}
\newcommand{\calE}{\mathcal{E}}
\newcommand{\calG}{\mathcal{G}}
\newcommand{\calI}{\mathcal{I}}
\newcommand{\calS}{\mathcal{S}}
\newcommand{\calR}{\mathcal{R}}
\newcommand{\calW}{\mathcal{W}}

\newcommand{\indora}{\mathrm{IND}_G}
\newcommand{\cindora}{\textup{cIND}\ensuremath{_G}\xspace} 
\newcommand{\coarsecountgeneral}{\textnormal{\texttt{HelperCoarse}}\xspace}
\newcommand{\coarsecount}{\textnormal{\texttt{ColourCoarse}}\xspace}
\newcommand{\impsamp}{\textnormal{\texttt{Reduce}}\xspace}
\newcommand{\halve}{\textnormal{\texttt{Refine}}\xspace}
\newcommand{\approxUncol}{\textnormal{\texttt{HelperCount}}\xspace}
\newcommand{\acc}{\textnormal{\texttt{Coarse}}\xspace}
\newcommand{\aau}{\textnormal{\texttt{Count}}\xspace}
\newcommand{\verifyguess}{\textnormal{\texttt{VerifyGuess}}\xspace}
\newcommand{\sample}{\textnormal{\texttt{HelperSample}}\xspace}
\newcommand{\realsample}{\textnormal{\texttt{Sample}}\xspace}
\newcommand{\samplecount}{\textnormal{\texttt{SampleCount}}\xspace}
\newcommand{\colourfulsamplecount}{\textnormal{\texttt{PartitionedSampleCount}}\xspace}

\newcommand{\yes}{\texttt{Yes}\xspace}
\newcommand{\no}{\texttt{No}\xspace}

\newcommand{\shortrule}{\begin{center}\rule{0.5\textwidth}{0.4pt}\end{center}}

\ifpdf
\hypersetup{
  pdftitle={Approximately counting and sampling small witnesses using a colourful decision oracle},
  pdfauthor={H. Dell, J. Lapinskas, K. Meeks}
}
\fi

\title{Approximately counting and sampling small witnesses using a colourful decision oracle%
\thanks{The research leading to these results has received funding from the European Research Council under the European Union's Seventh Framework Programme (FP7/2007-2013) ERC grant agreement no.~334828. The paper reflects only the authors' views and not the views of the ERC or the European Commission.  The European Union is not liable for any use that may be made of the information contained therein. The research was also supported by a Royal Society of Edinburgh Personal Research Fellowship, funded by the Scottish Government.  The collaboration began at Dagstuhl Seminar~17341.  An extended abstract~\cite{DBLP:conf/soda/DellLM20} was presented at SODA 2020.}%
}
\author{Holger Dell\thanks{Goethe University Frankfurt, Germany, IT University of Copenhagen and BARC, Copenhagen, Denmark} \email{dell@uni-frankfurt.de}
\and
John Lapinskas\thanks{University of Bristol, Bristol, UK (\email{john.lapinskas@bristol.ac.uk})}
\and
Kitty Meeks\thanks{University of Glasgow, Glasgow, UK (\email{Kitty.Meeks@glasgow.ac.uk})}
}

\begin{document}

\maketitle

\begin{abstract}
	In this paper, we design efficient algorithms to approximately count the number of edges of a given $k$-hypergraph, and to sample an approximately uniform random edge. The hypergraph is not given explicitly, and can be accessed only through its \emph{colourful independence oracle}:
	The colourful independence oracle returns yes or no depending on whether a given subset of the vertices contains an edge that is colourful with respect to a given vertex-colouring.
	Our results extend and/or strengthen recent results in the graph oracle literature due to Beame et al.~(ITCS 2018), Dell and Lapinskas (STOC 2018), and Bhattacharya et~al.~(ISAAC 2019).

    Our results have consequences for approximate counting/sampling:
    We can turn certain kinds of decision algorithms into approximate counting/sampling algorithms without causing much overhead in the running time.
    We apply this approximate-counting/sampling-to-decision reduction to key problems in fine-grained complexity (such as $k$-SUM, $k$-OV and weighted $k$-Clique) and parameterised complexity (such as induced subgraphs of size $k$ or weight-$k$ solutions to CSPs).
\end{abstract}

\section{Introduction}\label{sec:intro}

Many decision problems reduce to the question: Does a witness exist? Such problems admit a natural counting version: How many witnesses exist? For example, one may ask whether a bipartite graph contains a perfect matching, or how many perfect matchings it contains. As one might expect, the counting version is never easier than the decision version, and is often substantially harder; for example, deciding whether a bipartite graph contains a perfect matching is easy, and counting the number of such matchings is \#P-complete~\cite{Valiant-matchings}. However, even when the counting version of a problem is hard, it is often easy to approximate well. For example, Jerrum, Sinclair and Vigoda~\cite{JSV} gave a polynomial-time approximation algorithm for the number of perfect matchings in a bipartite graph. The study of approximate counting has seen amazing progress over the last two decades, particularly in the realm of trichotomy results for general problem frameworks such as constraint satisfaction problems, and is now a major field of study in its own right~\cite{dggj-approx,dgj-approx,ggj-approx,gllz-approx-local,gllz-approx}. In this paper, we explore the question of when approximating the counting version of a problem is not merely fast, but essentially as fast as solving the decision version.

We first recall the standard notion of approximation in the field: For all real $x,y > 0$ and $0 < \eps < 1$, we say that $x$ is an \emph{$\eps$-approximation} to $y$ if $|x-y| < \eps y$. Note in particular that any $\eps$-approximation to zero is itself zero, so computing an $\eps$-approximation to $N$ is always at least as hard as deciding whether $N > 0$ holds. For example, it is at least as hard to approximately count the number of satisfying assignments of a CNF formula (i.e.\ to $\eps$-approximate \#\pp{Sat}) as it is to decide whether it is satisfiable at all (i.e.\ to solve \pp{Sat}). 

Perhaps surprisingly, in many cases, the converse is also true. For example, Valiant and Vazirani~\cite{VV} proved that any polynomial-time algorithm to decide \pp{Sat} can be bootstrapped into a polynomial-time $\eps$-approximation algorithm for \#\pp{Sat}, or, more formally, that a size-$n$ instance of any problem in \#P can be $\eps$-approximated in time $\poly(n,\eps^{-1})$ using an NP-oracle. A similar result holds in the parameterised setting, where M\"uller~\cite{Muller} proved that a size-$n$ instance of any problem in \#W[$i$] with parameter $k$ can be $\eps$-approximated in time $g(k)\cdot \mbox{poly}(n,\eps^{-1})$ using a W[$i$]-oracle for some computable function $g\colon \N \to \N$. Another such result holds in the subexponential setting, where Dell and Lapinskas~\cite{DL} proved that the (randomised) Exponential Time Hypothesis is equivalent to the statement: There is no $\eps$-approximation algorithm for \#\pp{3-Sat} which runs on an $n$-variable instance in time $\eps^{-2}2^{o(n)}$.

We now consider the fine-grained setting, which is the focus of this paper. Here, we are concerned with the exact running time of an algorithm, rather than broad categories such as polynomial time, FPT time or subexponential time. The above reductions all introduce significant overhead, so they are not fine-grained. Here only one general result is known, again due to Dell and Lapinskas~\cite{DL}. Informally, if the decision problem reduces ``naturally'' to deciding whether an $n$-vertex bipartite graph contains an edge, then any algorithm for the decision version can be bootstrapped into an $\eps$-approximation algorithm for the counting version with only $\tildeOO(\eps^{-2})$ overhead. (Here and elsewhere, the $\tildeOO$-notation suppresses a factor of $C\log^C n$ for some constant $C>0$ depending only on the problem statement.) See \cref{sec:framework-intro} for more details.

The reduction of~\cite{DL} is general enough to cover core problems in fine-grained complexity such as \pp{Orthogonal Vectors}, \pp{3SUM} and \pp{Negative-Weight Triangle}, but it is not universal. In this paper, we substantially generalise it to cover any problem which can be ``naturally'' formulated as deciding whether a $k$-partite $k$-hypergraph contains an edge; thus we essentially recover the original result on taking $k=2$. For any problem which satisfies this property, our result implies that any new decision algorithm will automatically lead to a new approximate counting algorithm whose running time is at most a factor of $\log^{\OO(k)} n$ larger. Our framework covers several reduction targets in fine-grained complexity not covered by~\cite{DL}, including \pp{$k$-Orthogonal Vectors}, \pp{$k$-SUM} and \pp{Exact-Weight $k$-Clique}, as well as some key problems in parameterised complexity including weight-$k$ CSPs and size-$k$ induced subgraph problems. (Note that the overhead of $\log^{\OO(k)} n$ can be re-expressed as $k^{2k}n^{o(1)}$ using a standard trick, so an FPT decision algorithm is transformed into an FPT approximate counting algorithm; see \cref{sec:applications-intro}.)  

In fact, we get more than fast approximate counting algorithms --- we also prove that any problem in this framework has an algorithm for approximately-uniform sampling, again with $\log^{\OO(k)}n$ overhead over decision. There is a well-known reduction between the two for self-reducible problems due to Jerrum, Valiant and Vazirani~\cite{JVV}, but it does not apply in our setting since it adds polynomial overhead.

In the parameterised setting, our results have interesting implications. Here, the requirement that the hypergraph be $k$-partite typically corresponds to considering the ``colourful'' or ``multicolour'' version of the decision problem, so our result implies that uncoloured approximate counting is essentially equivalent to multicolour decision. We believe that our results motivate considerable further study of the relationship between multicolour parameterised decision problems and their uncoloured counterparts.

Finally, we note that the applications of our results are not just complexity-theoretic in nature, but also algorithmic. They give a ``black box'' argument that any decision algorithm in our framework, including fast ones, can be converted into an approximate counting or sampling algorithm with minimal overhead. Concretely, we obtain new algorithms for approximately counting and/or sampling zero-weight subgraphs, graph motifs, and satisfying assignments for first-order models, and our framework is sufficiently general that we believe new applications will be forthcoming.

In \cref{sec:framework-intro}, we set out our main results in detail as \cref{thm:main-counting,thm:main-sampling}, and discuss our edge-counting reduction framework (which is of independent interest). We describe the applications of our main results to fine-grained complexity and parameterised complexity in \cref{sec:applications-intro}.

\subsection{\texorpdfstring{The $k$-hypergraph framework}{The k-hypergraph framework}}\label{sec:framework-intro}

Given a $k$-hypergraph $G=(V,E)$, write $e(G) = |E|$, and let 
\[
	\calC(G) := \{(X_1, \dots, X_k) \colon \mbox{$X_1,\dots,X_k$ are disjoint subsets of $V$}\}.
\]
For any $(X_1,\ldots,X_k) \in \calC(G)$, we write $G[X_1, \dots, X_k]$ for the $k$-partite $k$-hypergraph on $X_1 \cup \dots \cup X_k$ whose edge set is $\{e \in E(G) \colon |e \cap X_i| = 1\mbox{ for all }i \in [k]\}$. We define the \emph{colourful independence oracle}\footnote{%
Arguably, \emph{colourful edge oracle} would be a more natural name, but we stick to the name established in the literature.}
of $G$ to be the function $\cindora\colon \calC(G)\to\{0,1\}$ such that $\cindora(X_1,\dots,X_k) = 1$ if $G[X_1, \dots, X_k]$ has no edges, and $\cindora(X_1, \dots, X_k) = 0$ otherwise. Informally, we think of elements of $\calC(G)$ as representing $k$-colourings of induced subgraphs of $G$, with $X_i$ being the $i$'th colour class; thus given a vertex colouring of an induced subgraph of $G$, the colourful independence oracle outputs 1 if and only if no colourful edge is present. We consider a computation model where the algorithm is given access to $V$ and $k$, but can only access $E$ via $\cindora$. We say that such an algorithm has \emph{colourful oracle access} to $G$, and for legibility we write it to have $G$ as an input. Note that given colourful oracle access to $G$, it is trivial to simulate the colourful independence oracle of $G[X]$ for any $X \subseteq V(G)$ as in~\cite{beame20}. Our main result is as follows.

\begin{restatable}{restatable-theorem}{statemaincounting}
\label{thm:uncol-approx}\label{thm:main-counting}
There is a randomised algorithm $\aau(G,\eps,\delta)$ with the following behaviour. Suppose $G$ is an $n$-vertex $k$-hypergraph, and that \aau has colourful oracle access to $G$. Suppose $\eps$ and $\delta$ are rational with $0 < \eps,\delta < 1$. Then, writing $T = \log(1/\delta)\eps^{-2}k^{6k}\log^{4k+7}n$: in time~$\OO(nT)$, and using at most~$\OO(T)$ queries to \cindora{}, $\aau(G,\eps,\delta)$ outputs a rational number~$\hat{e}$. With probability at least $1-\delta$, we have $\hat{e} \in (1\pm\eps)e(G)$.
\end{restatable}

As an example of how \cref{thm:main-counting} applies to approximate counting problems, consider the problem \#\pp{$k$-Clique} of counting the number of size-$k$ cliques in an $n$-vertex graph $H$. We take $G$ to be the $k$-hypergraph on vertex set $V(H)$ whose edges are precisely those size-$k$ sets which span cliques in~$G$. Thus, $\eps$-approximating the number of $k$-cliques in $H$ corresponds to $\eps$-approximating the number of edges in $G$. We may use a decision algorithm for $k$-Clique with running time $f(n,k)$ to evaluate $\cindora$ in time $f(n,k)$, by applying it to an appropriate subgraph of $G$ (in which we delete all edges within each colour class $X_i$). Thus, \cref{thm:main-counting} gives us an algorithm for $\eps$-approximating the number of $k$-cliques in $H$ in time $\OO(nT + Tf(n,k))$. Any decision algorithm for $k$-Clique must read a constant proportion of its input, so we have $f(n,k) = \Omega(n)$ and our overall running time is $\OO(Tf(n,k))$. It follows that any decision algorithm for $k$-clique yields an $\eps$-approximation algorithm for \#$k$-Clique with overhead only $T = \eps^{-2}(k\log n)^{\OO(k)}$.  

The polynomial dependence on $\eps$ in \cref{thm:main-counting} is not surprising, as by taking $\eps < 1/2n^k$ and rounding we can obtain the number of edges of $G$ exactly. Thus, if the dependence on $\eps$ were subpolynomial, \cref{thm:main-counting} would essentially imply a fine-grained reduction from exact counting to decision. This is impossible under SETH in our setting; see~\cite[Theorem~3]{DL} for a more detailed discussion.

We extend \cref{thm:main-counting} to approximately-uniform sampling as follows.
\begin{restatable}{restatable-theorem}{statemainsampling}\label{thm:main-sampling}
	There is a randomised algorithm $\realsample(G,\eps)$ which, given a rational number $\eps$ with $0 < \eps < 1$ and colourful oracle access to an $n$-vertex $k$-hypergraph $G$ containing at least one edge, outputs either a random edge $f \in E(G)$ or $\textnormal{\texttt{Fail}}$. For all $f \in E(G)$, $\realsample(G,\eps)$ outputs $f$ with probability $(1\pm \eps)/e(G)$; in particular, it outputs $\textnormal{\texttt{Fail}}$ with probability at most $\eps$. Moreover, writing $T = \eps^{-2}k^{7k}\log^{4k+11} n$, $\realsample(G,\eps)$ runs in time $\OO(nT)$ and uses at most $\OO(T)$ queries to $\cindora$.
\end{restatable}

We call the output of this algorithm an \emph{$\eps$-approximate sample}. Note that there is a standard trick using rejection sampling which, given an algorithm of the above form, replaces the $\eps^{-2}$ factor in the running time by a $\tildeOO(\eps^{-1})$ factor; see~\cite{JVV}. Unfortunately, it does not apply to \cref{thm:main-sampling}, as we do not have a fast way to compute the true distribution of $\realsample$'s output. 

By the same argument as above, \cref{thm:main-sampling} may be used to sample a size-$k$ clique from a distribution with total variation distance at most $\eps$ from uniformity with overhead only $T = \eps^{-2}(k\log n)^{\OO(k)}$ over decision. (We also note that it is easy to extend \cref{thm:main-counting,thm:main-sampling} to cover the case where the original decision algorithm is randomised, at the cost of an extra factor of $k\log n$ in the number of oracle uses; we discuss this below.)

\cref{thm:main-counting,thm:main-sampling} are also of independent interest, generalising known results in the graph oracle literature.
Colourful independence oracles are a natural generalisation of the bipartite independent set (BIS) oracles introduced in Beame et al.~\cite{beame20} to a hypergraph setting, and when $k=2$ the two notions coincide. They were first introduced in Bishnu et al.~\cite{DBLP:conf/isaac/BishnuGKM018} to solve various decision problems in parameterised complexity. The main result of Beame et al.~\cite[Theorem~4.9]{beame20} says that given BIS oracle access to an $n$-vertex graph $G$, one can $\eps$-approximate the number of edges of $G$ using $\OO(\eps^{-4}\log^{14} n)$ BIS queries (which they take as their measure of running time). The $k=2$ case of \cref{thm:main-counting,thm:main-sampling} give a total of $\OO(\eps^{-2}\log^{19} n)$ queries used, improving their running time for most values of $\eps$, and extending their algorithm to approximately-uniform sampling. 

When $k=3$, our colourful independence oracles are similar to the tripartite independent set (TIS) oracles of Bhattacharya et al.~\cite{BBGM}. (These oracles ask whether a 3-coloured graph $H$ contains a colourful triangle, rather than whether a 3-coloured 3-hypergraph $G$ contains a colourful edge. But if $G$ is taken to be the 3-hypergraph whose edges are the triangles of $H$, then the two notions coincide exactly.) Their main result, Theorem~1, says that given TIS oracle access to an $n$-vertex graph $G$ in which every edge belongs to at most $d$ triangles, one can $\eps$-approximate the number of triangles in $G$ using at most $\OO(\eps^{-12}d^{12}\log^{25}n)$ TIS queries; they have subsequently~\cite{BBGM-journal} improved this to $\OO(\eps^{-4}d^2\log^{18} n)$. Our \cref{thm:main-counting} gives an algorithm which requires only $\OO(\eps^{-2}\log^{19} n)$ TIS queries, with no dependence on $d$, and which also generalises to approximately counting $k$-cliques for all fixed $k$. Again, \cref{thm:main-sampling} extends the result to approximately-uniform sampling.

In an independent work made public shortly after this paper appeared on arXiv, Bhattacharya et al.~\cite{BBGM19} also prove a version of \cref{thm:main-counting} (but not \cref{thm:main-sampling}) for arbitrary $k$. Their generalised $d$-partite independent set oracles are essentially the same as our independence oracles, and their algorithm makes $\OO(\eps^{-4}\log^{5k+5}n)$ queries on an $n$-vertex $k$-uniform input hypergraph; for comparison, our algorithm makes $\OO(\eps^{-2}\log^{4k+7}n)$ queries. Our algorithm therefore trades a slightly worse dependence on $n$ for a significantly better dependence on $\eps$. As in~\cite{BBGM}, the authors are motivated by the theory of graph oracles rather than by applications to specific approximation algorithms.

We note in passing that the main result of~\cite{DL} doesn't quite fit into this setting, as it also makes unrestricted use of edge existence queries. It resembles a version of \cref{thm:main-counting} restricted to $k=2$ and with slightly lower overhead in $n$.

If the colourful independence oracle~$\cindora$ of a $k$-hypergraph~$G$ can be simulated by a deterministic algorithm~$A_G$, then \cref{thm:main-counting,thm:main-sampling} immediately yield algorithms for approximately counting and sampling edges of~$G$:
Whenever the algorithms in \cref{thm:main-counting,thm:main-sampling} would query $\cindora$, they instead run~$A_G$ as a black-box.
Suppose instead $A_G$ is a randomised algorithm whose error probability is bounded by a constant, say $1/3$. Then we can use standard probability amplification techniques to make sure that, during any execution of the algorithms in \cref{thm:main-counting,thm:main-sampling}, it is very likely that \emph{all} queries to $\cindora$ are answered correctly.
For convenience, we encapsulate this argument in the following corollary.
We defer the (simple) proof to~\cref{sec:corols}.
\begin{restatable}{restatable-corollary}{statecombinedmain}\label{thm:main-combined}
    There is a randomised algorithm \samplecount with the following behaviour. Let $G$ be an arbitrary $n$-vertex $k$-hypergraph for some $n$ and $k$, and let $A_G$ be a randomised implementation of the colourful independence oracle of $G$ with worst-case running time $T$ and error probability at most $1/3$. Let $\eps,\delta > 0$. Then \samplecount$(V(G),k,A_G,\eps,\delta)$ outputs an $\eps$-approximation of $e(G)$ with error probability at most $\delta$ and an $\eps$-approximate sample from $E(G)$ with error probability zero. Moreover, the running time of \samplecount is at most \[{\eps^{-2} \log^2(1/\delta)(k\log n)^{\OO(k)}(n+T)}\,.\]
\end{restatable}
We remark that the worst-case running time of~$A_G$ is measured as the maximum possible running time of executions~$A_G(X_1,\dots,X_k)$ over all internal random choices of~$A_G$ and inputs~$X_1,\dots,X_k$.
We also mention an analogous corollary for approximately counting and sampling edges of any $k$-partite hypergraph $G[X_1,\dots,X_k]$ even if~$G$ is not $k$-partite itself; note that we do not require $X_1 \cup \dots \cup X_k = V(G)$.

In the case where the hypergraph $G$ is $k$-partite, we can make do with a weaker form of the oracle. Given a $k$-hypergraph $G = (V,E)$, we define the \emph{uncoloured independence oracle} of $G$ to be the function $\indora\colon 2^V\to \{0,1\}$ such that $\indora(X) = 1$ if $G[X]$ contains no edges, and $\indora(X) = 0$ otherwise. As we show in Section~\ref{sec:corols}, on a $k$-partite graph it is not hard to simulate $\cindora$ given access to $\indora$.

\begin{restatable}{restatable-corollary}{statecombinedmaincolourful}\label{thm:main-combined-colourful}
	\mbox{}
    There is a randomised algorithm \colourfulsamplecount with the following behaviour. Let $G$ be an arbitrary $n$-vertex $k$-partite $k$-hypergraph for some $n$ and~$k$ with vertex classes $V_1,\dots,V_k$, and let $A_G$ be a randomised implementation of the uncoloured independence oracle of~$G$ with worst-case running time~$T$ and error probability at most $1/3$. Let $\eps,\delta > 0$. Then \colourfulsamplecount$(V_1,\dots,V_k,k,A_G,\eps,\delta)$ outputs an $\eps$-approximation of $e(G)$ with failure probability at most $\delta$ and an $\eps$-approximate sample from $E(G)$ with error probability zero. The running time of \colourfulsamplecount is at most ${\eps^{-2} \log^2(1/\delta)(k\log n)^{\OO(k)}(n+T)}$.
\end{restatable}

\subsection{Applications}\label{sec:applications-intro}

\cref{thm:main-combined,thm:main-combined-colourful} have algorithmic implications for many well-known problems in fine-grained and parameterised complexity. On a formal level, there is no reason not to simply apply \cref{thm:main-combined,thm:main-combined-colourful}. However, it is instructive to tie these corollaries back to the standard informal notion of the ``colourful version'' of a problem, which we do in the following pair of definitions. Recall that a counting problem is a function $\#\Pi\colon\{0,1\}^\ast\to\N$ and its corresponding decision problem is defined via ${\Pi=\{x\in\{0,1\}^\ast\colon \#\Pi(x)>0\}}$.

\begin{definition}\label{def:uwp}
    $\Pi$ is a \emph{uniform witness problem} if there is a function from instances $x\in\{0,1\}^\ast$ to uniform hypergraphs~$G_x$ such that:
    \begin{enumerate}[(i)]
        \item $\#\Pi(x)=e(G_x)$;
        \item $V(G_x)$ and the size of edges in $E(G_x)$ can be computed from~$x$ in time $\tildeOO(|x|)$;
        \item there exists an algorithm which, given $x$ and $S \subseteq V(G_x)$, in time $\tildeOO(|x|)$ prepares an instance $I_x(S)$ of $\Pi$ such that $G_{I_x(S)} = G_x[S]$ and $|I_x(S)| \in \OO(|x|)$.
    \end{enumerate}
    We say $\Pi$ is a \emph{colourful uniform witness problem} if $G_x$ is always $k$-partite, where $k$ is the edge size of $G_x$. The set $E(G_x)$ is the set of \emph{witnesses} of the instance~$x$. 
\end{definition}

For all the problems we consider, there will only be a single natural choice of hypergraph representation, and so we consider this representation to be a part of the problem statement. For example, $k$-\pp{Clique} is a uniform witness problem in which for a given instance $x = (H, k)$, $G_x$ is the hypergraph on $V(H)$ whose edges are the $k$-cliques of $H$. It is immediate that Definition~\ref{def:uwp}(i) and (ii) are satisfied, and Definition~\ref{def:uwp}(iii) is satisfied because ``induced subgraphs of $G_x$ correspond to sub-instances of $x$'' --- that is, for a given instance $x = (H,k)$ of $k$-\pp{Clique} and a given $S \subseteq V(G_x) = V(H)$, $I_x(S)$ is the instance $(H[S], k)$. As a second example, \pp{Orthogonal Vectors} is a colourful uniform witness problem in which the witnesses are pairs of orthogonal vectors from the given input sets, and again induced subgraphs of $G_x$ correspond to sub-instances of $x$.

\begin{definition}\label{def:colourful-problems}
Suppose $\Pi$ is a uniform witness problem. 
\pp{Colourful-$\Pi$} is defined as the problem of, given an instance $x\in\{0,1\}^\ast$ of $\Pi$ and a partition of $V(G_x)$ into disjoint sets $S_1,\dots,S_k$, deciding whether $\text{cIND}_{G_x}(S_1,\dots,S_k)=0$ holds. 
\end{definition}

In other words, in \pp{Colourful-$\Pi$}, the goal is to decide whether $G_x[S_1,\dots,S_k]$ has at least one edge containing one vertex from each colour class $S_i$. For example, in \pp{Colourful}-$k$-\pp{Clique}, we are given a graph $G$, an integer $k$, and a $k$-colouring of $G$, and wish to know whether $G$ contains a $k$-clique with one vertex of each colour. Observe that if $\Pi$ is already colourful as in Definition~\ref{def:uwp}, then \pp{Colourful-$\Pi$} reduces to $k^{O(k)}$ instances of $\Pi$, since any colourful edges must respect the existing vertex classes of~$G_x$ and since induced subgraphs of $G_x$ correspond to instances of $\Pi$. 
The colourful version of a problem is sometimes referred to as the \emph{multicolour} version \cite{meeksunbddtw}.

Given an instance of $\Pi$, we write $n_x$ for the number of vertices of $G_x$, $k_x$ for the edge size of $G_x$, and $W_x$ for the set of witnesses of $x$. Using this terminology, we can now restate (a slightly simplified form of) \cref{thm:main-combined,thm:main-combined-colourful} as follows, where we use part (iii) of Definition~\ref{def:uwp} to ensure that $\mbox{cIND}_{G_x}(S_1,\dots,S_k)$ can be computed efficiently even when $S_1 \cup \dots \cup S_k \subsetneq V(G_x)$. We prove this result in section~\ref{sec:corols}.

\begin{restatable}{restatable-theorem}{statemetainformal}\label{thm:meta-informal}
	Let $\Pi$ be a uniform witness problem, and let $T$ be any function from instances of $\Pi$ to the positive reals.
	Suppose that given an instance $x$ of $\Pi$, there is an algorithm to solve \pp{Colourful-$\Pi$} on $x$ with error probability at most~$1/3$ in time~$T(x)$.
	Suppose also that any such algorithm has running time~$\widetilde\Omega(|x|)$. Then there is a randomised algorithm which, given an instance $x$ of $\Pi$ and $\eps>0$, with running time 
	\begin{equation}\label{eq:meta-informal-runtime}
	    \eps^{-2}(k_x\log n_x)^{\OO(k_x)} \cdot \max\big\{T(I_x(S)) \colon S \subseteq V(G_x)\big\},
	\end{equation}
	outputs an $\eps$-approximation to $\#\Pi(x)$ with probability at least $2/3$ and an $\eps$-approximate sample from~$W_x$ with error probability zero.
\end{restatable}

In a typical application of \cref{thm:meta-informal}, the maximum in~\eqref{eq:meta-informal-runtime} will be $\tildeOO(T(x))$. Recall that if $\Pi$ is already colourful, then \pp{Colourful-$\Pi$} reduces to $k^{O(k)}$ instances of $\Pi$, so Theorem~\ref{thm:meta-informal} can be used to reduce from colourful approximate counting to colourful decision as well as from uncoloured approximate counting to colourful decision.

One large family of problems to which this meta-theorem can naturally be applied is that of \emph{self-contained $k$-witness problems} (see \cite{meeksenum}); these are essentially a form of uniform witness problems $\Pi$ in which for any instance $x$ of $\Pi$ and all $S \subseteq V(G_x)$, the instance $I_x(S)$ \emph{is} a sub-instance of $x$ in a well-defined sense. Thus, every self-contained $k$-witness problem is a uniform witness problem, but it seems likely that not every uniform witness problem is a self-contained $k$-witness problem.

\subsubsection*{Applications in fine-grained complexity}

In the fine-grained setting, $k$ is considered to be a fixed constant, so the running-time bound in \cref{thm:meta-informal} can be written as $\tildeOO(\eps^{-2} \cdot T(n,k))$.

In~\cite{DL}, fine-grained reductions from approximate counting to decision were shown for the problems \pp{Orthogonal Vectors}, \pp{3SUM}, and \pp{Negative-Weight Triangle}. In~\cref{sec:corols}, we generalise these reductions to \pp{$k$-Orthogonal Vectors}, \pp{$k$-SUM}, \pp{Zero-Weight \mbox{$k$-Clique}}, and other subgraph isomorphism problems, such as \pp{Colourful-$H$}, \pp{Exact-Weight $k$-Clique}, and \pp{Exact-Weight $H$}. Similarly, we also reduce approximate model counting to model checking with respect to $k$-variable first-order formulas. In each case, we also have a corresponding result for approximate sampling of witnesses. 

We prove these results in \cref{sec:corols} as a corollary of our main results. To do so, we observe that these problems are natural $k$-witness problems, and that any decision algorithm for each of these problems can be coaxed in a more or less standard way to also solve the colourful variant of the problem in roughly the same time.

\subsubsection*{Applications in parameterised complexity}

In the parameterised setting, we assume that $k$ is taken as the parameter and are therefore interested in the running time of our algorithms as a function of $k$.
We note that our reduction from approximate counting to decision involves only a ``fine-grained FPT overhead'': we can rewrite the overhead of $\log^{\OO(k)}n$ as an overhead of $k^{2k}n^{o(1)}$ by a standard calculation.\footnote{Indeed, if $k \le \log n / (\log \log n)^2$ then $\log^{\OO(k)}n = e^{\OO(\log n / \log \log n)} = n^{o(1)}$, and if $k \ge \log n / (\log \log n)^2$ then $\log^{\OO(k)}n = \OO(k^{2k})$.} Thus, whenever there is an FPT implementation of the colourful independence oracle, \cref{thm:meta-informal} gives us an FPTRAS (fixed parameter tractable randomised approximation scheme \cite{DBLP:conf/isaac/ArvindR02}) for the corresponding (colourful or uncoloured) counting problem.  For the formal definition of an FPTRAS, and other standard notions in parameterised (counting) complexity, we refer the reader to \cite{flumgrohe}.

The family of uniform witness problems from \cref{def:uwp} includes numerous problems in parameterised complexity, including weight-$k$ solutions to CSPs, size-$k$ solutions to database queries, and sets of $k$ vertices in a weighted or unweighted graph or hypergraph which induce a sub(hyper)graph with specific properties. We present three concrete applications of \cref{thm:meta-informal}.

Our first application of \cref{thm:meta-informal} is to the \pp{Graph Motif} problem, introduced by Lacroix, Fernandes and Sagot \cite{lacroix06motif} in the practical context of metabolic networks. While we defer a detailed discussion to Section~\ref{sec:graphMotif}, it will be immediate from the definition that \pp{Graph Motif} is a uniform witness problem, and that \pp{Colourful-Graph Motif} reduces to \pp{Graph Motif} (see \cref{lem:colourful-motif}); hence Theorem~\ref{thm:meta-informal} combined with the best-known algorithm for \pp{Graph Motif} yields an improved approximate counting algorithm (see \cref{cor:motif-algo}).

Our second application of \cref{thm:meta-informal} is to induced subgraph counting. Many problems in this area are special cases of \pp{Induced Subgraph With Property}($\Phi$), abbreviated to ISWP($\Phi$). This problem asks whether a given graph $G$ contains a $k$-vertex induced subgraph with the given property $\Phi$; the problem is then parameterised by $k$. For example, if $\Phi$ is the family of all cliques, then ISWP($\Phi$) is simply $k$-\pp{Clique}. There is also a multi-coloured variant of the problem, MISWP($\Phi$), in which every vertex in $G$ receives one of $k$ colours, and we seek a $k$-vertex induced subgraph with property $\Phi$ and one vertex of each colour. See~\cite{meeksunbddtw} for a survey of results on exact and approximate \#ISWP($\Phi$) and \#MISWP($\Phi$), and~\cite{fockeroth} for a more recent complexity classification of \#ISWP($\Phi$) whenever $\Phi$ is a hereditary property. Observe that ISWP is a uniform witness problem, where the witnesses are the $k$-vertex subsets of $V(G)$ which induce copies of graphs satisfying $\Phi$, and \pp{Colourful}-ISWP($\Phi$) is simply MISWP($\Phi$); hence Theorem~\ref{thm:meta-informal} immediately implies that there is an FPTRAS for \#MISWP($\Phi$) and \#ISWP($\Phi$) whenever there is an FPT decision algorithm for MISWP($\Phi$), and moreover that the running times of these algorithms are the same up to a sub-polynomial factor in the instance size. This improves on the best previously-known result (due to Meeks~\cite[Corollaries 4.8 and 4.10]{meeksenum}) in two ways: firstly, in~\cite{meeksenum} the result is proved only for properties $\Phi$ that are preserved under adding edges; and secondly, in~\cite{meeksenum} the FPTRAS is slower than the FPT decision algorithm by a polynomial factor.

Our third application of \cref{thm:meta-informal} is to the problems \pp{Colourful}-$H$ and \pp{Weighted}-$H$ for a given $k$-vertex graph $H$. In \pp{Colourful}-$H$, we are given a graph~$G$ whose vertices are coloured with $k$ colours, and we wish to decide whether~$G$ contains a subgraph copy of $H$ in which every colour is represented. In \pp{Weighted}-$H$, we are given an edge-weighted graph $G$, and we wish to decide whether $G$ contains a subgraph copy of $H$ with total weight zero. Unlike our other two applications, the application of \cref{thm:meta-informal} in this setting is not straightforward. The natural way to frame these problems as uniform witness problems would be, as with ISWP, to take the witnesses to be vertex sets which induce a copy of~$H$; however, the number of witnesses would then not in general agree with the number of colourful or zero-weighted copies of~$H$ as required by \cref{def:uwp}(i). Nevertheless, we are able to use \cref{thm:meta-informal} to give a fine-grained reduction from approximate \#\pp{Colourful}-$H$ to \pp{Colourful}-$H$ and from approximate \#\pp{Weighted}-$H$ to \pp{Weighted}-$H$ for all graphs $H$; see \cref{cor:colourful} and \cref{cor:exact-weight-H} for details.

One reason this third application is interesting is that it throws an existing research question into a new light. In \pp{Uncoloured}-$H$, we are given a graph~$G$, and we wish to decide whether~$G$ contains a subgraph copy of~$H$. There are easy and well-known fine-grained reductions from \pp{Uncoloured}-$H$ to \pp{Colourful}-$H$ and from approximate \#\pp{Uncoloured}-$H$ to approximate \#\pp{Colourful}-$H$, via colour coding~\cite{alon95colorcoding,alon10balancedhash}. For some graphs~$H$, such as cliques, there are also simple reductions in the other direction. However, a full proof of equivalence would imply the long-standing dichotomy conjecture for the parameterised embedding problem (see \cite{chen17grids} for recent progress on this conjecture). \cref{cor:colourful} shows that this question is strongly linked to the question of when there is a fine-grained FPT reduction from \pp{Colourful}-$H$ to \pp{Uncoloured}-$H$, and we hope this will spur further research in the area.

\textbf{Organisation.} In the following section, we set out our notation and quote some standard probabilistic results for future reference. We then prove \cref{thm:main-counting} in \cref{sec:main-algo}, using a weaker approximation algorithm which we set out in \cref{sec:coarse}. We then prove \cref{thm:main-sampling} (using \cref{thm:main-counting}) in \cref{sec:sampling}. Finally, we prove our assorted corollaries in \cref{sec:corols}; we emphasise that in general, the proofs in this section are easy and use only standard techniques. 

\section{Preliminaries}\label{sec:notation}
\subsection{Notation}\label{sec:actual-notation}

Let $k \ge 2$ and let $G = (V,E)$ be a $k$-hypergraph, so that each edge in $E$ has size exactly $k$. We write $e(G) = |E|$. For all $U \subseteq V$, we write $G[U]$ for the subgraph induced by $U$. For all $S \subseteq V$, we write $d_G(S) = |\{e \in E(G) \colon S \subseteq e\}|$ for the degree of $S$ in $G$. If $S = \{v_1, \dots, v_{|S|}\}$, then we will sometimes write $d_G(v_1,\dots,v_{|S|}) = d_G(S)$.

For all positive integers $t$, we write $[t] = \{1, \dots, t\}$. We write $\ln$ for the natural logarithm, and $\log$ for the base-2 logarithm. Given real numbers $x,y \ge 0$ and $0 < \eps < 1$, we say that $x$ is an \emph{$\eps$-approximation} to $y$ if $(1-\eps)x < y < (1+\eps)x$, and write $y \in (1\pm \eps)x$. We extend this notation to other operations in the natural way, so that (for example) $y \in xe^{\pm \eps}/(2\mp \eps)$ means that $xe^{-\eps}/(2+\eps) \le y \le xe^\eps/(2-\eps)$.

When stating quantitative bounds on running times of algorithms, we assume the standard randomised word-RAM machine model with logarithmic-sized words; thus given an input of size $N$, we can perform arithmetic operations on $\OO(\log N)$-bit words and generate uniformly random $\OO(\log N)$-bit words in $\OO(1)$ time.

Recall the definitions of $\calC(G)$ and the colourful independence oracle of $G$, and colourful oracle access from \cref{sec:framework-intro}. Note that for all $X \subseteq V(G)$, $\textup{cIND}_{G[X]}$ is a restriction of \cindora{}. Thus, an algorithm with colourful oracle access to $G$ can safely call a subroutine that requires colourful oracle access to $G[X]$.

\subsection{Probabilistic results}
  
We use some standard results from probability theory, which we collate here for reference. The following lemma is commonly known as Hoeffding's inequality.
\begin{lemma}[{\cite[Theorem~2.8]{BGM}}]\label{lem:hoeffding}
 	Let $X_1,\dots,X_m$ be independent real random variables, and suppose there exist $a_1,\dots,a_m, b_1,\dots,b_m \in \R$ be such that $X_i \in [a_i,b_i]$ with probability 1. Let $X = \sum_{i=1}^m X_i$. Then for all $t \ge 0$, we have
 	\[
 		\pr\big(|X-\E(X)| \ge t\big) \le 2e^{-2t^2/\sum_{i=1}^m (b_i-a_i)^2}.
 	\]
\end{lemma}

The next lemma is a form of Bernstein's inequality.
\begin{lemma}\label{lem:bernstein}
 	Let $X_1, \dots, X_k$ be independent real random variables. Suppose there exist $\nu$ and $M$ such that with probability 1, $\sum_i \E(X_i^2) \le \nu$ and $|X_i| \le M$ for all $i \in [k]$. Let $X = \sum_{i=1}^k X_i$. Then for all $z \ge 0$, we have
 	\[
 		\pr\big(|X-\E(X)| \ge z\big) \le 2\exp\Big({-}\frac{3z^2}{6\nu + 2Mz}\Big).
 	\]
\end{lemma}
\begin{proof}
 	Apply~\cite[Corollary~2.11]{BGM} to both $X$ and ${-}X$, taking $c = M/3$ and $t = z$, then apply a union bound.
\end{proof}

The next lemma collates two standard Chernoff bounds.

\begin{lemma}[{\cite[Corollaries~2.3-2.4]{JLR}}]\label{lem:chernoff}
	Suppose $X$ is a binomial or hypergeometric random variable with mean $\mu$. Then:
\begin{enumerate}[(i)]
  \item for all $0 < \eps \le 3/2$, $\pr(|X - \mu| \ge \eps\mu) \le 2e^{-\eps^2\mu/3}$;
  \item for all $t \ge 7\mu$, $\pr(X \ge t) \le e^{-t}$.\qed
\end{enumerate}
\end{lemma}

The next lemma is a standard algebraic bound.

\begin{lemma}\label{lem:bin-bound}
  Let $N,k\in\Z_{>0}$. If $N \ge 2k^2$, then $\binom{2N-k}{N-k}/\binom{2N}{N} \ge 2^{-k-1}$.
\end{lemma}
\begin{proof}
	We have
	\begin{align*}
	\binom{2N-k}{N-k}\Big/\binom{2N}{N} &= \frac{(2N-k)!N!}{(2N)!(N-k)!} = \prod_{i=0}^{k-1} (N-i) \Big/ \prod_{j=0}^{k-1} (2N-j)\\
	&\ge \Big(\frac{N-k+1}{2N-k+1}\Big)^k = \Big(\frac{1}{2} - \frac{k-1}{2(2N-k+1)}\Big)^k\\
	&\ge 2^{-k}\Big(1 - \frac{k}{N}\Big)^k \ge 2^{-k}\Big(1 - \frac{k^2}{N}\Big) \ge 2^{-k-1}.
	\end{align*}
\end{proof}	

We now prove a technical lemma, which should be read as follows. We are given the ability to sample from bounded probability distributions $\calD_1,\dots,\calD_q$ on $[0,\infty)$. We wish to estimate the sum of their means using as few samples as possible, and we are given access to a crude estimate of the mean of each $\calD_i$ with multiplicative error $b$ (for ``bias''). \cref{lem:dist-samp-2} says that we can do so to within relative error $\xi$, with failure probability at most $\delta$, by sampling $t_i$ times from $\calD_i$ for each $i \in [q]$. This lemma will be important for reducing the running time of our algorithm; see the sketch proof in section~\ref{sec:sketch} for more details.

\begin{lemma}\label{lem:dist-samp-2}
	Let $0 < \xi,\delta < 1$, let $b \ge 1$, and let $M_1,\dots,M_q > 0$. For all $i \in [q]$, let $\calD_i$ be a probability distribution on $[0,M_i]$ with mean $\mu_i$. For all $i \in [q]$, let $\hmu_i$ satisfy $0 < \hmu_i \le \mu_i b$, and let 
	\[
		t_i = \bigg\lceil\frac{4bM_i\log(2/\delta)}{\xi^2\sum_\ell\hmu_\ell}\bigg\rceil.
	\]
	Let $\{X_{i,j} \colon i \in [q], j \in [t_i]\}$ be independent random variables with $X_{i,j} \sim \calD_i$. Then with probability at least $1-\delta$,
	\[
		\sum_{i=1}^q\sum_{j=1}^{t_i}\frac{X_{i,j}}{t_i} \in (1\pm\xi)\sum_{i=1}^q\mu_i.
	\]
\end{lemma}
Note that while \cref{lem:dist-samp-2} does not require a lower bound on $\hmu_1,\dots,\hmu_q$, without one it is useless as $\sum_i t_i$ may be arbitrarily large. When we apply \cref{lem:dist-samp-2}, we will do so with $\mu_i/b \le \hmu_i \le \mu_ib$ for all $i \in [q]$.
\begin{proof}
	We will apply a form of Bernstein's inequality (\cref{lem:bernstein}). Let
	\[
		X = \sum_{i=1}^q\sum_{j=1}^{t_i}\frac{X_{i,j}}{t_i},\qquad x = \sum_{i=1}^q \mu_i.
	\]
	Thus, we seek to prove $\Pr(X \in (1\pm\xi)x) \ge 1-\delta$. Note that $\E(X) = x$, and that 
	\[
		\sum_{i=1}^q\sum_{j=1}^{t_i} \E\big((X_{i,j}/t_i)^2\big) 
		\le \sum_{i=1}^q\sum_{j=1}^{t_i}\frac{1}{t_i^2}\E(M_iX_{i,j}) 
		= \sum_{i=1}^q\frac{M_i\mu_i}{t_i}.
	\]
	Let $M = \max\{M_i/t_i \colon i \in [q]\}$, so that $X_{i,j}/t_i \le M$ for all $i,j$. Then by \cref{lem:bernstein}, applied to the variables $X$ and $X_{i,j}/t_i$ with $z = \xi x$, it follows that 
	\begin{align}\nonumber
		\Pr(|X-x|\ge\xi x) 
		&\le 2\exp\bigg({-}\frac{3\xi^2x^2}{6\sum_i\frac{M_i\mu_i}{t_i} + 2M\xi x}\bigg)\\
		\nonumber
		&\le 2\exp\bigg({-}\frac{3\xi^2x^2}{2\max\big\{6\sum_i\frac{M_i\mu_i}{t_i},\ 2M\xi x\big\}}\bigg)\\\label{eqn:dist-samp-2a}
		&= \max\bigg\{2\exp\Big({-}\frac{\xi^2 x^2}{4\sum_i\frac{M_i\mu_i}{t_i}}\Big),\ 2\exp\Big({-}\frac{3\xi x}{4M}\Big) \bigg\}.
	\end{align}
	
	We now bound the exponents of each term in the max. By our choice of $t_i$'s, we have
	\[
		\frac{\xi^2x^2}{4\sum_i\frac{M_i\mu_i}{t_i}} \ge \xi^2x^2\Big/\bigg(4\sum_iM_i\mu_i\cdot\frac{\xi^2\sum_j\hmu_j}{4bM_i\log(2/\delta)}\bigg) = \frac{bx^2\log(2/\delta)}{\sum_i \mu_i \cdot \sum_j \hmu_j} = \frac{bx\log(2/\delta)}{\sum_j\hmu_j}.
	\]
	Since $\hmu_i \le \mu_ib$ for all $i$, we have $x \ge \sum_j\hmu_j/b$, so
	\begin{equation}\label{eqn:dist-samp-2b}
		\frac{\xi^2x^2}{4\sum_i\frac{M_i\mu_i}{t_i}} \ge \log(2/\delta) \ge \ln(2/\delta).
	\end{equation}
	Moreover, again by our choice of $t_i$'s we have
	\[
		M=\max\Big\{\frac{M_i}{t_i} \colon i \in [q] \Big\} \le \max\Big\{\frac{\xi^2\sum_j \hmu_j}{4b\log(2/\delta)} \colon i \in [q] \Big\} \le \frac{\xi^2x}{4\log(2/\delta)},
	\]
	so 
	\begin{equation}\label{eqn:dist-samp-2c}
		\frac{3\xi x}{4M} \ge \frac{3\log(2/\delta)}{\xi} > \ln(2/\delta).
	\end{equation}
	The result therefore follows from~\eqref{eqn:dist-samp-2a},~\eqref{eqn:dist-samp-2b} and~\eqref{eqn:dist-samp-2c}.
\end{proof}

\section{The main algorithm}\label{sec:main}

In this section we prove our main approximate counting result, \cref{thm:main-counting}. We will make use of an algorithm with a weaker approximation guarantee, whose properties are stated in \cref{lem:acc}; we will prove this lemma in \cref{sec:coarse}.

\subsection{Sketch proof}\label{sec:sketch}

Let $G$ be the input $n$-vertex $k$-hypergraph and let $\eps$ be the input error tolerance, so that we wish to find an $\eps$-approximation of $e(G)$. The overall aim of our algorithm is to iteratively construct a list $L$ of random induced sub-hypergraphs $G[X_1],\dots,G[X_t]$, together with associated weights $w_1,\dots,w_t$, such that with high probability:
\begin{enumerate}
    \item[(A1)] $\sum_i w_i\cdot e(G[X_i])$ is an $\eps$-approximation to the total number of edges in $G$; and
    \item[(A2)] each sub-hypergraph $G[X_i]$ has few enough edges so that we can efficiently compute $e(G[X_i])$ exactly by brute force using the coloured independence oracle; and
    \item[(A3)] the list $L$ is short enough so that its length does not significantly affect the running time.
\end{enumerate}
Once a list~$L$ satisfying (A1)--(A3) is constructed, we can efficiently approximate~$e(G)$ by computing each $e(G[X_i])$ using (A2) and taking the weighted sum $\sum_i w_i e(G[X_i])$.

\subsubsection*{Constructing the list \boldmath$L$}
We represent $L$ as a list of pairs $(w_i, X_i)$. Initially~$L$ contains the single pair $(1,V(G))$, which naturally satisfies (A1). In order to make progress towards a list satisfying~(A2) and~(A3), we repeatedly modify it in two ways:

\medskip
\begin{enumerate}
    \item[\texttt{\halve}:] To make the sub-hypergraphs smaller, we replace each hypergraph $G[X]$ in the list~$L$ by several random sub-hypergraphs $G[X_{i}]$ induced by half of the vertices of~$X$.
    That is, each pair $(w, X)$ is replaced by $(w_1, X_1),\allowbreak\dots, (w_{t}, X_{t})$ where $|X_{i}| = |X|/2$ holds for all $i$. The weights $w_{i}$ are set to maintain (A1) in expectation. This step gets us closer to achieving (A2) and farther from achieving~(A3) while maintaining (A1).
    \item[\texttt{\impsamp}:] To make the list~$L$ shorter, we discard randomly-selected elements from it and re-weight the remaining elements to maintain (A1) in expectation. This step gets us closer to achieving (A3) while maintaining (A1) and (A2).
\end{enumerate}

\medskip\noindent
We alternate between applying \halve and \impsamp to the list until all of (A1)--(A3) are satisfied, at which point we compute and return $\sum_i w_i e(G[X_i])$.

The overall structure of our algorithm for \cref{thm:main-counting} is very similar to that of Beame et al.~\cite{beame20}.
The \texttt{Cleanup} step of~\cite{beame20} roughly corresponds to the calculation of the weighted sum at the end, and is similar.
However, our implementation of \halve and \impsamp differs in two major ways. Firstly, a key component of both \halve and \impsamp is the ability to compute a coarse estimate of the number of edges in each graph $G[X]$, with multiplicative error $\log^{\Theta(k)} n$, and the technique for this used in~\cite{beame20} does not generalise easily to hypergraphs. We provide such an algorithm as \acc and state its properties as~\cref{lem:acc}, but we defer both the sketch proof and the full proof of this lemma to Section~\ref{sec:coarse}. Secondly, by better incorporating these estimates into \halve and \impsamp{}, we improve the running time's dependence on $\eps$ from~$\eps^{-4}$ to~$\eps^{-2}$. 
In order to expand on these differences, we now sketch our implementation of \halve and \impsamp given \acc{}.

\subsubsection*{Sketch of \texttt{\textbf{Reduce}}}
Suppose we have a long list $L$ of pairs $(w,X)$ satisfying~(A1), and we wish to make it shorter by randomly discarding and re-weighting elements while maintaining (A1). (In fact, we start by maintaining something a little stronger than (A1) since the quality of approximation will degrade as the algorithm runs, but we ignore this subtlety in our sketch.) We do this by exploiting a statistical technique called importance sampling, previously applied to the $k=2$ case by Beame et al.~\cite{beame20}. The idea is to use the coarse estimates of each $e(G[X])$ given by \acc to divide the pairs $(w,X) \in L$ into ``bins'', with one bin for pairs with $w e(G[X]) \in [1,2)$, one bin for pairs with $w e(G[X]) \in [2,4)$, one bin for pairs with $w e(G[X]) \in [4,8)$, and so on for a total of $\OO(k\log n)$ bins. We then choose a suitably large $t$ and sample $t$ pairs from each bin uniformly at random (or take the whole bin if it contains less than $t$ pairs), then re-weight each element to ensure that~(A1) is maintained in expectation; we can then prove concentration via e.g.\ Hoeffding's inequality (\cref{lem:hoeffding}) to conclude that~(A1) is maintained with high probability. 

The larger we take $t$ to be in \impsamp, the fewer pairs we can discard from $L$; thus to minimise our running time, we should take $t$ to be as small as possible while still being able to apply Hoeffding's inequality, and this is essentially the approach of~\cite{beame20}. The key difference in our approach is that we make use of \cref{lem:dist-samp-2}, allowing $t$ to vary between bins. If our estimates from \acc say that pairs $(w,X)$ in some bin account for about half of the quantity $\sum_{(w,X) \in L} w e(G[X])$ we wish to preserve, then we will take about half our samples from that bin, and correspondingly fewer from the other bins. This allows our algorithm for \impsamp to maintain a substantially shorter list than the algorithm used in~\cite{beame20}, and this is one of the two reasons for our improved running time.

\subsubsection*{Sketch of \texttt{\textbf{Refine}}}
In~\cite{beame20}, \halve was implemented by keeping the number of vertices the same but halving the number of edges using random colourings.
Our implementation is different and works as follows. Suppose we have a list $L$ of pairs~$(w,X)$ satisfying~(A1), and we wish to halve the size of all sets~$X$ while maintaining~(A1). Then for each $X$ and any integer $t \ge 1$, we independently choose $t$ uniformly random subsets $X_1,\dots,X_t \subseteq X$ subject to $|X_i|=|X|/2$ for all $i$. It is not hard to show using \cref{lem:bin-bound} that $\E(e(G[X_i])) \approx e(G)/2^k$ holds for all $i$. Thus, using Hoeffding's inequality for large enough~$t$, we can show that the total number of edges $\sum_{i=1}^t e(G[X_i])$ is concentrated around its mean of roughly $te(G)/2^k$. With high probability, we now have $(2^k/t)\sum_{i=1}^t e(G[X_i]) \approx e(G)$, so that replacing $(w,X)$ with $\{(2^kw/t, X_i) \colon i \in [t]\}$ maintains~(A1). Similarly to \impsamp{}, we use \cref{lem:dist-samp-2} along with our estimates from \acc to tailor the value of $t$ to each pair $(w,X) \in L$ and avoid making the list too long in a single iteration of \halve{}, and this is the second reason for our improved running time.

\subsubsection*{Organisation}
The remainder of this section is organised as follows.  We begin by setting out the precise format of our list $L$ in \cref{def:L}, which (unlike our sketch) also stores the estimates of edge counts from \acc{}. We then state the pruning algorithm \impsamp and prove its correctness in \cref{lem:impsamp}. After this, we state the behaviour we expect from our coarse counting subroutine \acc in \cref{lem:acc} (deferring the proof to \cref{sec:coarse}), state the expansion algorithm \halve using \acc as a subroutine, and prove its correctness in \cref{lem:halve}. We then state \approxUncol{}, which is essentially our main algorithm, and prove its correctness in \cref{lem:uncol-count}. Finally, we prove \cref{thm:main-counting}. The purpose of \approxUncol is to move some uninteresting bookkeeping details out of the main proof for readability; namely, the differences between \approxUncol and \aau are that \approxUncol requires the number of vertices $n$ to be a power of 2, has a constant failure probability (rather than $\delta$), and could (with vanishing probability) run more slowly than expected.

\subsection{The main algorithm}\label{sec:main-algo}

Recall from our sketch proof in \cref{sec:sketch} that our algorithm will maintain a weighted list $L$ of induced subgraphs of steadily decreasing size. For convenience, we will also include coarse estimates of the edge count of each graph in $L$. Rather than set out the format of this list each time we use it, we define it formally now.

\begin{definition}\label{def:L}
	Let $G$ be a hypergraph, let $i>0$ be an integer, and let $b \ge 1$ be rational. Then a \emph{$(G,b,y)$-list} is a list of triples $(w,S,\hat{e})$ such that $w$ and $\hat{e}$ are positive rational numbers, $S \subseteq V(G)$ with $|S|=2^y$, and $\hat{e}/b \le e(G[S]) \le \hat{e}b$. For any $(G,b,y)$-list $L$, we define
	\begin{align*}
		Z(L) &:= \sum_{(w,S,\hat{e})\in L}we(G[S]).
	\end{align*}
\end{definition}

Initially, we will take $L = ((1,V(G),\hat{e}))$ where $\hat{e}/b \le e(G) \le \hat{e}b$, so that $Z(L) = e(G)$. As the algorithm progresses, $Z(L)$ will remain a good approximation to $e(G)$, and eventually we will be able to compute it efficiently. We are now ready to set out our importance sampling algorithm, \impsamp, which we will use to keep the length of $L$ small.

\begin{framed}
	\noindent
	\textbf{Algorithm $\impsamp(G,b,y,L,\xi,\delta)$.}
	
 	\smallskip\noindent\textbf{Input:} $G$ is an $n$-vertex $k$-hypergraph, where $n$ is a power of 2, to which \impsamp has (only) colourful oracle access. $b$ is a rational number with $b \ge 1$, and $y$ is a positive integer. $L$ is a $(G,b,y)$-list with $1/2 \le Z(L) \le 2n^k$ and $|L| \le n^{11k}$. $\delta$ is a rational number with $0 < \delta < 1$, and $\xi$ is a rational number with $n^{-2k} \le \xi < 1$. 
 	
	\smallskip\noindent\textbf{Behaviour:} $\impsamp(G,b,y,L,\xi,\delta)$ outputs a $(G,b,y)$-list $L'$ satisfying the following \mbox{properties}.
	\begin{enumerate}[(a)]
		\item $|L'| \le 33k\log(4nb) + 32b^2\log(2/\delta)/\xi^2$.
		\item With probability at least $1-\delta$, we have $Z(L') \in (1\pm\xi)Z(L)$.
	\end{enumerate}
	\shortrule 
	\begin{enumerate}[({T}1)]
		\item Calculate $a \leftarrow \floor{15k\log (4nb)}+1$ and 
		\begin{align*}
			L_i &\leftarrow \{(w,S,\hat{e}) \in L \colon 2^{i-1} \le w\hat{e} < 2^i\}\mbox{ for each }{-}a \le i \le a.
		\end{align*}
		\textit{(We will show that every significant entry of $L$ will be contained in exactly one $L_i$, and entries $(w,S,\hat{e}) \in L_i$ satisfy $w\hat{e} \approx 2^i$.)}
		\item For each ${-}a \le i \le a$, calculate
		\[
			t_i \leftarrow \Big\lceil\frac{16b^22^i|L_i|\log(2/\delta)}{\xi^2W} \Big\rceil,\mbox{ where } W := \sum_{(w,S,\hat{e}) \in L} w\hat{e}.
		\]
		\item For each ${-}a \le i \le a$, calculate a multiset $L_i'$ as follows. If $|L_i| \le t_i$, let $L_i' \leftarrow L_i$. Otherwise, sample $t_i$ entries $(w_{i,1},S_{i,1},\hat{e}_{i,1}),\allowbreak \dots, (w_{i,t_i},S_{i,t_i},\hat{e}_{i,t_i})$ from $L_i$ independently and uniformly at random, let $w_{i,j}' \leftarrow w_{i,j}|L_i|/t_i$, and let $L_i' \leftarrow \{(w_{i,j}',S_{i,j},\hat{e}_{i,j}) \colon j \in [t_i]\}$.
		\item Form $L'$ by concatenating the multisets $\{L_i' \colon {-}a \le i \le a\}$ in an arbitrary order, and return~$L'$.
	\end{enumerate}
\end{framed}

The algorithm \impsamp improves significantly on the summation reduction algorithm of~\cite[Lemma~2.5]{beame20}, which in our notation outputs a list of length \[\Omega(kb^4\log(\tfrac{1}{\delta})\log(nb)/\xi^2)\] compared to our length of $O(k\log(nb) + b^2\log(\tfrac{1}{\delta})/\xi^2)$. We obtain this improvement by defining the number $t_i$ of elements to sample from each ``bin'' $L_i$ of tuples in creating~$L_i'$ to depend on the approximate value $w\hat{e} \approx 2^i$ of tuples in $L_i$. By contrast,~\cite[Lemma~2.5]{beame20} defined a single threshold $\alpha$ and sampled $\min\{\alpha, |L_i|\}$ tuples from each~$L_i$.

\begin{lemma}\label{lem:impsamp}
	$\impsamp(G,b,y,L,\xi,\delta)$ behaves as claimed above, has running time $\OO(|L|k^3\log(nb/\delta))$, and does not invoke \cindora.
\end{lemma}
\begin{proof}
	\textbf{Running time.}
	It is clear that $\impsamp(G,b,y,L,\xi,\delta)$ does not invoke \cindora. Recall that we work with the word-RAM model, so we can carry out elementary arithmetic operations on $\OO(k\log n)$-bit numbers in $\OO(k^2)$ time. Thus step (T1) takes time $\OO(k^2a|L|)$, and step (T2) takes time $\OO(k^2a(\log(1/\delta)+|L|))$. Since $|L_i'| = \min\{|L_i|,t_i\}$, steps (T3) and (T4) take time $\OO(k^2a|L|)$.	The required bounds follow.
	
	\medskip\noindent\textbf{Correctness.}
	In forming $L'$ from $L$, $\impsamp$ only updates the first elements (i.e.\ the weights) of entries of $L$; since $L$ is a $(G,b,y)$-list, and the definition of a $(G,b,y)$-list does not depend on these weights, $L'$ is also a $(G,b,y)$-list. We next prove (a). We have
	\begin{align}\nonumber
		|L'| &= \sum_{|i| \le a} |L_i'| \le \sum_{|i| \le a} t_i 
		\le \sum_{|i| \le a} \Big(1 + \frac{16b^22^i|L_i|\log(2/\delta)}{\xi^2W}\Big)\\\label{IS-1}
		&= 2a + 1 + \frac{16b^2\log(2/\delta)}{\xi^2W}\sum_{|i| \le a}2^i|L_i|.
	\end{align}
	Recall from the definition of $L_i$ that, for all $(w,S,\hat{e}) \in L_i$, we have $w\hat{e} \ge 2^{i-1}$, so
	\[
		\sum_{|i|\le a}2^i|L_i| 
		\le \sum_{|i|\le a}\Big(2\sum_{(w,S,\hat{e}) \in L_i}w\hat{e}\Big) 
		= 2W.
	\]
	It therefore follows from~\eqref{IS-1} that $|L'| \le 2a + 1 + 32b^2\log(2/\delta)/\xi^2$, so property (a) holds. 
	
	It remains to prove property (b). This will follow easily from \cref{lem:dist-samp-2}. Before we can apply it, however, we must set our notation and prove the conditions of the lemma hold. Let
	\[
		\calI := \{{-}a \le i \le a \colon |L_i| > t_i\};
	\]
	thus in step (T3) of \impsamp, for each $i \notin \calI$ we choose $L_i' = L_i$, and for each $i \in \calI$ we choose $L_i'$ by sampling $t_i$ elements $(w_{i,j},S_{i,j},\hat{e}_{i,j})$ from $L_i$.
	For each $i \in \calI$ and $j \in [t_i]$, let 
	\begin{alignat*}{4}
		X_{i,j} &:= w_{i,j} e(G[S_{i,j}])|L_i|,\qquad 
		&M_i &:= 2^ib|L_i|,\\
		\mu_i &:= \sum_{(w,S,\hat{e}) \in L_i} we(G[S]),\qquad
		&\hmu_i &:= \sum_{(w,S,\hat{e}) \in L_i} w\hat{e}.
	\end{alignat*}
	For all $i$ and $j$, it is clear that $X_{i,j} \ge 0$. Moreover, since $L$ is a $(G,b,y)$-list and $(w_{i,j},S_{i,j},\hat{e}_{i,j}) \in L$, we have $e(G[S_{i,j}]) \le b\hat{e}_{i,j}$; thus by the definitions of $L_i$ and $X_{i,j}$ we have 
	\[
		X_{i,j} \le bw_{i,j}\hat{e}_{i,j}|L_i| \le 2^ib|L_i| = M_i.
	\]
	It is also true that $\E(X_{i,j}) = \mu_i$, that $0 \le \hmu_i \le \mu_ib$, that the $X_{i,j}$'s are independent, and that
	\[
		\Big\lceil \frac{4bM_i\log(2/\delta)}{(\xi/2)^2\sum_\ell \hmu_\ell} \Big\rceil = \Big\lceil \frac{16b^22^i|L_i|\log(2/\delta)}{\xi^2W} \Big\rceil = t_i.
	\]
	It therefore follows from \cref{lem:dist-samp-2} that with probability at least $1-\delta$,
	\begin{equation}\label{eq:IS-2}
		\sum_{i\in \calI} \sum_{j=1}^{t_i} \frac{X_{i,j}}{t_i} \in (1\pm\xi/2)\sum_{i \in \calI}\mu_i.
	\end{equation}
	Suppose this event occurs; then we will show that $Z(L') \in (1\pm\xi)Z(L)$, as in (b).
	
	Plugging our definitions into~\eqref{eq:IS-2}, we see that
	\[
		\sum_{i\in \calI} \sum_{j=1}^{t_i} \frac{X_{i,j}}{t_i} = \sum_{i \in \calI}\sum_{j=1}^{t_i} \frac{w_{i,j}|L_i|}{t_i}e(G[S_{i,j}]) = \sum_{i \in \calI}\sum_{(w,S,\hat{e}) \in L_i'} we(G[S]),
	\]
	and
	\[
		\sum_{i \in \calI}\mu_i = \sum_{i\in \calI}\sum_{(w,S,\hat{e}) \in L_i} we(G[S]),
	\]
	so
	\[
		\sum_{i \in \calI}\sum_{(w,S,\hat{e}) \in L_i'} we(G[S])\in (1\pm\xi/2)\sum_{i\in \calI}\sum_{(w,S,\hat{e}) \in L_i} we(G[S]).
	\]
	We have $L_i' = L_i$ for all $i \in \{-a,\dots,a\}\setminus\calI$, so it follows that
	\begin{equation}\label{eqn:trim-1}
		\sum_{|i| \le a}\sum_{(w,S,\hat{e}) \in L_i'} we(G[S])\in (1\pm\xi/2)\sum_{|i| \le a}\sum_{(w,S,\hat{e}) \in L_i} we(G[S]).
	\end{equation}
	
	For all $(w,S,\hat{e}) \in L$, since $L$ is a $(G,b,y)$-list with $Z(L) \le 2n^k$, we have 
	\[
		w\hat{e} \le bwe(S) \le bZ(L) \le 2bn^k < 2^a.
	\]
	Thus, for all $(w,S,\hat{e}) \in L \setminus \bigcup_{|i|\le a} L_i$, we have $w\hat{e} \le 2^{-a}$. It follows from~\eqref{eqn:trim-1} that
	\begin{align*}
		Z(L') = \sum_{|i| \le a}\sum_{(w,S,\hat{e}) \in L_i'} we(G[S]) \in (1\pm\xi/2)Z(L) \pm 2^{-a}|L|.
	\end{align*}
	Observe that by hypothesis, $\xi Z(L)/2 \ge n^{-2k}/4$ and $2^{-a}|L| \le n^{-2k}/4$. Hence, $Z(L') \in (1\pm \xi)Z(L)$, as required.
\end{proof}

We next state the behaviour of our coarse approximate counting algorithm; we will prove the following lemma in \cref{sec:coarse}.

\begin{restatable}{restatable-lemma}{stateACC}\label{lem:acc}
	There is a randomised algorithm $\acc(G,\delta)$ with the following behaviour. Suppose $G$ is an $n$-vertex $k$-hypergraph to which \acc has (only) colourful oracle access, where $n$ is a power of two, and suppose $0 < \delta < 1$. Then in time $\OO(\log(1/\delta)k^{3k}n\log^{2k+2}n)$, and using at most $\OO(\log(1/\delta)k^{3k}\log^{2k+2}n)$ queries to \cindora, $\acc(G,\delta)$ outputs a rational number $\hat{e}$. Moreover, with probability at least $1-\delta$, 
	\[
	\frac{e(G)}{2(4k\log n)^k} \le \hat{e} \le e(G)\cdot 2(4k\log n)^k.
	\]
\end{restatable}

Using \acc, we now describe our algorithm for turning our $(G,b,y)$-list $L$ into a $(G,b,y-1)$-list $L'$ with $Z(L') \approx Z(L)$.

\begin{framed}
	\noindent
	\textbf{Algorithm $\halve(G,b,y,L,\xi,\delta)$.}
 	
 	\smallskip\noindent
	\textbf{Input:} $G$ is an $n$-vertex $k$-hypergraph, where $n$ is a power of 2, to which \halve has (only) colourful oracle access. $b$ is a rational number with $b \ge 2(4k\log n)^k$, and $y$ is a positive integer with $2^{y-1} \ge 2k^2$. $L$ is a $(G,b,y)$-list. $\xi$ and $\delta$ are rational numbers with $0 < \xi,\,\delta < 1$.
	
	\smallskip\noindent
	\textbf{Behaviour:} $\halve(G,b,y,L,\xi,\delta)$ outputs a list $L'$, which satisfies the following properties with probability at least $1-\delta$.
	\begin{enumerate}[(a)]
		\item $L'$ is a $(G,b,y-1)$-list.
		\item $|L'| \le |L| + 2^{k+3}b^2\log(4/\delta)/\xi^2$.
		\item $Z(L') \in (1\pm\xi)Z(L)$.
	\end{enumerate} 
	\shortrule
	\begin{enumerate}[(H1)]
		\item Write $L =: \{(w_i,S_i,\hat{e}_i) \colon 1\le i \le |L|\}$. Calculate
		\[
			p \leftarrow \binom{2^y-k}{2^{y-1}-k}\Big/\binom{2^y}{2^{y-1}},\qquad\qquad W \leftarrow \sum_{i=1}^{|L|} w_i\hat{e}_i,
		\]
		\[
			\mbox{and }t_i \leftarrow \Big\lceil\frac{4b^2w_i\hat{e}_i\log(4/\delta)}{p\xi^2W} \Big\rceil \mbox{ for all }1 \le i \le |L|.
		\]
		\item For all $1 \le i \le |L|$, sample subsets $S_{i,1}, \dots, S_{i,t_i} \subseteq S_i$ independently and uniformly at random subject to $|S_{i,j}| = 2^{y-1}$. Then calculate $w_i' \leftarrow w_i/pt_i$ and 
		\[	
			L_i' \leftarrow \Big\{\Big(w_i', S_{i,j}, \acc\big(G[S_{i,j}],\delta/(2\textstyle{\sum_it_i})\big)\Big) \colon 1 \le i \le |L|,\ j \in [t_i] \Big\}.
		\]
		\item Form $L'$ by concatenating the multisets $\{L_i' \colon 1 \le i \le |L|\}$ in an arbitrary order and removing any entries $(w,S,\hat{e})$ with $\hat{e} = 0$, and return~$L'$.
	\end{enumerate}	
\end{framed}

\begin{lemma}\label{lem:halve}
	$\halve(G,b,y,L,\xi,\delta)$ behaves as claimed above. Moreover, writing 
	\[
		\lambda = |L| + \frac{2^kb^2\log(1/\delta)}{\xi^2},\qquad\qquad T = \lambda\log(\lambda/\delta)k^{3k}\log^{2k+2}n,
	\]
	$\halve(G,b,y,L,\xi,\delta)$ has running time $\OO(nT)$ and invokes \cindora at most $\OO(T)$ times.
\end{lemma}
\begin{proof}
	\textbf{Running time.} The running time and oracle usage are both dominated by the invocations of \acc in step (H2). We first bound the number $\sum_i t_i$ of such invocations. We have
	\begin{equation}\label{eqn:halve-tis}
		\sum_{i=1}^{|L|} t_i 
		\le |L| + \sum_{i=1}^{|L|} \frac{4b^2w_i\hat{e}_i\log(4/\delta)}{p\xi^2W} 
		= |L| + \frac{4b^2\log(4/\delta)}{p\xi^2}.
	\end{equation}
	Since $2^{y-1} \ge 2k^2$, by a standard binomial coefficient bound (\cref{lem:bin-bound}), we have $p \ge 2^{-k-1}$. Thus,~\eqref{eqn:halve-tis} implies
	\begin{equation}\label{eqn:halve-tis-2}
		\sum_{i=1}^{|L|} t_i \le |L| + \frac{2^{k+3}b^2\log(4/\delta)}{\xi^2} = \Theta(\lambda).
	\end{equation}
	By \cref{lem:acc}, writing $T' = \log(\lambda/\delta)k^{3k}\log^{2k+2}2^{y-1}$, \acc has running time $\OO(2^{y-1}T')$ and invokes \cindora $\OO(T')$ times. Since $L$ is a $(G,b,y)$-list, we have $2^{y-1} \le n$ and so the claimed bounds follow from~\eqref{eqn:halve-tis-2}.
	
	\medskip\noindent\textbf{Correctness.} Let $\calE_1$ be the event that $Z(L') \in (1\pm\xi)Z(L)$, and let $\calE_2$ be the event that every invocation of $\acc$ in step (H2) succeeds. We will show that $\Pr(\calE_1 \cap \calE_2) \ge 1-\delta$, and that properties (a)--(c) hold whenever $\calE_1 \cap \calE_2$ occurs.
	
	\textbf{Bounding $\boldsymbol{\Pr(\calE_1 \cap \calE_2)}$:} To bound $\Pr(\calE_1)$ below, we will apply \cref{lem:dist-samp-2}. We first explain the purpose of $p$, as defined in \halve. Since $L$ is a $(G,b,y)$-list, we have $|S_i|=2^y$ and $|S_{i-1}|=2^{y-1}$, so there are $\binom{2^y}{2^{y-1}}$ possible choices of each set $S_{i,j}$. Moreover, for any given edge $e \in E(G[S_i])$, there are $\binom{2^y-k}{2^{y-1}-k}$ possible choices of $S_{i,j}$ containing $e$. It follows that any given edge of $G[S_i]$ is present in $S_{i-1}$ with probability precisely $p$. 
	
	We now set up our notation and show that the relevant assumptions hold in order to apply \cref{lem:dist-samp-2}. For all $1 \le i \le |L|$ and all $j \in [t_i]$,~let
	\begin{alignat*}{4}
		X_{i,j} &:= w_i e(G[S_{i,j}])/p,\qquad & M_i &:= w_i\hat{e}_ib/p,\\
		\mu_i &:= w_ie(G[S_i]),\qquad& \hmu_i &:= w_i\hat{e}_i.
	\end{alignat*}
	For all $i$ and $j$, it is clear that $X_{i,j} \ge 0$. Moreover, since $(w_i,S_i,\hat{e}_i) \in L$, we have $e(G[S_{i,j}]) \le e(G[S_i]) \le \hat{e}_ib$, so $X_{i,j} \le M_i$. Since $p$ is the probability that any given edge in $G[S_i]$ survives in $G[S_{i,j}]$, we have $\mu_i = \E(X_{i,j})$. The $X_{i,j}$'s are independent, we have $0 \le \hmu_i \le \mu_ib$, and we have
	\[
		\Big\lceil \frac{4bM_i\log(4/\delta)}{\xi^2\sum_\ell\hmu_\ell}\Big\rceil = \Big\lceil\frac{4b^2w_i\hat{e}_i\log(4/\delta)}{p\xi^2W} \Big\rceil = t_i.
	\]
	It therefore follows from \cref{lem:dist-samp-2} that with probability at least $1-\delta/2$,
	\begin{equation}\label{eqn:halve}
		\sum_{i=1}^{|L|}\sum_{j=1}^{t_i}\frac{X_{i,j}}{t_i} \in (1\pm\xi)\sum_{i=1}^{|L|}\mu_i.
	\end{equation}
	Plugging our definitions in, we see~\eqref{eqn:halve} implies that $Z(L') \in (1\pm\xi)Z(L)$. Thus,
	\begin{equation}\label{eqn:halve-E1}
		\Pr(\calE_1) \ge 1-\delta/2.
	\end{equation}
	
	By the correctness of \acc (\cref{lem:acc}) and a union bound over all $1 \le i \le |L|$ and all $j \in [t_i]$, we have $\Pr(\calE_2) \ge 1-\delta/2$. By a union bound with~\eqref{eqn:halve-E1}, we therefore have $\Pr(\calE_1 \cap \calE_2) \ge 1-\delta$ as claimed.
	
	\textbf{Properties (a)--(c) hold:} Suppose $\calE_1 \cap \calE_2$ occurs. For every entry $(w,S,\hat{e})$ of $L'$, $w$ and $\hat{e}$ are positive rational numbers and $S \subseteq V(G)$ with $|S| = 2^{y-1}$. Since $\calE_2$ occurs and $b \ge 2(4k\log n)^k$, by the correctness of \acc (\cref{lem:acc}) we have $\hat{e}/b \le e(G[S]) \le \hat{e}b$. Thus, $L$ is a $(G,b,y-1)$-list as required by property (a). We have $|L'| = \sum_i t_i$, so~\eqref{eqn:halve-tis-2} implies that property (b) holds. Finally, since $\calE_1$ occurs, property (c) holds. Thus, properties (a)--(c) all hold whenever $\calE_1 \cap \calE_2$ occurs, which we have already shown happens with probability at least~$1-\delta$.
\end{proof}

We now state our main algorithm.

\begin{framed}
	\noindent
	\textbf{Algorithm $\approxUncol(G,\epsilon)$.}
	
	\smallskip\noindent
	\textbf{Input:} $G$ is an $n$-vertex $k$-hypergraph, where $n$ is a power of 2. \approxUncol only has colourful oracle access to $G$, and $\eps$ is a rational number with $0 < \eps < 1/2$.
	
	\smallskip\noindent\textbf{Behaviour:} $\approxUncol(G,\epsilon)$ outputs a rational number $\hat{e}$ such that with probability at least 2/3, $\hat{e} \in (1\pm \eps)e(G)$.
	\shortrule
	\begin{enumerate}[({A}1)]
		\item If $\eps < n^{-k}$, or if $n \le 500$, then return $\sum_{Y \subseteq V(G),\ |Y|=k}(1-\cindora(Y))$.
		\item If $\acc(G,\delta) = 0$, return $0$.
		Otherwise, let
		\begin{alignat*}{4}
		I &\leftarrow \log n - \ceil{\log(2k^2)},&
		\qquad b &\leftarrow 2(4k\log n)^k,\\
		\xi &\leftarrow \eps/4I,&
		\qquad\delta &\leftarrow 1/3(2I+1),
		\end{alignat*}
		\[
		L \leftarrow \big\{\big(1,V(G),\acc(G,\delta)\big)\big\}.
		\]
		\item For $i = 1$ to $I$:\\
		\textit{(We will maintain the invariant that $L$ is a $(G,b,\log n - (i-1))$-list with $Z(L) \in (1\pm \xi)^{2i}e(G)$, and that $|L|$ is suitably small (see proof of \cref{lem:uncol-count}). Note that this is trivially satisfied at the start of the loop.)}
		\begin{enumerate}
			\item[(A4)] Update $L \leftarrow \halve(G,b,\log n - (i-1),L,\xi,\delta)$.\\
			\textit{(This step turns $L$ into a $(G,b,\log n-i)$-list.)} 
			\item[(A5)] Update $L \leftarrow \impsamp(G,b,\log n - i,L,\xi,\delta)$.\\
			\textit{(This step reduces the length of $L$.)}
		\end{enumerate}
		\item[(A6)] For each entry $(w,S,\hat{e}) \in L$, calculate 
		\[
		e_S \leftarrow \sum_{\substack{Y \subseteq S\\|Y|=k}}(1 - \cindora(Y)).
		\]
		\item[(A7)] Output $\sum_{(w,S,\hat{e}) \in L} we_S$.
	\end{enumerate}
\end{framed}

\begin{lemma}\label{lem:uncol-count}
	With probability at least $2/3$, $\approxUncol(G,\eps)$ outputs a rational number $\hat{e} \in (1\pm\eps)e(G)$ as claimed above, has running time $\OO(\eps^{-2}k^{6k}n\log^{4k+7}n)$, and invokes \cindora at most $\OO(\eps^{-2}k^{6k}\log^{4k+7}n)$ times.
\end{lemma}
\begin{proof}
	\textbf{Correctness.} \cref{lem:acc} implies that whenever $\approxUncol(G,\eps)$ outputs on step (A1) or (A2), correctness holds, so suppose that this does not occur. 
	For all integers $i \ge 0$, let $\pi(i)$ be the statement that $L$ satisfies the following properties.
	\begin{enumerate}[(i)]
		\item $L$ is a $(G,b,\log n - i)$-list.
		\item $Z(L) \in (1 \pm \xi)^{2i}e(G)$.
		\item $|L| \le 33k\log(4nb) + 32b^2\log(2/\delta)/\xi^2$.
	\end{enumerate}
	We will prove that with probability at least $2/3$, $\pi(i)$ holds at the end of the $i$th iteration of loop (A3) for all $i \in [I]$. Suppose this is true: we will show that correctness follows. With probability at least $2/3$, $\pi(I)$ holds when we exit the loop. In this case, the final value of $L$ satisfies $Z(L) \in (1\pm \xi)^{2I}e(G)$ by (ii). We have $(1-\xi)^{2I} \ge 1-2I\xi$, and $(1+\xi)^{2I} \le e^{2I\xi} \le 1+4I\xi$ (since $4I\xi = \eps < 1$), so 
	\[
		Z(L) \in (1\pm 4I\xi)e(G) = (1\pm\eps)e(G).
	\]
	Moreover, in step (A6) we have $e_S = e(G[S])$ for all $(w,S,\hat{e}) \in L$, so $Z(L)$ is the output.
	
	It remains to prove that with probability at least $2/3$, $\pi(i)$ holds at the end of the $i$th iteration of loop (A3) for all $i \in [I]$. Let $\calE_0$ be the event that \acc behaves correctly in step (A2); note that $\pr(\calE_0) \ge 1-\delta$ by the correctness of \acc (\cref{lem:acc}). For all $i \in [I]$, let $\calE_i$ be the event that \halve behaves correctly in the $i$th iteration of step (A4) and \impsamp behaves correctly in the $i$th iteration of step (A5). (If the input restrictions of \halve or \impsamp are violated on the $i$th iteration, then $\calE_i$ occurs automatically.) By correctness of \halve and \impsamp (\cref{lem:halve,lem:impsamp}), we have $\Pr(\calE_i \mid \calE_0,\dots,\calE_{i-1}) \ge 1 - 2\delta$. Thus, by a union bound over all $0 \le i \le I$, we have
	\[
		\Pr\Big(\bigcap_{i=0}^{I}\calE_i\Big) \ge 1 - (2I+1)\delta = 2/3.
	\]
	It therefore suffices to show that when $\bigcap_j \calE_j$ occurs, $\pi(i)$ holds at the end of the $i$th iteration of loop (A3) for all $i \in [I]$. 
	
	At the start of the first iteration of loop (A3), when $i=1$, $L$ is a $(G,b,\log n)$-list since $\calE_0$ occurs, $Z(L) = e(G)$, and $|L| = 1$. Thus, $\pi(0)$ holds. Let $i \in [I]$, and suppose that $\pi(i-1)$ holds at the start of the $i$th iteration of loop (A3). Let $L_i$ be the value of $L$ at the start of the $i$th iteration, let $L_i'$ be the value of $L$ after executing step (A4), and let $L_{i+1}$ be the value of $L$ after executing step (A5). 
	
	By property (i) of $\pi(i)$, $L_i$ is a $(G,b,\log n-(i-1))$-list, where by our choice of $I$ we have $2^{\log n - i} \ge 2k^2$. Since $\calE_i$ occurs, it follows by the correctness of \halve (\cref{lem:halve}) that $L_i'$ is a $(G,b,\log n - i)$-list with 
	\begin{align}\label{eq:uncol-count}
		Z(L_i') &\in (1\pm \xi)Z(L_i),\\\nonumber
		|L_i'| &\le |L_i| + 2^{k+3}b^2\log(4/\delta)/\xi^2.
	\end{align} 
	We next show that $1/2 \le Z(L_i') \le 2n^k$, that $|L_i'| \le n^{11k}$, and that $\xi \ge n^{-2k}$, as required by \impsamp. Since property (ii) of $\pi(i)$ holds for $Z(L_i)$, we have $Z(L_i') \in (1\pm\xi)^{2i+1}e(G) \subseteq (1\pm\eps) e(G)$. Since $\approxUncol(G,\eps)$ did not halt at (A2), we have $1 \le e(G) \le n^k$; since $\eps < 1/2$, it follows that $1/2 \le Z(L_i') \le 2n^k$. Since $\approxUncol(G,\eps)$ did not halt at (A1), we have $n \ge 500$ and hence $n\ge 50\log n$. We also have $\eps \ge n^{-k}$ and $k \le n$. Hence:
	\begin{alignat*}{3}
		&b^2 \le n^{4k};\qquad\qquad 2^k\le n^k; \qquad &\log(4/\delta) \le \log(24\log n) \le n;\\
		&\log(4nb)\le 6k\log n \le n^2; \qquad &1/\xi^2 \le 16(\log n)^2n^{2k} \le n^{4k}.
	\end{alignat*}
	Since property (iii) of $\pi(i)$ holds for $L_i$, it follows that
	\begin{align*}
		|L_i'| &\le 33k\log(4nb) + 40\cdot 2^kb^2\log(4/\delta)/\xi^2 \le 33n^3 + 40n^{9k+1} \le n^{10k}.
	\end{align*}
	We have therefore shown that $L_i'$ and $\xi$ satisfy the input restrictions of \impsamp. Since $\calE_i$ occurs, by the correctness of \impsamp (\cref{lem:impsamp}) and by~\eqref{eq:uncol-count} it follows that $L_{i+1}$ is a $(G,b,\log n - i)$-list with
	\begin{align*}
		Z(L_{i+1}) &\in (1\pm\xi)Z(L_i') \subseteq (1\pm\xi)^2Z(L_i) \subseteq (1\pm\xi)^{2(i+1)}e(G),\\
		|L_{i+1}| &\le 33k\log(4nb)+32b^2\log(2/\delta)/\xi^2.
	\end{align*} 
	Thus, properties (i)--(iii) hold for $L_{i+1}$, as required.
	
	\medskip\noindent\textbf{Running time and oracle queries.} If step (A1) is executed, so that $\eps < n^{-k}$ or $n \le 500$, then the algorithm runs in time $\OO(n^k) = \OO(\eps^{-1})$ and uses $\OO(n^k) = \OO(\eps^{-1})$ oracle queries, so our claimed bounds hold. Suppose instead step (A1) is not executed, so that $\eps \ge n^{-k}$. Recall that $\bigcap_i\calE_i$ holds with probability at least $2/3$. Suppose this occurs. The bottleneck in both running time and oracle invocations is then step (A4). For legibility, we give the time and oracle requirements of the other steps in the following table, giving justifications in the paragraph below. We write $\Lambda = 33k\log(4nb) + 32b^2\log(2/\delta)/\xi^2$ for the upper bound on $|L|$ in property (iii) of our invariant~$\pi$.
	
	\begin{center}\begin{tabular}{|l|l|l|}
		\hline
		Step number & Running time & Oracle calls\\
		\hline
		(A1) & $\OO(k^2)$ & None\\
		(A2) & $\OO\big(k^{3k+2}n\log^{2k+2}n\log(1/\delta)\big)$ & $\OO\big(k^{3k}\log^{2k+2}n\log(1/\delta)\big)$\\
		(A3) & $\OO(I)$ & None \\
		(A5) & $\OO\Big(I\big(\Lambda + 2^kb^2\xi^{-2}\log(1/\delta)\big)k^4\log n\Big)$ & None\\
		(A6) & $\OO(\Lambda (4k^2)^k)$ & $\OO(\Lambda (4k^2)^k)$ \\
		(A7) & $\OO(k^2\Lambda)$ & None\\
		\hline
	\end{tabular}\end{center}
	For step (A1), we use the fact that $\eps \ge n^{-k}$ and so the conditional does not trigger; we verify this fact in time $\OO(k^2)$. For step (A2), we use the fact that $n$ is a power of 2 (so computing $\log n$ is easy) and the time bounds on \acc (\cref{lem:acc}). For step (A5), we first observe that the step is executed $I$ times. We then apply the time bounds on \impsamp (\cref{lem:impsamp}), together with property (iii) of $\pi(i)$ and the fact that \halve adds at most $2^{k+3}b^2\xi^{-2}\log(4/\delta)$ to the length of $L$ (see \cref{lem:halve}). (Note that $k^3\log(nb/\delta) = \OO(k^4\log n)$.) For step (A6), we use the fact that after the loop of (A3), $L$ is a $(G,b,\log n - I)$-list, so each entry $(w,S,\hat{e})$ of $L$ has $|S| = 2^{\log n-I} \le 4k^2$.
	
	We now consider step (A4), which is executed $I$ times. From the time bounds of \halve (\cref{lem:halve}), it follows that the total running time of step (A4) is $\OO(nT)$ and the total number of oracle accesses is $\OO(T)$ times, where
	\[
		T = \OO\big(I\lambda\log(\lambda/\delta)k^{3k}\log^{2k+2}n\big),\qquad \lambda = \Lambda + 2^kb^2\xi^{-2}\log(1/\delta).
	\]
	This clearly dominates everything in the table. Observe that $\Lambda = \OO(2^kb^2\xi^{-2}\log(1/\delta))$, so $\lambda = \OO(2^kb^2\xi^{-2}\log(1/\delta))$ also. Since $\eps \ge n^{-k}$,
	\begin{align*}
		\log(\lambda/\delta) = \OO\Big(k+k\log(k\log n)+\log I+\log(1/\eps)+\log(1/\delta) \Big) = \OO(k\log n).
	\end{align*}
	Thus
	\begin{align*}
		T &= \OO\Big(I2^kb^2\xi^{-2}\log(1/\delta)k^{3k+1}\log^{2k+3}n\Big)\\
		&= \OO\Big(\log n \cdot 2^k\cdot (k\log n)^{2k} \cdot \eps^{-2}\log^2 n \cdot \log \log n \cdot k^{3k+1}\log^{2k+3}n \Big)\\
		&= \OO\big(\eps^{-2}k^{6k}\log^{4k+7}n\big),
	\end{align*}
	and the claimed bounds follow.
\end{proof}

We now recall \cref{thm:uncol-approx} and then prove it.
\statemaincounting*
\begin{proof}
	To evaluate $\aau(G,\eps,\delta)$, we first make $n$ into a power of two by adding at most $n$ isolated vertices to $G$; note that this does not impede the evaluation of $\cindora$. We then run $\approxUncol(G,\min\{\eps,1/3\})$ a total of $36\ceil{\ln(2/\delta)}$ times and return the median result $\hat{e}$. If some invocation of $\approxUncol(G,\min\{\eps,1/3\})$ takes more than $\Theta(\eps^{-2}k^{6k}n\log^{4k+7}n)$ time, or invokes $\cindora$ more than $\Theta(\eps^{-2}k^{6k}\log^{4k+7}n)$ times, we halt execution and consider the output to be $-1$. 
	
	It is immediate that this algorithm satisfies our stated time bounds. Moreover, $\hat{e} \in (1\pm\eps)e(G)$ unless at least half our invocations of \approxUncol fail. The number of such failures is dominated above by a binomial variable $N$ with mean $12\ceil{\ln(2/\delta)}$, so by a standard Chernoff bound (namely \cref{lem:chernoff}(i)) we have
	\begin{align*}
		\pr\big(\hat{e} \notin (1\pm\eps)e(G)\big) &\le \pr\big(N \ge 18\ceil{\ln(2/\delta)}\big)\\
		&\le \pr\Big(\big|N - \E(N)\big| \ge \frac{1}{2}\E(N)\Big)
		\le 2e^{-\ceil{\ln(2/\delta)}} \le \delta,
	\end{align*}
	as required.
\end{proof}

\section{Coarse approximate counting}
\label{sec:coarse}

In this section, we prove \cref{lem:acc}.
We fix the input graph $G$ to be an $n$-vertex $k$-hypergraph to which we have (only) colourful oracle access, where $n$ is a power of two.

\subsection{Sketch proof}

The heart of our algorithm will be a subroutine to solve the following simpler ``gap-version'' of the problem. Given a $k$-partite $k$-hypergraph $G$ and a guess $M \ge 0$, we ask: Does $G$ have more than $M$ edges?
We wish to answer correctly with high probability provided that either $G$ has at least $M$ edges, or $G$ has significantly fewer than $M$ edges, namely at most $\gamma M$ edges with $\gamma = 1/(2^{3k+1}k^{2k}\log^k n)$.
Suppose we can solve this problem probabilistically,
 perhaps outputting \yes with probability at least $1/50$ if $e(G) \ge M$ (which we call \emph{completeness}) and outputting $\yes$ with probability at most $1/100$ if $e(G) \le \gamma M$ (which we call \emph{soundness}). We then apply probability amplification to substantially reduce the failure probability, and use binary search to find the least $M$ such that our output is \yes --- with high probability, this will approximate $e(G)$ when our input $k$-hypergraph is $k$-partite. We then generalise our algorithm to arbitrary inputs using random colour-coding. These parts of the algorithm are fairly standard, so in this sketch proof we will only solve the gap-problem. (We implement this sketch below as the \verifyguess algorithm.)

Let $G$ be a $k$-partite $k$-hypergraph with vertex classes $X_1, \dots, X_k$. The basic idea of the algorithm is to randomly remove vertices from $G$ to form a new graph $H$ in such a way that each edge survives with probability roughly $1/M$, and then query the colourful independence oracle and output \yes if and only if at least one edge remains. If $G$ has at most $\gamma M$ edges, then a union bound implies we are likely to output \no (soundness); if $G$ has at least $M$ edges, then in expectation at least one edge survives the removal, so we hope to output \yes (completeness). Unfortunately, the number of edges remaining in $H$ need not be concentrated around its expectation --- for example, if every edge of $G$ is incident to a single vertex $v$ --- so we must be very careful if this hope is to be realised. 

Suppose for the moment that $k=2$, so that $G$ is a bipartite graph with vertex classes $X_1$ and $X_2$. Then we will form $X_1' \subseteq X_1$ by including each vertex independently with probability $p_1$, and $X_2' \subseteq X_2$ by including each vertex independently with probability $p_2$. Each edge survives with probability $p_1p_2$, so we require $p_1 p_2 \le 1/M$ to ensure soundness. To ensure completeness, we would then like to choose $p_1$ and $p_2$ such that $G[X_1',X_2']$ is likely to contain an edge whenever $e(G) \geq M$.

To see that such a pair $(p_1,p_2)$ exists, we first partition the vertices in $X_1$ according to their degree: For $1 \le d \le \log n$, let $X_1^d$ be the set of vertices $v$ with $2^{d-1} \le d(v) < 2^d$. By the pigeonhole principle, there exists some $D$ such that $X_1^D$ is incident to at least $e(G)/\log n$ edges. Then we take $p_1 = 2^D/M$ and $p_2 = 1/2^D$. We certainly have $p_1p_2 \le 1/M$. Suppose $e(G) \ge M$. Since $X_1^D$ is incident to at least $e(G)/\log n$ edges, we have $|X_1^D| \ge M/2^D\log n$, so with reasonable probability $X_1'$ contains a vertex $v_1 \in X_1^D$. Then $v_1$ has degree roughly $2^D$ in $X_2$, so again with reasonable probability $X_2'$ contains a vertex adjacent to it.

There is one remaining obstacle: Since we only have colourful oracle access to $G$, we do not know what $D$ is! Fortunately, since there are only $\OO(\log n)$ possibilities, we can simply try them all in turn, and output \yes if any one of them yields a pair $X_1',\,X_2'$ such that $G[X_1',X_2']$ contains an edge. (It is not hard to tune the parameters so that this doesn't affect soundness.) This is essentially the argument used by Beame et al.~\cite{beame20}.

When we try to generalise this approach to $k$-hypergraphs, we hit a problem. For illustration, take $k=3$ and suppose $e(G) \ge M$. Then we wish to guess a vector $(p_1,p_2,p_3)$ such that $p_1p_2p_3 \le 1/M$ and, with reasonable probability, $G[X_1',X_2',X_3']$ contains an edge. As in the $k=2$ case, we can guess an integer $0 \le D \le 2\log n$ such that a large proportion of $G$'s edges are incident to a vertex in $X_1$ of degree roughly $2^D$. Also, as in the $k=2$ case, if we take $p_1 = 2^D/M$ then it is reasonably likely that $X_1'$ will contain a vertex of degree roughly $2^D$, say $v_1$. But we cannot iterate this process --- the structure of $G[v_1, X_2, X_3]$, and hence the ``correct'' value of $p_2$, depends very heavily on $v_1$. So for example, when we test the two guesses $(2^D/M, 1/2^D, 1)$ and $(2^D/M, 1, 1/2^D)$, we wish to ensure that the value of $v_1$ is the same in each test. This is the reason for step (C1) in the following algorithm; it is important that we do not choose new random subsets of $X_1, \dots, X_k$ independently with each iteration of step~(C2).

\subsection{Solving the gap problem}

 \begin{framed}
 \noindent
 \textbf{Algorithm $\verifyguess(G,M,X_1, \dots, X_k)$.}
 
 \smallskip\noindent\textbf{Input:}
$G$ is an $n$-vertex $k$-hypergraph to which \verifyguess has (only) colourful oracle access. $n$ and $M$ are positive powers of two, and $X_1, \dots, X_k \subseteq V(G)$ are disjoint.

 \smallskip\noindent\textbf{Behaviour:}
 Let $\pout=(8k\log n)^{-k}$.\\
 \textit{Completeness:} If $e(G[X_1,\dots,X_k]) \ge M$, then \verifyguess outputs \yes with probability at least $\pout$.\\
 \textit{Soundness:} If $e(G[X_1,\dots,X_k]) < M \cdot\pout/2(k\log n)^k$, then \verifyguess outputs \yes with probability at most $\pout/2$.
 \shortrule
   \begin{enumerate}[(C1)]
	 \item[(C1)] For each $i \in [k]$ and each $0 \le j \le k\log n$, construct a subset $Y_{i,j}$ of $X_i$ by including each vertex independently with probability $1/2^j$. Construct the finite set $A$ of all tuples $(a_1, \dots, a_k)$ with $0 \le a_1, \dots, a_k \le k\log n$ and $a_1 + \dots + a_k \ge \log M$.
	 \item[(C2)] For each tuple $(a_1, \dots, a_k) \in A$: If $\cindora(Y_{1,a_1},\dots,\allowbreak Y_{k,a_k}) = 0$, then halt and output~\yes.
	 \item[(C3)] We have not halted yet, but do so now and output \no. 
 \end{enumerate}
\end{framed}

\begin{lemma}\label{lem:verifyguess}
  \verifyguess behaves as stated, runs in time $\OO(nk^k \log^k n)$, and makes at most $\OO(k^k\log^k n)$ oracle queries.
\end{lemma}
\begin{proof}
 Let $G,M,X_1,\dots,X_k$ be the input for \verifyguess, and write \[H = G[X_1, \dots, X_k]\,.\]
 For notational convenience, we denote the gap in the soundness case by $\gamma$, that is, we set $\gamma:=\pout/2(k\log n)^k = 1/2^{3k+1}k^{2k}\log^{2k} n$.

 \medskip\noindent\textbf{Running time and oracle queries.} Step (C1) takes $\OO(nk^2\log n + k^k\log^k n)$ time and no oracle queries; step (C2) takes $\OO(k^k\log^k n)$ time and $\OO(k^k\log^k n)$ oracle queries; and step (C3) takes $\OO(1)$ time and no oracle queries. The claimed bounds follow, and it remains to prove that the soundness and completeness properties hold.

 \medskip\noindent\textbf{Soundness.}
 We next prove soundness, as this is the easier part of proving correctness. So suppose $e(H)\le \gamma M$. Let $(a_1, \dots, a_k) \in A$, and let $H' \subseteq H$ denote the random induced subgraph $G[Y_{1,a_1}, \dots, Y_{k,a_k}]$. Then for all $e \in E(H)$, we have
 \[
 	\pr(e \in E(H'))
 	= \prod_{j=1}^k 2^{-a_j}
 	\le \frac{1}{M}
 	\le \frac{\gamma}{e(H)}
 	\le \frac{\pout}{2|A|e(H)}\,.
 \]
 By a union bound over all $e \in E(H)$ and all $(a_1, \dots, a_k) \in A$, it follows that the probability that \verifyguess outputs \yes is at most $\pout/2$.  This establishes the soundness of the algorithm, so it remains to prove completeness.

   \medskip\noindent\textbf{Completeness.}
   Suppose now that $e(H)\ge M$ holds.
   We show 
   that \verifyguess outputs \yes with probability at least $\pout$.
   It suffices to show that with probability at least $\pout$, there is at least one setting of the vector $(a_1, \dots, a_k)\in A$ such that $G[Y_{1,a_1},\dots,Y_{k,a_k}]$ contains at least one edge. 
   
   We will define this setting iteratively. First, with reasonable probability, we will find an integer $a_1$ and a vertex $v_1 \in Y_{1,a_1}$ such that $G[v_1,X_2, \dots, X_k]$ contains roughly $2^{-a_1}e(H)$ edges. In the process, we expose $Y_{1,a_1}$. We then, again with reasonable probability, find an integer $a_2$ and a vertex $v_2 \in Y_{2,a_2}$ such that $G[v_1, v_2, X_3, \dots, X_k]$ contains roughly $2^{-a_1 - a_2}e(H)$ edges. Continuing in this vein, we eventually find $(a_1, \dots, a_k) \in A$ and vertices $v_i \in Y_{i,a_i}$ such that $\{v_1, \dots, v_k\}$ is an edge in $G[Y_{1,a_1}, \dots, Y_{k,a_k}]$, proving the result.
   
   More formally, recall from Section~\ref{sec:actual-notation} that for all $i \in [k]$ and $v_1,\dots,v_i \in V(G)$, $d_H(v_1,\dots,v_i)$ is the number of edges in $H$ containing $\{v_1,\dots,v_i\}$ as a subset. For all $i \in [k]$, let $\calE_i$ be the event that there exist $0 \le a_1, \dots, a_i \le k\log n$ and $v_1, \dots, v_i \in V(H)$ such that:
   \begin{enumerate}[(a)]
   		\item for all $j \in [i]$, $v_j \in Y_{j,a_j}$;
   		\item we have $d_H(v_1,\dots,v_{i}) \ge e(H)/ \prod_{j=1}^{i}2^{a_j}$.
   \end{enumerate}
   We make the following \textbf{Claim}: $\Pr(\calE_1) \ge 1/(8k\log n)$ and, for all $2 \le i \le k$, $\Pr(\calE_i \mid \calE_{i-1}) \ge 1/(8k\log n)$.
   
   \textbf{Proof of \cref{lem:verifyguess} from Claim:} Suppose $\calE_k$ occurs, and let $a_1, \dots, a_k$ and $v_1, \dots, v_k$ be as in the definition of $\calE_k$. By (b), $d_H(v_1, \dots, v_k) > 0$, so $\{v_1, \dots, v_k\}$ is an edge in $H$; it follows by (a) that it is also an edge in $G[Y_{1,a_1}, \dots, Y_{k,a_k}]$. Also by (b), since $d_H(v_1, \dots, v_k) = 1$, we have $\prod_{j=1}^k 2^{a_j} \ge e(H) \ge M$, so $a_1 + \dots + a_k \ge \log M$. Thus, $(a_1, \dots, a_k) \in A$, so whenever $\calE_k$ occurs, \verifyguess outputs \yes on reaching $(a_1, \dots, a_k)$ in step (C2). By the Claim, we have
   \[
	   	\pr(\calE_k) = \pr(\calE_1)\prod_{j=2}^k \pr(\calE_j \mid \calE_1, \dots, \calE_{j-1}) = \pr(\calE_1)\prod_{j=2}^k \pr(\calE_j \mid \calE_{j-1}) \ge 1/(8k\log n)^k = \pout,
   \]
   so completeness follows. The lemma statement therefore follows as well.
   
   \textbf{Proof of Claim:} We first prove the claim for $\calE_1$. We will choose $a_1$ depending on the degree distribution of vertices in $X_1$. For all integers $1 \le d \le k\log n$, let 
   \[
   		X_1^d := \{v \in X_1 \colon 2^{d-1} \le d_H(v) < 2^d\}
   \] 
   be the set of vertices in $X_1$ with degree roughly $2^d$. Every edge in $H$ is incident to exactly one vertex in exactly one set $X_1^d$, so there exists $D$ such that $X_1^{D}$ is incident to at least $e(H)/k\log n$ edges of $H$. We take $a_1 := \ceil{\log e(H)} - D + 1$. Note that $0 \le a_1 \le k\log n$, since $X_1^D \ne \emptyset$ and so $e(H) \ge 2^{D-1}$. 
   
   We would like to take $v_1 \in Y_{1,a_1} \cap X_1^D$, so we next bound the probability that this set is non-empty. We have 
   \begin{align*}
   		\pr(X_1^D \cap Y_{1,a_1} \ne \emptyset) 
   		&= 1 - (1 - 2^{-a_1})^{|X_1^D|} 
   		\ge 1 - \exp({-}2^{-a_1}|X_1^D|).
   	\end{align*}
   	Since every vertex in $X_1^{D}$ has degree at most $2^D$, by the definition of $D$ we have $2^D|X_1^D| \ge e(H)/(k\log n)$. Moreover, we have $a_1 \le \log e(H)-D+2$.
   	It follows that
   	\begin{align*}
   		\pr(X_1^D \cap Y_{1,a_1} \ne \emptyset) &\ge 1 - \exp\Big({-}\frac{2^{D-2}}{e(H)}\cdot\frac{e(H)}{k2^D\log n}\Big)
   		\\
   		&= 1 - \exp\Big({-}\frac{1}{4k\log n}\Big) 
   		\ge \frac{1}{8k\log n}.
   \end{align*}
   Suppose $X_1^D \cap Y_{1,a_1} \ne \emptyset$, and take $v_1 \in X_1^D \cap Y_{1,a_1}$. Then $v_1$ certainly satisfies (a), and by the definitions of $a_1$ and $X_1^D$ we have $e(H)/2^{a_1} \le 2^{D-1} \le d_H(v_1)$, so $v_1$ also satisfies (b). We have therefore shown $\pr(\calE_1) \ge 1/(8k\log n)$ as required.
   
   Now let $2 \le i \le k$. The argument is similar, but we include it explicitly for the benefit of the reader. We first expose $Y_{1,a_1}, \dots, Y_{i-1,a_{i-1}}$: Let $\mathcal{F}$ be a possible filtration of these variables consistent with $\calE_{i-1}$, and let $a_1, \dots, a_{i-1}$ and $v_1, \dots, v_{i-1}$ be as in the definition of $\calE_{i-1}$. It then suffices to show that $\pr(\calE_i \mid \mathcal{F}) \ge 1/(8k\log n)$. 
   
   Similarly to the $i=1$ case, for all integers $1 \le d \le k\log n$, let 
   \[
   		X_i^d := \{v \in X_i \colon 2^{d-1} \le d_H(v_1, \dots, v_{i-1}, v) < 2^d\}.
   \]
   Every edge in $H[v_1, \dots, v_{i-1}, X_i, \dots, X_k]$ is incident to exactly one vertex in exactly one set $X_i^d$, so there exists $D_i$ such that $X_i^{D_i}$ is incident to at least \[d_H(v_1, \dots, v_{i-1})/k\log n\] edges of $H[v_1,\dots,v_{i-1},X_i,\dots,X_k]$. We take $a_i := \ceil{\log d_H(v_1, \dots, v_{i-1})} - D_i + 1$; note that $0 \le a_i \le k\log n$.
   
   As in the $i=1$ case, we would like to take $v_i \in Y_{i,a_i} \cap X_i^{D_i}$, so we next bound the probability that this set is non-empty. Since every vertex $v \in X_i^{D_i}$ satisfies $d_H(v_1, \dots, v_{i-1}, v) \le 2^{D_i}$, we have $2^{D_i}|X_i^{D_i}| \ge d_H(v_1, \dots, v_{i-1})/k\log n$. It follows that
   \begin{align*}
   		\pr(X_i^{D_i} \cap Y_{i,a_i} \ne \emptyset \mid \mathcal{F}) 
   		&= 1 - (1 - 2^{-a_i})^{|X_i^{D_i}|} 
   		\ge 1 - \exp({-}2^{-a_i}|X_i^{D_i}|)\\
   		&\ge 1 - \exp\Big({-}\frac{2^{D_i-2}}{d_H(v_1, \dots, v_{i-1})}\cdot \frac{d_H(v_1, \dots, v_{i-1})}{k2^{D_i}\log n} \Big)\\
   		&= 1 - \exp\Big({-}\frac{1}{4k\log n}\Big)
   		\ge \frac{1}{8k\log n}.
   \end{align*}
   Suppose $X_i^{D_i} \cap Y_{i,a_i} \ne \emptyset$, and take $v_i \in X_i^{D_i} \cap Y_{i,a_i}$. Then $v_i$ certainly satisfies (a). By the definitions of $a_i$ and $X_i^{D_i}$, and the fact that $v_1, \dots, v_{i-1}$ satisfy (b), we have
   \begin{align*}
   		e(H)/\prod_{j=1}^i2^{a_j}&\le d_H(v_1, \dots, v_{i-1})/2^{a_i} \le 2^{D_i-1} \le d_H(v_1, \dots, v_i).
   \end{align*}
   Thus, (b) is satisfied, and we have shown $\pr(\calE_i \mid \mathcal{F}) \ge 1/(8k\log n)$ as required. 
\end{proof}

\subsection{Proving \cref{lem:acc}}

  We next turn \verifyguess into a crude approximation algorithm for $k$-partite $k$-hypergraphs in the natural way. 
  
  \begin{framed}
  \noindent
  \textbf{Algorithm $\coarsecount(G,X_1,\dots,X_k)$.}
  
  \smallskip\noindent\textbf{Input:}
  $G$ is an $n$-vertex $k$-hypergraph, where $n$ is a power of two, to which \coarsecount has colourful oracle access (only). $X_1, \dots, X_k$ form a partition of $V(G)$.
  
  \smallskip\noindent\textbf{Behaviour:} Let $b := (4k\log n)^k$. Then \coarsecount($G$) outputs a non-negative integer $m$ such that, with probability at least $2/3$, $m/b \le e(G[X_1,\dots,X_k]) \le mb$.
  \shortrule
  \begin{enumerate}[(D1)]
      \item Set $\pout := (8k\log n)^{-k}$ and $N := \ceil{48\ln(6k\log n)/\pout}$. 
      \item For each $M$ in $\{1,2,4,8,\dots,n^k \}$: Execute \verifyguess$(G,M,X_1,\dots,X_k)$ a total of $N$ times, and let $S_M\in\{0,\dots,N\}$ be the number of executions that returned \yes. (Naturally we use independent randomness for each value of $M$.) 
      \item If $\cindora(X_1,\dots,X_k) = 1$, let $m = 0$. Otherwise, if there exists $M$ such that $S_M \ge \frac{3}{4}\pout N$, let $m$ be the least such $M$. Otherwise, let $m=n^k$. Output $m\big(\pout/2k^k\log^k n\big)^{1/2}$. 
    \end{enumerate}
  \end{framed}

  \begin{lemma}\label{lem:coarse-count}
    \coarsecount{} behaves as stated, runs in time $\OO((8k\log n)^{2k+2}n)$, and requires $\OO((8k\log n)^{2k+2})$ oracle queries.
  \end{lemma}
  
  \begin{proof}
    Let $G$ and $X_1, \dots, X_k$ be the inputs, so that $G$ is an $n$-vertex $k$-hypergraph and $X_1,\dots,X_k$ partition $V(G)$.

    \textbf{Running time and oracle queries.}
    Observe $N = \OO((8k\log n)^{k+1})$.
    The algorithm \coarsecount executes \verifyguess at most $\OO(\log(n^k) N)$ times.
    By \cref{lem:verifyguess}, each execution takes $\OO(nk^k\log^kn)$ time and makes $\OO(k^k \log^k n)$ oracle queries.
    Thus the claimed bounds on the running time and number of oracle queries of \coarsecount follow.

    \textbf{Correctness.}
    Let $M\in\{1,2,4,8,\dots,n^{k}\}$, and let $H = G[X_1, \dots, X_k]$.
    For this fixed~$M$, the algorithm invokes \verifyguess $N$ times, so the random variable~$S_M$ is the sum of~$N$ independent indicator variables.
    By a standard Chernoff bound (\cref{lem:chernoff}(i) taking $\eps = 1/4$),
    \begin{align}\label{eqn:coarsecount-chernoff}
    	\Pr\big(|S_M - \E(S_M)| \ge \E(S_M)/4\big) \le 2\exp\big({-}\tfrac{1}{48}\E(S_M)\big)
    \end{align}
    If $e(H) \ge M$, then the completeness of \verifyguess implies $\E(S_M)\ge N\pout$.
    Thus, \eqref{eqn:coarsecount-chernoff} and our choice of $N$ imply that
    \[
    	\pr\big(S_M \le \tfrac{3}{4}N\pout\big) \le 2\exp\big({-}\tfrac{1}{48}N\pout \big) \le 1/(3k\log n).
    \]
    Similarly, if $e(H) < M\pout/(2(k\log n)^k)$, then the soundness of \verifyguess implies $\E(S_M)\le N\pout/2$.
    But then~\eqref{eqn:coarsecount-chernoff} implies that
    \[
    	\pr\big(S_M \ge \tfrac{3}{4}N\pout \big) \le \pr\big(S_M \ge \tfrac{5}{4}\E(S_M)\big) \le 2\exp\big({-}\tfrac{1}{48}N\pout \big) \le 1/(3k\log n).
    \]
    
    Finally, we perform a union bound over all~$M\in\{1,2,4,8,\dots,n^{k}\}$ that satisfy either $e(H) \ge M$ or $e(H) \le M\pout/(2k^k\log^k n)$. (Note that no value of $M$ satisfies both inequalities.) There are at most $k\log n$ such $M$'s, so with probability at least~$2/3$, we see $S_M > 3N\pout/4$ whenever $e(H) \ge M$ and $S_M < 3N\pout/4$ whenever $e(H) \le M\pout/(2k^k\log^kn)$. By the definition of $m$, it follows that in this case 
    \[
    	\frac{\pout}{2k^k\log^k n}m \le e(H) \le m.
    \]
    Hence, writing $x = m(\pout/2k^k\log^k n)^{1/2}$ for the output of \coarsecount,
    \[
    	x\sqrt{\frac{\pout}{2k^k\log^k n}} \le e(H) \le x\Big/\sqrt{\frac{\pout}{2k^k\log^k n}}.
    \]
    Since $\pout=1/(8k \log n)^k$, the output of \coarsecount approximates~$e(H)$ up to a factor of~$(4k \log n)^k$ as required. 
  \end{proof}

  We now combine our algorithm for coarsely approximately counting edges in $k$-partite $k$-hypergraphs with colour-coding to obtain an algorithm for general $k$-hypergraphs.
  
  \begin{framed}
  \noindent\textbf{Algorithm $\coarsecountgeneral(G)$.}
  
  \smallskip\noindent\textbf{Input:}
  $G$ is an $n$-vertex $k$-hypergraph, where $n$ is a power of two, to which \coarsecountgeneral has colourful oracle access (only).
  
  \smallskip\noindent\textbf{Behaviour:} $\coarsecountgeneral(G)$ outputs a non-negative integer $\hat{e}$ which, with probability at least~$2/3$, satisfies $\hat{e}/2(4k\log n)^k \le e(G) \le \hat{e}\cdot 2(4k\log n)^k$.
  \shortrule
    \begin{enumerate}[(E1)]
      \item Let $t=12e^{2k}$, and let $T = \ceil{72\ln t}+3$.
      \item For each~$i\in[t]$:
      \begin{enumerate}[(E1)]
        \item[(E3)] Sample a uniformly random function~$c_i:V(G)\to[k]$, which yields a random $k$-partition $X_1,\dots,X_k$ of $V(G)$.
        \item[(E4)] Execute $\coarsecount(G,X_1,\dots,X_k)$ exactly~$T$ times and let $M_i$ be the median output produced by these executions.
      \end{enumerate}
      \item Output $\frac{k^k}{tk!}\sum_{i=1}^t M_i$.
    \end{enumerate}
  \end{framed}

  \begin{lemma}\label{lem:coarse-count-general}
  		\coarsecountgeneral{} behaves as stated, runs in time $\OO(k^{3k}n\log^{2k+2}n)$, and requires $\OO(k^{3k}\log^{2k+2}n)$ oracle queries.
  \end{lemma}
  \begin{proof}
    Let $G$ be an $n$-vertex $k$-hypergraph input for \coarsecountgeneral.

    \textbf{Running time and oracle queries.}
    It is clear that the bottleneck in both the running time and the number of oracle queries is the $Tt$ total invocations of \coarsecount. Recall from \cref{lem:coarse-count} that each such invocation runs in time $\OO(n(8k\log n)^{2k+2})$ and requires $\OO((8k\log n)^{2k+2})$ oracle queries. Since $t = \OO(e^{2k})$ and $T = \OO(k)$, the claimed bounds follow.  
    
    \textbf{Correctness.}
    Let $b := (4 k \log n)^k$ be the approximation ratio of \coarsecount. For all $i \in [t]$, let $G_i = G[c_i^{-1}(1),\dots,c_i^{-1}(k)]$ be the $i$th hypergraph we consider, and let $m_i = e(G_i)$. Let $x_{i,j}$ be the output of the $j$th call to \coarsecount in evaluating $M_i$, and let $\calE_{i,j}$ be the event that $x_{i,j}/b \le m_i \le x_{i,j}b$. 
    
    Note that the $\calE_{i,j}$'s are independent conditioned on $c_i$, and that the correctness of \coarsecount (\cref{lem:coarse-count}) implies that $\pr(\calE_{i,j}) \ge 2/3$ for all $j \in [T]$. Moreover, for all $i \in [t]$, if at least half the $\calE_{i,j}$'s occur, then $M_i/b \le m_i \le bM_i$. Thus, by a Chernoff bound (\cref{lem:chernoff}(i) applied with $\eps=1/4$ and $\mu=2T/3$), we have
    \[
    	\pr(\tfrac{1}{b}M_i \le m_i \le bM_i \mid c_i) \ge 1 - 2e^{-T/72} \ge 1 - 2e^{-\ln t - 3} > 1 - 1/(6t).
    \]
    It follows by a union bound that, with probability at least $5/6$, $M_i/b \le m_i \le bM_i$ for all $i \in [t]$.
    
    Now observe that $\E(\sum_i m_i) = t(k!/k^k)e(G)$, and that each $m_i$ lies in $[0,e(G)]$. It follows by Hoeffding's inequality (\cref{lem:hoeffding}) that
    \begin{align*}
      \Pr\Big(\Big|\frac{k^k}{tk!}\sum_i m_i - e(G) \Big| > \frac{1}{2} e(G) \Big) &= \Pr\Big(\Big|\sum_i m_i - \frac{tk!\cdot e(G)}{2k^k} \Big| > \frac{tk!}{2k^k} e(G) \Big)\\
      &\le 2\exp\bigg({-}2\Big(\frac{tk!}{k^k}e(G)\Big)^2/te(G)^2 \bigg)
      \\
      &= 2\exp\big({-}2t(k!/2k^k)^2\big).
    \end{align*}
    By Stirling's formula and our definition of $t$, it follows that
    \[
	    \Pr\Big(\Big|\frac{k^k}{tk!}\sum_i m_i - e(G) \Big| > \frac{1}{2} e(G) \Big) \le 2\exp\big({-}te^{-2k}/4\big) \le 1/6.
    \]
	Thus, with probability at least $5/6$, we have $e(G)/2 \le \tfrac{k^k}{tk!}\sum_i m_i \le 2e(G)$. 
    
    It now follows by a union bound that with probability at least $2/3$, $\frac{1}{2b}\cdot e(G) \le \tfrac{k^k}{tk!}\sum_i M_i \le 2b\cdot e(G)$ as required.
  \end{proof}

	We now restate \cref{lem:acc} and prove it via the usual probability amplification argument.
	\stateACC*
	\begin{proof}
		Given $G$ and $\delta>0$, we simply invoke $\coarsecountgeneral(G)$ a total of $T := \ceil{36\ln(2/\delta)}$ times and return the median output. By the correctness of \coarsecountgeneral (\cref{lem:coarse-count-general}), each invocation returns a valid approximation of $e(G)$ with probability at least $2/3$, and if at least $T/2$ invocations return valid approximations then the median is also a valid approximation. It follows by Chernoff bounds (\cref{lem:chernoff}(i) with $\eps = 1/2$ and $\mu = T/3$) that we output a valid approximation with probability at least $1 - 2e^{-T/36} \ge 1-\delta$, as required, and our bounds on running time and oracle usage are immediate from \cref{lem:coarse-count-general}.
	\end{proof}

\section{Approximately uniform sampling}\label{sec:sampling}

In this section we demonstrate that we can use our approximate counting algorithm to sample an edge almost uniformly at random, proving \cref{thm:main-sampling}. We begin by sketching our algorithm.

Suppose for the moment that we are given access to a \textit{deterministic} algorithm $\texttt{ExactCount}(G)$ which, given colourful oracle access to a $k$-hypergraph $G$, returns the \textit{exact} value of $|E(G)|$. Let $G$ be an $n$-vertex $k$-hypergraph for which we have access to a colourful independence oracle, and suppose further that $n$ is a power of two and that $G$ contains at least one edge. In this hypothetical scenario, we could proceed to sample an edge of $G$ from the \textit{exact} uniform distribution using iterated rejection sampling as follows.

Let $X_1 = V(G)$ and $M_1 = \texttt{ExactCount}(G)$. Choose a uniformly random subset $X \subset X_1$ of size $n/2$, and let $M = \texttt{ExactCount}(G[X])$. With probability $M/M_1$, we ``accept'' $X$, setting $X_2 = X$ and $M_2 = M$. Otherwise, we ``reject'' $X$, reinitialising it to a new uniformly random size-$(n/2)$ subset, and repeat the process until we accept some set $X$. Likewise, we then choose a uniformly-random subset $X \subset X_2$ of size $n/4$ and let $M = \texttt{ExactCount}(G[X])$; with probability $M/M_2$, we ``accept'' $X$, setting $X_3 = X$ and $M_3 = M$, and otherwise we ``reject'' $X$ and resample. Continuing in this way, we generate two sequences $X_1,\dots,X_I$ and $M_1,\dots,M_I$ with $|X_i| = n/2^{i-1}$ for all $i \le I$, where $I$ is chosen so that $k \le |X_I| \le 2k$. We then enumerate the edges of $G[X_I]$ directly using $O(2^k)$ calls to the colourful independence oracle, and output a uniformly random such edge $F$.

Observe that by a standard rejection sampling argument (see e.g.~Florescu~\cite[Proposition 3.3]{florescu}), for any possible value $(Y_1,\dots,Y_I,e)$ of $(X_1,\dots,X_I,F)$, we have
\begin{align*}
    \pr\big((X_1,\dots,X_I,F) = (Y_1,\dots,Y_I,e)\big) &= \frac{1}{|E(G[Y_I])|} \cdot \prod_{i=1}^{I-1} \frac{|E(G[Y_{i+1}])|}{\sum_{\substack{Z \subseteq Y_i\\|Z| = n/2^i}} |E(G[Z])|}.
\end{align*}
Since for each $i$, each edge of $G[Y_i]$ contributes 1 to exactly $\binom{|Y_i|-k}{|Y_i|/2-k}$ terms in the sum above and 0 to the rest, it follows that
\begin{align*}
    \pr\big((X_1,\dots,X_I,F) = (Y_1,\dots,Y_I,e)\big) &= \frac{1}{|E(G[Y_I])|} \cdot \prod_{i=1}^{I-1} \frac{|E(G[Y_{i+1}])|}{|E(G[Y_i])|}\binom{|Y_i|-k}{|Y_i|/2-k}^{-1}\\
    &= \frac{1}{|E(G)|}\prod_{i=1}^{I-1}\binom{|Y_i|-k}{|Y_i|/2-k}^{-1}.
\end{align*}
On summing over all possible sequences $Y_1,\dots,Y_I$, we recover that $\pr(F=e) = 1/|E(G)|$ as required. Moreover, each edge of $G[X_i]$ appears in $G[X_{i+1}]$ with probability roughly $1/2^k$, so in expectation we will reject roughly $2^k$ samples before accepting each $X_i$, for a total of $\OO(2^k\log n)$ calls to $\texttt{ExactCount}$.

Of course, \aau is neither a deterministic nor an exact counting algorithm. However, with careful bookkeeping, essentially the same algorithm turns out to yield an approximate sample. We do the bulk of the work with the following ancillary algorithm $\sample(G,\eps)$, which assumes that $|V(G)|$ is a power of two and that~$G$ contains at least one edge; we then address these assumptions (which will be easy) in the main proof of \cref{thm:main-sampling} later in the section.

  \begin{framed}
    \noindent
    \textbf{Algorithm $\sample(G,\eps)$.}
    
    \smallskip\noindent
    \textbf{Input:}
	$G$ is an $n$-vertex $k$-hypergraph containing at least one edge, where $n$ is a power of two, to which \coarsecountgeneral has colourful oracle access (only). $0 < \eps < 1/2$ is a rational number.
	
    \smallskip\noindent\textbf{Behaviour:} With probability at least $1-\eps/n^k$, $\sample(G,\eps)$ outputs a sample from a distribution $\hat{U}$ on $E(G)$ such that, for all $e \in E(G)$, $\hat{U}(e) \in (1\pm \eps)/e(G)$.
    
    \shortrule
    \begin{enumerate}[(S1)]
    	\item If $\eps \le n^{-k}$, then enumerate the edges of $G$ using $\binom{n}{k}$ invocations of $\cindora$ and return a uniformly-sampled edge. 
    	\item Let $I = \log n - \ceil{\log (8k^2)}$, $\xi=\eps/(100\log n)$, and $\delta = \xi/2^{k+8}n^{2k}$. If $I \le 1$, enumerate the edges of $e(G)$ using $\cindora$ and return a uniformly random sample.
	    \item Let $X_1 \leftarrow V(G)$, $M_1 \leftarrow \aau(X_1,\xi,\delta)$, and $i\leftarrow 2$.
	    While $i \le I$:
	    \begin{enumerate}
		    \item Choose a size-$(|X_{i-1}|/2)$ set $X \subseteq X_{i-1}$ uniformly at random, and let $M \leftarrow \aau(G[X],\xi,\delta)$. 
		    \item If $M_{i-1} = 0$, output \texttt{Fail}. Otherwise, with probability $\max\{0,1 - M/M_{i-1}\}$, go to (a) (i.e.\ reject $X$ and resample).
		    \item Accept $X$ by setting $X_i \leftarrow X$, $M_i \leftarrow M$ and $i \leftarrow i+1$.
		 \end{enumerate}
    	\item Enumerate the edges of $G[X_I]$ using \cindora and return a uniformly random sample.
	\end{enumerate}
  \end{framed}

  \newcommand{\filt}{\mathcal{F}}
  \begin{lemma}\label{lem:sampling-works}
      $\sample(G,\eps)$ behaves as claimed. With probability at least $1-\eps/n^k$, writing $T = \eps^{-2}k^{7k}\log^{4k+11}n$, its running time is $\OO(nT)$, and it invokes $\cindora$ at most $\OO(T)$ times. 
  \end{lemma}
  \begin{proof}
	  If $\eps \le n^{-k}$ or $I \le 1$, then both correctness and the stated time bounds are clear, so suppose $\eps > n^{-k}$ and $I \ge 2$ (which implies $n \ge 32k^2$). We first carefully bound the probability that something goes wrong over the course of the algorithm's execution. 
	  
	  We define events as follows for all $r \in [I]$:
	  \begin{itemize}
	      \item Let $\calE_r$ be the event that in step (S3) $\aau$ is called at most $2^{k+3}\ln(\tfrac{8 I n^k}{\eps})$ times in calculating $M_r$, that each time it is called it returns $M$ satisfying $M \in (1\pm \xi)e(G[X])$, and that $M_r > 0$.
	      \item Let $\calE_{r,1}$ be the event that when $i=r$, we accept a set $X_r$ within $2^{k+3}\ln(\tfrac{8 I n^k}{\eps})$ iterations of (S3a)--(S3c).
	      \item Let $\calE_{r,2}$ be the event that when $i=r$, we accept a set $X_r$ without $\aau$ ever returning an inaccurate estimate $M \notin (1\pm\xi)e(G[X])$.
	  \end{itemize}
	  Intuitively, $\calE_r$ is the event that the algorithm behaves as we expect in determining $X_r$. We will use $\calE_{r,1}$ and $\calE_{r,2}$ to bound $\Pr(\calE_r \mid \calE_1,\dots,\calE_{r-1})$ below for all $r \in [I]$, and hence bound $\Pr(\calE_1,\dots,\calE_r)$ below. When $r=1$, it follows from \cref{thm:main-counting} and the fact that $e(G) > 0$ that $\Pr(\calE_1) \ge 1-\delta$.
	  
	  Let $2 \le r \le I$, let $\filt$ be a possible filtration of $X_1,\dots,X_{r-1}$ and $M_1,\dots,M_{r-1}$ compatible with $\calE_1,\dots,\calE_{r-1}$, and let $Y$ be the value of $X_{r-1}$ determined by $\filt$. Note that if $\calE_{r,2} \cap \filt$ occurs, then $M_r \in (1\pm\xi)e(G[X_r])$; moreover, since we accepted $X_r$ with probability at most $M_r/M_{r-1}$, we must have $M_r > 0$. Thus
	  \begin{equation}\label{eqn:sampling-works-1}
	  	\Pr(\calE_r \mid \filt) \ge \Pr(\calE_{r,1} \cap \calE_{r,2} \mid \filt).
	  \end{equation}
	  
	  To bound $\Pr(\calE_{r,1} \cap \calE_{r,2} \mid \filt)$ below, consider the first iteration of (S3a)--(S3c) with $i=r$. In this iteration, let $\mathcal{A}_1$ be the event that $\aau$ returns an inaccurate estimate $M$, and let $\mathcal{A}_2$ be the event that $\aau$ returns an accurate estimate $M$ and we subsequently accept~$X$; note that each of these events is independent of past samples, and that their probability does not depend on the value of $i$. By \cref{thm:main-counting}, we accept $X$ in any given round with probability at least $\Pr(\mathcal{A}_2\mid\filt)-\delta$; hence
	  \begin{equation}\label{eqn:sampling-works-1a}
	      \Pr(\calE_{r,1} \mid \filt) \ge 1 - \big(1 - \Pr(\mathcal{A}_2 \mid \filt) + \delta\big)^{2^{k+3}\ln(8In^k/\eps)}.
	  \end{equation}
	  Let $T$ be the number of iterations of (S3a)--(S3c) before either $\aau$ returns an inaccurate estimate or we accept an accurate estimate. By Bayes' theorem, for all $t \ge 0$, we have
	  \begin{align*}
	      \pr(\calE_{r,2}\mid\filt\mbox{ and }T=t) &= \frac{\pr(\calE_{r,2}\mbox{ and }T=t\mid\filt)}{\pr(T=t\mid\filt)} = \frac{\pr(\calE_{r,2}\mbox{ and }T=t\mid\filt\mbox{ and }T\ge t)}{\pr(T=t\mid\filt\mbox{ and }T\ge t)}\\
	      &= \frac{\pr(\mathcal{A}_2\mid\filt)}{\pr(\mathcal{A}_2\mid\filt)+\pr(\mathcal{A}_1\mid\filt)}.
	  \end{align*}
	  Summing over all values of $t$ yields
	  \begin{align}
	  	\Pr(\calE_{r,2}\mid\filt) &= \frac{\Pr(\mathcal{A}_2\mid\filt)}{\Pr(\mathcal{A}_2\mid\filt)+\Pr(\mathcal{A}_1\mid\filt)}\,.\label{eqn:sampling-works-2}
	  \end{align}
	  
	  We now bound $\Pr(\mathcal{A}_1 \mid \filt)$ above and $\Pr(\mathcal{A}_2 \mid \filt)$ below. \cref{thm:main-counting} implies that 
	  \begin{equation}\label{eqn:sampling-works-3}
	  	\Pr(\mathcal{A}_1\mid\filt) \le \delta.
	  \end{equation}
	  Moreover, for all $S \subset Y$ with $|S| = |Y|/2$, \cref{thm:main-counting} implies that
	  \begin{align*}
	  	\Pr\big(\mathcal{A}_2 \mbox{ occurs and }X = S\mid\filt\big)
	  	&\ge \binom{|Y|}{|Y|/2}^{-1} \cdot (1-\delta) \cdot \frac{(1-\xi)e(G[S])}{M_{r-1}}\,.
	  \end{align*}
	  By the definition of $\filt$ we have $M_{r-1} \in (1\pm\xi)e(G[Y])$, so it follows that
	  \[
		\Pr\big(\mathcal{A}_2 \mbox{ occurs and }X = S\mid\filt\big) \ge \frac{1}{2}\binom{|Y|}{|Y|/2}^{-1}\frac{e(G[S])}{e(G[Y])}\,.
	  \]
	  On summing both sides over $S$, since each edge of $G[Y]$ appears in exactly $\binom{|Y|-k}{|Y|/2-k}$ sets $S\subset Y$ with $|S| = |Y|/2$, we obtain
	  \[
	  	\Pr(\mathcal{A}_2 \mid \filt) \ge \frac{1}{2}\binom{|Y|}{|Y|/2}^{-1}\binom{|Y|-k}{|Y|/2-k}.
	  \]
	  Since $r \le I$, we have $|Y| \ge n/2^{I-1} \ge 4k^2$, so by \cref{lem:bin-bound} it follows that $\Pr(\mathcal{A}_2 \mid \filt) \ge 2^{-k-2}$. It therefore follows from~\eqref{eqn:sampling-works-1a},~\eqref{eqn:sampling-works-2} and~\eqref{eqn:sampling-works-3} that 
	  \begin{align*}
		  \Pr(\calE_{r,1} \mid \filt) &\ge 1 - (1-2^{-k-2}+\delta)^{2^{k+3}\ln(8In^k/\eps)} \ge 1 - (1-2^{-k-3})^{2^{k+3}\ln(8In^k/\eps)}\\
		  & \ge 1 - e^{-\ln(8In^k/\eps)} = 1 - \eps/8In^k,\\
		  \Pr(\calE_{r,2} \mid \filt) &\ge 2^{-k-2}/(2^{-k-2}+\delta) \ge 1 - 2^{k+2}\delta \ge 1-\eps/8In^k.
	  \end{align*}
	  By~\eqref{eqn:sampling-works-1}, it follows that $\Pr(\calE_r \mid \filt) \ge 1 - \eps/4In^k$ for all $r \in [I]$. Thus
	  \begin{align}\label{eqn:sampling-works-4-pre}\Pr(\calE_r\mid\calE_1,\dots,\calE_{r-1}) &\ge 1-\eps/4In^k \mbox{ for all }r \in [I], \mbox{ and}\\
	  \label{eqn:sampling-works-4}
	  	\Pr(\calE_1, \dots, \calE_I) &\ge 1 - \eps/4n^k.
	  \end{align}
	  With these equations in hand, we are now ready to proceed with the main proof.
	  
	  \textbf{Running time and oracle queries.} Since $|X_I| = n/2^{I-1} \le 32k^2$ by our choice of $I$, the bottleneck in the running time is the invocations of \aau in (S3b). By \cref{thm:main-counting}, each invocation takes time \[\OO(\log(1/\delta)\xi^{-2}k^{6k}n\log^{4k + 7}n)\] and requires $\OO(\log(1/\delta)\xi^{-2}k^{6k}\log^{4k + 7}n)$ invocations of the oracle. Since $1/\xi = \OO((\log n)/\eps)$, $\log(1/\delta) = \OO(k\log n + \log(1/\eps))$ and $1/\eps = \OO(n^k)$, each invocation takes time $\OO(\eps^{-2}k^{6k+1}n\log^{4k+10}n)$ and requires $\OO(\eps^{-2}k^{6k+1}\log^{4k+10}n)$ oracle invocations. As by \eqref{eqn:sampling-works-4}, with probability at least $1-\eps/n^k$ there are at most $2^{k+3}\ln(8In^{k}/\eps) = \OO(k2^k\log n)$ such invocations, the claimed bounds follow. 
	  
	  \textbf{Correctness.} Let $F$ be the output of $\sample(G,\eps)$, or $\texttt{Fail}$ if it does not halt. Since $e(G) > 0$, by~\eqref{eqn:sampling-works-4}, $\sample(G,\eps)$ outputs a sample from $E(G)$ with probability at least $1-\eps/n^k$; thus to prove \cref{lem:sampling-works}, it suffices to show that for all $f \in E(G)$ we have $\Pr(F = f) \in (1\pm\eps)/e(G)$.
	  
	  Let $S_1 = V(G)$ and, for all $S_1 \supset S_2 \supset \dots \supset S_I \supset f$ with $|S_r| = n/2^{r-1}$ for all $r \in [I]$, let
	  \[
  	  	p(S_1,\dots,S_I,f) = \Pr(X_r=S_r\mbox{ for all }r \in [I],\ F = f,\mbox{ and }\calE_1,\dots,\calE_I\mbox{ occur}).
	  \]
	  Thus for all $f \in E(G)$, we have
	  \begin{align*}
		  \Pr(F = f) &\ge \sum_{\substack{S_1 \supset \dots \supset S_I\supset f\\|S_r|=n/2^{r-1}}} p(S_1,\dots,S_I,f),\\
		  \Pr(F = f) &\le \sum_{\substack{S_1 \supset \dots \supset S_I\supset f\\|S_r|=n/2^{r-1}}} p(S_1,\dots,S_I,f) + \big(1 - \Pr(\calE_1 \cap \dots \cap \calE_I)\big).
	  \end{align*}
	  By~\eqref{eqn:sampling-works-4}, it follows that 
	  \begin{equation}\label{eqn:sampling-works-4a}
	  	  \sum_{\substack{S_1 \supset \dots \supset S_I\supset f\\|S_r|=n/2^{r-1}}} p(S_1,\dots,S_I,f) \le \Pr(F = f) \le \sum_{\substack{S_1 \supset \dots \supset S_I\supset f\\|S_r|=n/2^{r-1}}} p(S_1,\dots,S_I,f) + \frac{\eps}{4n^k}\,.
	  \end{equation}
	  
	  To bound each term $p(S_1,\dots,S_I,f)$, we first derive estimates for the probability that $X_r = S_r$, conditioned on $X_t = S_t$ for all $t \in [r-1]$ and on $\calE_1,\dots,\calE_{r-1}$. Let $\filt_r$ be an arbitrary filtration of $X_1, \dots, X_{r-1}$ and $M_1,\dots,M_{r-1}$ compatible with these events. For all $S \subset S_{r-1}$ with $|S|=|S_{r-1}|/2$, let $q(S)$ be the probability that we accept $S$ on a given iteration of (S3a)--(S3c) conditioned on $X=S$ and $\filt_r$. Then by \cref{thm:main-counting} and the definition of $\filt_r$,
	  \begin{equation}\label{eqn:sampling-works-5}
  	  	q(S) \ge (1-\delta) \frac{(1-\xi)e(G[S])}{(1+\xi)e(G[S_{r-1}])} \mbox{ and }q(S) \le \frac{(1+\xi)e(G[S])}{(1-\xi)e(G[S_{r-1}])}+\delta.
	  \end{equation}
	   Moreover, by a standard rejection sampling argument (see e.g.\ Florescu~\cite[Proposition 3.3]{florescu}), we have
	   \begin{equation}\label{eqn:sampling-works-6}
	   	\Pr(X_r = S_r \mid \filt_r) = \frac{q(S_r)}{\sum_{\substack{T \subset S_{r-1}\\|T|=|S_{r-1}|/2}} q(T)}\,.
	   \end{equation}
	   
	   By~\eqref{eqn:sampling-works-5} and~\eqref{eqn:sampling-works-6}, and using the fact that $e(G[S_r]) \ge 1$, we have
	   \begin{align*}
		   	\Pr(X_r = S_r \mid \filt_r) 
		   	&\le \Big(\frac{(1+\xi)e(G[S_r])}{(1-\xi)e(G[S_{r-1}])}+\delta\Big)\Big/\Big(\sum_{\substack{T \subset S_{r-1}\\|T|=|S_{r-1}|/2}}\frac{(1-\delta)(1-\xi)e(G[T])}{(1+\xi)e(G[S_{r-1}])}\Big)\\
		   	&= \frac{(1+\xi)^2}{(1-\xi)^2(1-\delta)}\cdot\frac{e(G[S_r])+\delta e(G[S_{r-1}])}{\sum_{\substack{T \subset S_{r-1}\\|T|=|S_{r-1}|/2}} e(G[T])}\\
		   	&\le \frac{(1+\xi)^3}{(1-\xi)^3}\cdot\frac{e(G[S_r])}{\sum_{\substack{T \subset S_{r-1}\\|T|=|S_{r-1}|/2}} e(G[T])}\,.
	   \end{align*}
	   Since each edge of $G[S_{r-1}]$ appears exactly $\binom{|S_{r-1}|-k}{|S_{r-1}|/2-k}$ times in this sum, it follows that
	   \begin{align}\nonumber
	   		\Pr(X_r = S_r \mid \filt_r) 
	   		&\le \frac{(1+\xi)^3}{(1-\xi)^3}\cdot\frac{e(G[S_r])}{\binom{|S_{r-1}|-k}{|S_{r-1}|/2-k}e(G[S_{r-1}])} \\\label{eqn:sampling-works-6a}
	   		&\le (1+8\xi)\frac{e(G[S_r])}{\binom{|S_{r-1}|-k}{|S_{r-1}|/2-k}e(G[S_{r-1}])}\,.
	   \end{align}
	   (Here the last inequality follows since $\xi < 1/20$.)
	   
	   Also by~\eqref{eqn:sampling-works-5} and~\eqref{eqn:sampling-works-6}, we have 
	   \begin{align*}
	   	\Pr(X_r = S_r \mid \filt_r) &\ge \frac{(1-\delta)(1-\xi)e(G[S_r])}{(1+\xi)e(G[S_{r-1}])} \Big/ \bigg(\sum_{\substack{T \subset S_{r-1}\\|T|=|S_{r-1}|/2}} \Big(\frac{(1+\xi)e(G[T])}{(1-\xi)e(G[S_{r-1}])}+\delta\Big)\bigg)\\
	   	&\ge \frac{(1-\delta)(1-\xi)^2}{(1+\xi)^2}\cdot\frac{e(G[S_r])}{\sum_{\substack{T \subset S_{r-1}\\|T|=|S_{r-1}|/2}} \big(e(G[T]) + \delta e(G[S_{r-1}]) \big)}.
	   	\end{align*}
        Since there are $\binom{|S_{r-1}|}{|S_{r-1}|/2}$ terms in the sum, and since each edge of $G[S_{r-1}]$ contributes 1 to $\binom{|S_{r-1}|-k}{|S_{r-1}|/2-k}$ terms of the sum and zero to the rest, it follows that
	   	\begin{align*}
	   	\Pr(X_r = S_r \mid \filt_r) &\ge \frac{(1-\xi)^3}{(1+\xi)^2}\cdot\frac{e(G[S_r])}{\binom{|S_{r-1}|-k}{|S_{r-1}|/2-k}e(G[S_{r-1}]) + \binom{|S_{r-1}|}{|S_{r-1}|/2}\delta e(G[S_{r-1}])}\,.\\
	   \end{align*}
	   Since $|S_{r-1}| = n/2^{r-1} \ge 4k^2$, by \cref{lem:bin-bound} we have 
	   \[
	   	\binom{|S_{r-1}|}{|S_{r-1}|/2} \le 2^{k+1}\binom{|S_{r-1}|-k}{|S_{r-1}|/2-k}.
	   \]
	   It follows that
	   \begin{align}\nonumber
	   	\Pr(X_r = S_r \mid \filt_r) &\ge \frac{(1-\xi)^3}{(1+\xi)^2}\cdot \frac{e(G[S_r])}{(1+2^{k+1}\delta)\binom{|S_{r-1}|-k}{|S_{r-1}|/2-k}e(G[S_{r-1}])}\\\label{eqn:sampling-works-6b}
	   	&\ge (1-6\xi)\frac{e(G[S_r])}{\binom{|S_{r-1}|-k}{|S_{r-1}|/2-k}e(G[S_{r-1}])}\,.
	   \end{align}
	   
	   With upper and lower bounds on $\Pr(X_r = S_r \mid \filt_r)$ now in place, we return to the task of bounding $p(S_1,\dots,S_I,f)$. Observe that for all $f \in E(G)$,
	   \[
	   	\sum_{\substack{S_1 \supset \dots \supset S_I\supset f\\|S_r|=n/2^{r-1}}} p(S_1,\dots,S_I,f) 
	   	= \sum_{\substack{S_1 \supset \dots \supset S_I\supset f\\|S_r|=n/2^{r-1}}}\frac{1}{e(G[S_I])}\prod_{r=1}^I \Pr\big(X_r = S_r \mbox{, }\calE_r\mbox{ occurs} \mid \filt_r\big).
	   \]
	   It therefore follows from~\eqref{eqn:sampling-works-6a} that
	   \begin{align*}
	   	\sum_{\substack{S_1 \supset \dots \supset S_I \supset f\\|S_r| = n/2^{r-1}}} p(S_1,\dots,S_I,f) 
	   	&\le \sum_{\substack{S_1 \supset \dots \supset S_I \supset f\\|S_r| = n/2^{r-1}}} \frac{1}{e(G[S_I])}\prod_{r=2}^I \frac{(1+8\xi)e(G[S_r])}{\binom{|S_{r-1}|-k}{|S_{r-1}|/2-k}e(G[S_{r-1}])}.
	   \end{align*}
	   (Recall that $e(G[S_r])\ge 1$ for all $r\in[I]$.) We have $(1+8\xi)^I \le e^{8I\xi} \le 1 + 16I\xi$, so on collapsing the telescoping product we obtain
	   \begin{equation*}
	   	\sum_{\substack{S_1 \supset \dots \supset S_I \supset f\\|S_r| = n/2^{r-1}}} p(S_1,\dots,S_I,f) 
	   	\le \sum_{\substack{S_1 \supset \dots \supset S_I \supset f\\|S_r| = n/2^{r-1}}} \frac{1+16I\xi}{e(G)}\prod_{r=2}^I \binom{n/2^{r-2}-k}{n/2^{r-1}-k}^{-1}.
	   \end{equation*}
	   All terms of this sum are equal, and there are precisely $\prod_{r=0}^{I-2} \binom{n/2^r-k}{n/2^{r+1}-k}$ terms, so 
	   \begin{equation*}
	    \sum_{\substack{S_1 \supset \dots \supset S_I \supset f\\|S_r| = n/2^{r-1}}} p(S_1,\dots,S_I,f)  \le \frac{1 + 16I\xi}{e(G)} \le \frac{1+\eps/2}{e(G)}.
	   \end{equation*}
	   Hence by~\eqref{eqn:sampling-works-4a}, we have $\Pr(F=f) \le (1+\eps)/e(G)$, as required.
	   
	   Similarly, it follows from~\eqref{eqn:sampling-works-6b} that
	   \begin{align*}
	   	\sum_{\substack{S_1 \supset \dots \supset S_I\supset f\\|S_r|=n/2^{r-1}}} p(S_1,\dots,S_I,f) 
	   	&\ge \sum_{\substack{S_1 \supset \dots \supset S_I\supset f\\|S_r|=n/2^{r-1}}} \frac{1-\delta}{e(G[S_I])}\prod_{r=2}^I \frac{(1-6\xi)e(G[S_r])}{\binom{|S_{r-1}|-k}{|S_{r-1}|/2-k}e(G[S_{r-1}])}\\
	   	&\qquad -\, \sum_{r=1}^I \big(1 - \pr(\calE_r \mid \calE_1,\dots,\calE_{r-1})\big).
	   \end{align*}
	   By~\eqref{eqn:sampling-works-4-pre}, the last term is bounded above by $\eps/4n^k$; it follows that
	   \begin{align*}
	   	\sum_{\substack{S_1 \supset \dots \supset S_I\supset f\\|S_r|=n/2^{r-1}}} p(S_1,\dots,S_I,f) 
	   	&\ge \sum_{\substack{S_1 \supset \dots \supset S_I\supset f\\|S_r|=n/2^{r-1}}} \frac{1-6I\xi}{e(G)}\prod_{r=2}^I\binom{n/2^{r-2}-k}{n/2^{r-1}-k}^{-1} - \frac{\eps}{4n^k}\\
	   	&= \frac{1-6I\xi}{e(G)} - \frac{\eps}{4n^k} \ge \frac{1-\eps}{e(G)}.
	   \end{align*}
	   It therefore follows from~\eqref{eqn:sampling-works-4a} that $\Pr(F=f) \ge (1-\eps)/e(G)$ as required.
	\end{proof}
  
    It is now easy to prove \cref{thm:main-sampling} from \cref{lem:sampling-works}.
    \statemainsampling*
    \begin{proof}
    	To evaluate $\realsample(G,\eps)$, we first make $n$ into a power of two by adding at most $n$ isolated vertices to $G$; note that this does not impede the evaluation of $\cindora$. We then call $\sample(G,\eps/3)$. If it returns $\texttt{Fail}$, or does not return a value within $\OO(nT)$ time and $\OO(T)$ oracle queries, then we return $\texttt{Fail}$. Otherwise, we return its output. Writing $F$ for our output, by \cref{lem:sampling-works}, for all $f \in E(G)$, we have $\Pr(F = f) \le (1+\eps)/e(G)$ and
    	\[
    		\Pr(F = f) \ge \frac{1 - \eps/3}{e(G)} - \frac{2\eps}{3n^k} \ge \frac{1 - \eps}{e(G)},
    	\]
    	as required.
    \end{proof}

\section{Corollaries of \cref{thm:main-counting,thm:main-sampling}}\label{sec:corols}

In this section, we restate and prove some fairly straightforward consequences of our main results, which allow us to turn decision algorithms for various problems in parameterised and fine-grained complexity into approximate counting and sampling algorithms.

\statecombinedmain*
\begin{proof}
    We may assume without loss of generality that $\eps > n^{-k}$, since otherwise we can count or sample by brute force in time $\OO(\eps^{-1}T)$ by applying the (simulated) colourful independence oracle to each $k$-element subset of $V(G)$. 
    
    We next reduce the failure probability of $A_G$ with a standard probability amplification technique. Let $A_G'$ be the algorithm with runs $A_G$ on the given inputs $\lceil 600k\log (n/\delta)\rceil$ times and returns the most common answer. By Lemma~\ref{lem:chernoff}, the probability that this answer is incorrect is at most $2e^{-600k\log(n/\delta)/36} < (n/\delta)^{-14k}$, and each invocation of $A_G'$ takes $\OO(Tk\log n)$ time. We now simply run $\aau(G,\eps,\delta/2)$ and $\realsample(G,\eps/2)$ and return the results, using $A_G'$ to simulate each call to the colourful independence oracle of $G$. 
    
    Recall the properties of $\aau$ and $\realsample$ from \cref{thm:main-counting,thm:main-sampling}. In both algorithms, $A_G'$ is invoked $\OO(\eps^{-2}k^{7k}\log^{4k+11}n)$ times; by a union bound, for sufficiently large $n$, the probability it returns the correct answer on every invocation is at least
    \[
        1 - \OO\big(\eps^{-2}k^{7k}\log(1/\delta)\log^{4k+11}n \cdot (n/\delta)^{-14k}\big) \ge 1 - (n/\delta)^{-10k}.
    \]
    Conditioned on all oracle calls returning the correct answer, \aau returns a valid $\eps$-approximation with probability at least $1-\delta/2$ and \realsample returns a valid $(\eps/2)$-approximate sample. It follows by union bounds that without this conditioning, \aau returns a valid $\eps$-approximation with probability at least $1-\delta/2-n^{-10k}$, and \realsample returns a valid $(n^{-10k}+\eps/2)$-approximate sample with no possibility of failure. Since $(n/\delta)^{-10k} \le \delta/2$ and $(n/\delta)^{-10k} \le \eps/2$, it follows that \texttt{SampleCount} behaves as desired. Moreover, by our running time bounds on \aau, \realsample and $A_G'$, the running time of \texttt{SampleCount} is $\eps^{-2}(k\log n)^{\OO(k)}(n+T)$ as required.
\end{proof}

We believe that \cref{thm:main-combined-colourful} does follow from a careful reading of the proofs of \cref{thm:main-counting,thm:main-sampling} in the way one might expect, replacing all randomly-applied colourings of the graph by the colouring induced by $X_1,\dots,X_k$. However, for the benefit of the reader, we give a slightly less efficient but far more easily-checkable algorithm by means of a standard colour coding argument.

\statecombinedmaincolourful*
\begin{proof}
    By \cref{thm:main-combined}, it suffices to simulate the colourful independence oracle of $G$ in time $(k\log n)^{\OO(k)}(n+T)$. Let $X_1,\dots,X_k \subseteq V(G)$ be an input to the colourful independence oracle of $G$, and let $X_{i,j} = X_i \cap V_j$ for all $i,j \in [k]$. Since $G$ is $k$-partite with vertex classes $V_1,\dots,V_k$, every edge $e$ of $G[X_1 \cup \dots \cup X_k]$ contains one vertex from each of $X_{1,\sigma_e(1)},X_{2,\sigma_e(2)},\dots,X_{k,\sigma_e(k)}$ for some map $\sigma_e\colon[k]\to[k]$. Moreover, $e$ is colourful under $X_1,\dots,X_k$ if and only if $\sigma_e$ is a bijection. Motivated by this, for each bijection $\sigma:[k]\to[k]$, let $G_\sigma = G[X_{1,\sigma(1)} \cup \dots \cup X_{k,\sigma(k)}]$; then every colourful edge of $G$ appears in exactly one $G_\sigma$, and every edge appearing in any $G_\sigma$ is colourful. It follows that
    \[
        \mbox{cIND}_{G}(X_1,\dots,X_k) = \min\big\{\indora(X_{1,\sigma(1)} \cup \dots \cup X_{k,\sigma(k)})\colon \sigma\mbox{ a bijection } [k]\to [k] \big\}.
    \]
    We can compute each tuple $(X_{1,\sigma(1)},\dots,X_{k,\sigma(k)})$ in $\OO(n)$ time, each oracle invocation takes $T$ time, and there are $k!$ such tuples to compute; thus overall we can compute $\mbox{cIND}_{G}(X_1,\dots,X_k)$ in $(k\log n)^{\OO(k)}(n+T)$ time as required.
\end{proof}

Finally, we give a formal proof of Theorem~\ref{thm:meta-informal}, which is straightforward from \cref{thm:main-combined} and \cref{def:uwp}.

\statemetainformal*
\begin{proof}
    Let $x$ be an instance of $\Pi$, and let $\eps > 0$. For all disjoint $X_1,\dots,X_{k_x} \subseteq V(G_x)$, we may compute $\mbox{cIND}_{G_x}(X_1,\dots,X_{k_x})$ as follows:
    \begin{enumerate}[(i)]
        \item Compute $I_x(X_1 \cup \dots \cup X_{k_x})$ using the algorithm of Definition~\ref{def:uwp}(iii);
        \item Run the assumed \pp{Colourful}-$\Pi$ algorithm on $y=I_x(X_1 \cup \dots \cup X_{k_x})$ with input partition $X_1,\dots,X_{k_x}$ of $G_y = G_x[X_1 \cup \dots \cup X_{k_x}]$.
        \item Return $1$ if \pp{Colourful}-$\Pi$ returns \no and $0$ if \pp{Colourful}-$\Pi$ returns \yes.
    \end{enumerate}
    Denote this implementation by $A$. For brevity, let $T^+(x) = \max\big\{T(I_x(S)) \colon S \subseteq V(G_x)\big\}$; then $A$ has running time $\tildeOO(|x| + T^+(x))$. Since $T^+(x) \ge T(x) = \widetilde\Omega(|x|)$ by hypothesis, $A$ therefore has running time $\tildeOO(T^+(x))$.
    
    Our algorithm now simply computes $V(G_x)$ and $k_x$ from $x$ using the algorithm of Definition~\ref{def:uwp}(ii), then returns $\samplecount(V(G_x), k_x, A, \eps, 2/3)$; by Corollary~\ref{thm:main-combined}, this will yield an $\eps$-approximation to $|E(G_x)|$ and the desired $\eps$-approximate sample from $E(G_x)=W_x$ with the desired error probability. By Definition~\ref{def:uwp}(i) we have $|E(G_x)| = \#\Pi(x)$, and $E(G_x) = W_x$ by definition, so the algorithm performs as required. 
    
    We now bound the running time. Observe from Definition~\ref{def:uwp}(ii) that $n_x = \tildeOO(|x|)$, and recall that by hypothesis we have $T^+(x) \ge T(x) = \widetilde\Omega(|x|)$; hence by \cref{thm:main-combined}, the overall running time is
    \begin{align*}
        &\tildeOO(|x|) + \eps^{-2}\log^2(1/\delta)(k_x\log n_x)^{\tildeOO(k)}(n_x + T^+(x))\\
        &\qquad\qquad\qquad\qquad\qquad\qquad\qquad\qquad = \eps^{-2}\log^2(1/\delta)(k_x\log n_x)^{\tildeOO(k)}T^+(x),
    \end{align*}
    as required.
\end{proof}

\subsection{Application to Graph Motif}
\label{sec:graphMotif}

We now describe an approximate counting problem in parameterised complexity for which our results imply an immediate improvement over the best stated running time bound in the literature:
the \pp{Graph Motif} problem, introduced by Lacroix, Fernandes and Sagot~\cite{lacroix06motif} in the context of metabolic networks.
This problem takes as input an $n$-vertex $m$-edge graph with a (not necessarily proper) vertex-colouring, together with a multiset $M$ of colours. We write $k=|M|$. A \emph{motif witness} for~$M$ is a set $U\subseteq V(G)$ of $k$ vertices such that the induced subgraph~$G[U]$ is connected and the colour multiset of $U$ is exactly $M$. If~$M$ has a motif witness, then~$M$ is called a \emph{motif}.
Without loss of generality, let us assume that the set~$C$ of allowed colours always satisfies $C=\{1,\dots,n\}$.
The problem is formally stated as follows.
\begin{framed}
	\noindent
	\pp{Graph Motif}\\
	\textit{Input:} A graph $G$ on $n$ vertices and $m$ edges, a colouring ${c\colon V(G) \rightarrow C}$, and a multiset $M$ consisting of elements of $C$, with $|M| = k$.\\
	\textit{Question:} Is there a set $U \subseteq V(G)$ with $|U| = k$ such that $U$ induces a connected subgraph of $G$ and the multiset $\{c(u): u \in U\}$ is equal to $M$?
\end{framed}

There has been substantial progress in recent years on improving the running-time of decision algorithms for \pp{Graph Motif}~\cite{BBFKN-motifs,bjorklund16motif,FFHV-motifs,guillemot13,K-motifs}.
Björklund, Kaski, and Kowalik \cite{bjorklund16motif} gave the fastest known randomised algorithm to solve (a generalisation of) this problem; in the following theorem, $\mu = \OO(\log k \log \log k \log \log \log k)$ accounts for the time required to carry out arithmetic operations in a finite field of size $\OO(k)$ and characteristic 2.

\begin{theorem}[Björklund, Kaski, and Kowalik~\cite{bjorklund16motif}]\label{thm:graph-motif-uncol}
	There exists a Monte Carlo algorithm for \pp{Graph Motif} that runs in $\OO(2^kk^2m\mu)$  time and in polynomial space, with the following guarantees: (i) the algorithm always returns \textnormal{\texttt{No}} when given a \textnormal{\texttt{No}}-instance as input, (ii) the algorithm returns \textnormal{\texttt{Yes}} with probability at least 1/2 when given a \textnormal{\texttt{Yes}}-instance as input.
\end{theorem}

For the counting version of \pp{Graph Motif}, Guillemot and Sikora \cite{guillemot13} addressed the related problem of counting $k$-vertex sub\textit{trees} whose multiset of vertex colours equals~$M$. This problem is equivalent to counting motif witnesses~$U$ for~$M$ weighted by the number of trees spanned by $U$. When~$M$ is a set, this exact counting problem admits an FPT algorithm, but is \#W[1]-hard otherwise. Subsequently, Jerrum and Meeks~\cite{jerrummeeksconnected} addressed the more natural counting analogue of \pp{Graph Motif} in which the goal is to count all motif witnesses for $M$ (without weights).  They demonstrated that this problem is \#W[1]-hard to solve exactly even if $M$ is a set, but gave an FPTRAS to solve it approximately. The goal in \cite{jerrummeeksconnected} was simply to demonstrate the existence of an FPTRAS rather than to optimise the running time; the algorithm as described has running time $\Omega(n^3)$.
We believe that, with sufficient care, the general strategy described in \cite{jerrummeeksconnected} could be adapted to give a running time similar to that obtained in the following corollary; however, \cref{thm:meta-informal} allows us to deduce this improvement immediately from \cref{thm:graph-motif-uncol}. Moreover, it provides a method for translating any future improvement to the decision algorithm into an improved algorithm for approximate counting or sampling.
\begin{corollary}\label{cor:motif-algo}
	Given an $n$-vertex, $m$-edge instance $(G,c,M)$ of \pp{Graph Motif} with $k=|M|$ and $0 < \eps < 1$, there is a randomised algorithm to $\eps$-approximate the number of motif witnesses or to draw an $\eps$-approximate sample from the set of motif witnesses in time $\varepsilon^{-2}k^{\OO(k)}m\log^{\OO(k)}n$.
\end{corollary}

In order to apply \cref{thm:meta-informal}, we need to show that \pp{Graph Motif} is a uniform witness problem according to \cref{def:uwp}.  Given an instance $x=(G,c,M)$, we let~$G_x$ be the $k$-hypergraph on the vertex-set $V(G)$ whose edges are all sets $U\subseteq V(G)$ that are motif witnesses for~$M$. It is clear that~$G_x$ satisfies the conditions of \cref{def:uwp} on taking $I_x(S) = (G[S], c|_S, M)$, and so \pp{Graph Motif} is a uniform witness problem.
Another precondition of \cref{thm:meta-informal} is that any algorithm for the problem requires at least time~$\widetilde{\Omega}(n)$; this is true for \pp{Graph Motif}, because any algorithm for it must read a constant proportion of the input. Finally, before we can apply \cref{thm:meta-informal} to immediately obtain \cref{cor:motif-algo}, we must state an algorithm for the problem \pp{Colourful-Graph Motif} of Definition~\ref{def:colourful-problems}; to avoid confusion with the colouring that already exists, we instead call this \pp{Partitioned Graph Motif}.  It is clear that the maximum running time for this algorithm on any sub-instance will be dominated by that for the original instance.

\begin{lemma}\label{lem:colourful-motif}
    There exists a randomised algorithm for \pp{Partitioned Graph Motif} that runs in time $k^{\OO(k)}m$, with error probability at most $1/3$.
\end{lemma}
\begin{proof}
    The input to \pp{Partitioned Graph Motif} consists of $(G,c,M)$ and a partition of $V(G)$ into disjoint sets $S_1,\dots,S_k$. We now describe an algorithm that decides whether there is a motif witness~$U\subseteq V(G)$ that intersects each set~$S_i$ in exactly one vertex.

    Write $M = \{c_1,\ldots,c_k\}$. For each possible bijection $\pi:[k] \rightarrow [k]$ we will determine, with probability at least $1 - \frac{1}{3k!}$, whether there is a motif witness $U$ with $U=\{u_1,\ldots,u_k\}\subseteq V(G)$ such that, for all $i \in [k]$, we have $u_i \in S_i$ and $c(u_i) = c_{\pi(i)}$.
	We achieve this by solving a new instance $(G,c',M')$ of \pp{Graph Motif} using the algorithm of \cref{thm:graph-motif-uncol}: we use the same input graph, but define a new colouring $c': V(G) \rightarrow C \times [k]$, where $c'(v) = (c,i)$ if and only if $c(v) = c$ and $v \in V_i$.
	Moreover, we set $M' = \{(c_{\pi(1)},1),\ldots,(c_{\pi(k)},k)\}$.
	We silently replace $C\times [k]$ with $[n]$ by discarding unused colours and renaming the rest. To achieve the claimed success probability, it suffices to call the randomised algorithm for \pp{Graph Motif} a total of $\ceil{3 k \log k}$ times, returning \texttt{Yes} if any of the calls returns \texttt{Yes}, and \texttt{No} otherwise.

	We return \textnormal{\texttt{Yes}} if any of our trials over all possibilities for $\pi$ returns \textnormal{\texttt{Yes}}; otherwise we return \textnormal{\texttt{No}}.  By a union bound, the probability that we obtain the correct answer for all $k!$ choices of $\pi$ is at least 2/3, and in this case we output the correct answer.

	In total, we invoke the randomised \pp{Graph Motif} algorithm $k! \ceil{3 k \log k}$ times, so the total running time is $\OO(k! k \log k \cdot 2^k k^2 m \log k \log \log k \log \log \log k) = k^{\OO(k)}m$.
\end{proof}

\cref{cor:motif-algo} is now immediate from \cref{lem:colourful-motif} and \cref{thm:meta-informal}. (Recall that, for any $k$-colour $m$-edge instance $x = (G,c,M)$ of \pp{Graph Motif} and any $S \subseteq V(G_x) = V(G)$, we have $I_x(S) = (G[S], c|_S, M)$; thus the maximum in the right-hand side of~\eqref{eq:meta-informal-runtime} is simply $k^{\OO(k)}m$.)

\subsection{Application to \texorpdfstring{{\boldmath$k$-SUM}}{k-SUM}}\label{sec:ksum-proof}
The $k$-SUM problem has been studied since the 1990s as it arises naturally in the context of computational geometry (see for example~\cite{DBLP:journals/comgeo/GajentaanO12}), and it has become an important problem in fine-grained complexity theory~\cite{DBLP:conf/iwpec/Williams15}.
For all integers $k \ge 3$, the \pp{$k$-SUM} problem asks, given a set of integers, whether some $k$ of them sum to zero. Each $k$-subset of integers that does sum to zero is called a \emph{witness}.
While Kane, Lovett, and Moran~\cite{DBLP:journals/jacm/KaneLM19} very recently developed almost linear-size linear decision trees for $k$-SUM, the fastest known algorithm for this problem still runs in time $\tildeOO(n^{\lceil k/2 \rceil})$.
A running time of $n^{o(k)}$ as $k\to\infty$ is ruled out under the exponential-time hypothesis~\cite{DBLP:conf/soda/PatrascuW10}. We prove that any sufficiently non-trivial improvement over the best known decision algorithm carries over to approximate counting and witness sampling.

For any integer $k \ge 3$, the \pp{$k$-SUM} problem is stated formally as follows.

\begin{framed}
	\noindent
	\textbf{\pp{\boldmath$k$-SUM}}\\
	\textit{Input:} A set $X$ of integers.\\
	\textit{Question:} Is there a set $S \subseteq X$ with $|S| = k$ such that $\sum_{x \in S}x = 0$? We call such sets \emph{witnesses} for $X$.
\end{framed}

Observe that \pp{$k$-SUM} is a uniform witness problem by Definition~\ref{def:uwp}: The $k$-hyper\-graph has vertex set~$X$, its edges are precisely the witnesses for $X$, and for all $S \subseteq X$, the instance $I_X(S)$ is simply $S$. Thus, the problem \pp{Colourful-$k$-SUM} of Definition~\ref{def:colourful-problems} is given as follows: The input consists of $k$ disjoint sets~$X_1,\dots,X_k$ of integers, and the goal is to determine whether there exist $x_1 \in X_1,\dots,x_k \in X_k$ such that $\sum_i x_i = 0$. Note that any algorithm for \pp{Colourful-$k$-SUM} requires time $\widetilde{\Omega}(|X|)$ to read the input.

In order to apply \cref{thm:meta-informal}, we first reduce \pp{Colourful $k$-SUM} to \pp{$k$-SUM} with the following lemma (which is well-known folklore).

\begin{lemma}\label{lem:colourful-k-sum}
    If an $n$-integer instance of \pp{$k$-SUM} can be solved in time $T(n)$, then there is an $\OO(T(n))$-time algorithm for \pp{Colourful $k$-SUM}.
\end{lemma}
\begin{proof}
	Let $X_1, \dots, X_k$ be the input for \pp{Colourful $k$-SUM}. For each $i \in [k]$, define injections $f_i\colon \Z \to \Z$~by 
	\begin{align*}
		f_i(x) &= (k+1)^kx + (k+1)^{i-1} \mbox{ for all }i \in [k-1],\\
		f_k(x) &= (k+1)^kx - \sum_{i=1}^{k-1}(k+1)^{i-1}.
	\end{align*}
	For all $i \in [k]$, let $Y_i = \{f_i(x) \colon x \in X_i\}$, and let $Y = Y_1 \cup \dots \cup Y_k$.
	Now we run the assumed \pp{$k$-SUM} decision algorithm on $Y$ and output the result.
	Preparing the set~$Y$ and running the algorithm takes time $\OO(n) + T(n)$.
	Since any algorithm for \pp{$k$-SUM} must read a constant proportion of its input, we have $T(n) \ge \Omega(n)$; thus, the overall running time of this algorithm is $\OO(T(n))$.

	To prove correctness, let $x_i\in X_i$ for all $i\in[k]$ such that $\sum_i x_i = 0$.
	Then 
	\[
		\sum_{i=1}^k f_i(x_i) = (k+1)^k\sum_{i=1}^kx_i + \sum_{i=1}^{k-1} (k+1)^{i-1} - \sum_{i=1}^{k-1} (k+1)^{i-1} = 0,
	\]
	and so there are $k$ distinct numbers in~$Y$ whose sum is zero.
	Conversely, suppose $y_1, \dots, y_k\in Y$ are distinct numbers whose sum is zero.
	Then by the uniqueness of base-$(k+1)$ expansions, we must have $y_i \in Y_{\sigma(i)}$ for some permutation $\sigma\colon [k]\to[k]$; moreover, $\sum_i f_{\sigma(i)}^{-1}(y_i) = 0$. Thus $\{f_{\sigma(1)}^{-1}(y_1), \dots, f_{\sigma(k)}^{-1}(y_k)\}$ is a witness for \pp{Colourful $k$-SUM} as required.
	We conclude that the reduction is correct.
\end{proof}

The following corollary is now immediate from \cref{lem:colourful-k-sum} and \cref{thm:meta-informal}. (Recall that, for any $n$-element instance $X$ of \pp{$k$-SUM} and any $S \subseteq V(G_X) = X$, we have $I_X(S) = S$; thus the maximum in the right-hand side of~\eqref{eq:meta-informal-runtime} is simply $T(n)$.)

\begin{corollary}\label{cor:ksum}
  Fix $k \ge 3$, suppose an $n$-integer instance of \pp{$k$-SUM} can be solved in time $T(n)$, and write $W$ for the set of witnesses. Then there is a randomised algorithm to $\eps$-approximate $|W|$, or draw an $\eps$-approximate sample from $W$, in time $\eps^{-2}\cdot\tildeOO(T(n))$.
\end{corollary}

\subsection{Application to \texorpdfstring{\pp{\boldmath$k$-Orthogonal Vectors}}{k-Orthogonal Vectors}}

As is standard, we abbreviate \pp{$k$-Orthogonal vectors} to \pp{$k$-OV}. The \pp{$k$-OV} problem has connections to central conjectures in fine-grained complexity theory~\cite{DBLP:conf/stoc/AbboudBDN18,DBLP:journals/talg/GaoIKW19}.
Clearly, \pp{$k$-OV} can be solved in time $O(N^kD)$ using exhaustive search.
Gao et al.~\cite{DBLP:journals/talg/GaoIKW19} stated the Moderate-Dimension \pp{$k$-OV} Conjecture, which says that \pp{$k$-OV} cannot be solved in time $O(N^{k-\varepsilon} \operatorname{poly}(D))$ for any $\varepsilon>0$.
We show that every superlogarithmic improvement over exhaustive search for \pp{$k$-OV} carries over to approximate counting and sampling of $k$-OV witnesses.
Note that such an improvement is already known for \pp{2-OV}, namely \pp{2-OV} has an $N^{2-1/\OO(\log(D/\log N))}$-time algorithm~\cite{DBLP:conf/soda/AbboudWY15}. Chan and Williams~\cite{DBLP:conf/soda/ChanW16} already generalised this \text{2-OV} decision algorithm to an exact counting algorithm. 
For all integers $k \ge 2$, the \pp{$k$-OV} problem is stated formally as follows.

\begin{framed}
	\noindent
	\textbf{\pp{\boldmath$k$-OV}}\\
	\textit{Input:} Sets $X_1,\dots,X_k$ of vectors from $\{0,1\}^D$.\\
	\textit{Question:} Do there exist $x_1 \in X_1, \dots, x_k \in X_k$ with $\sum_{j=1}^D\prod_{i=1}^k (x_i)_j = 0$? We call such tuples $(x_1,\dots,x_k)$ \emph{witnesses} for $X_1,\dots,X_k$.
\end{framed}
We remark that the sum and product in the problem definition refer to the usual arithmetic operations over $\Z$. Observe that \pp{$k$-OV} is a uniform witness problem according to Definition~\ref{def:uwp}: for a \pp{$k$-OV} instance $x = (X_1,\dots,X_k)$, the $k$-hypergraph $G_x$ has vertex-set $X_1 \cup \dots \cup X_k$, its edges are precisely the witnesses of the \pp{$k$-OV} instance, and for any $S \subset X_1 \cup \dots \cup X_k$, the instance $I_x(S)$ is $(S \cap X_1, \dots, S \cap X_k)$.  Notice that \pp{$k$-OV} is in fact a colourful uniform witness problem, since $G_x$ is always $k$-partite, and so \pp{Colourful-$k$-OV} reduces to $k^{\OO(k)}$ instances of \pp{$k$-OV}.

The following corollary is now immediate by \cref{thm:meta-informal}, since any algorithm for \pp{$k$-OV} requires time $\widetilde{\Omega}(|X_1 \cup \dots \cup X_k|)$ to read the input and, by the definition of $I_x(S)$ and our choice of $T$, it is immediate that the maximum in the right-hand side of \eqref{eq:meta-informal-runtime} is simply $T(x)$.

\begin{corollary}\label{cor:ov}
  Fix $k \ge 2$, suppose an $N$-vector $D$-dimensional instance of \pp{$k$-OV} can be solved in time $T(N,D)$, and write $W$ for the set of witnesses. Then there is a randomised algorithm to $\eps$-approximate $|W|$, or draw an $\eps$-approximate sample from~$W$, in time $\eps^{-2}\cdot\tildeOO(T(N,D))$. 
\end{corollary}

\subsection{Application to first-order model checking}

\def\FOqr{\ensuremath{k\textrm{-FO}}}
\newcommand{\klogic}{\pp{\FOqr{} Property Testing}\xspace}
In this section, we apply our results to the \emph{property testing problem} for $\FOqr$.
Informally, the input to this problem consists of a formula and a structure (e.g., the edge relation of a graph), to decide whether the formula is satisfiable in the structure, that is, whether there is an assignment to the free variables that makes the formula true.
Correspondingly, the \emph{property counting problem} is to count all satisfying assignments.
In order to formally introduce this problem, we must introduce some standard notation from logic; for the convenience of the reader, we will also give a brief introduction to the associated concepts.

\paragraph{Syntax} The class \FOqr{} is the set of all first-order formulas~$\varphi$ in prenex normal form with at most $k$ variables.
Here is an example of a formula $\Phi \in 5$-FO:
\begin{equation*}
	\Phi(x_1,x_2,x_3)=\quad \forall x_{4} \exists x_{5}\,.\, R_1(x_1,x_2) \wedge (R_2(x_5,x_3,x_1) \Rightarrow R_1(x_4,x_5))\,.
\end{equation*}
Because all the quantifiers are at the front, the formula is in \emph{prenex normal form}.
The formula~$\Phi$ has \emph{free} variables $x_1,x_2,x_3$ and uses two \emph{relation symbols} $R_1$ and $R_2$. The \emph{arity} of $R_1$ is $2$ and the arity of $R_2$ is $3$.
Therefore, we say that $\Phi$ uses the \emph{vocabulary} $\nu_\Phi=(R_1,R_2,2,3)$.

In general, a \emph{vocabulary} is a tuple $\nu=(R_1,\dots,R_r,\alpha_1,\dots,\alpha_r)$, where $R_1, \dots, R_r$ are \emph{relation symbols} and $\alpha_1, \dots, \alpha_r$ are positive integers denoting the respective arities of the relation symbols.
A formula $\varphi\in\FOqr{}$ over the vocabulary $\nu$ satisfies
\begin{equation*}
    \varphi(x_1,\dots,x_\ell)=\quad Q_1 x_{\ell+1} Q_2 x_{\ell+2} \dots Q_{k - \ell} x_k\,.\,\psi(x_1,\dots,x_k)%
\,,\end{equation*}
where the variables~$x_1,\dots,x_\ell$ are the \emph{free} variables of $\varphi$, each $Q_i\in\{\exists,\forall\}$ is a quantifier, and $\psi$ is a quantifier-free Boolean formula over the variables $x_1,\dots,x_k$ such that each atom of $\psi$ is of the form $R_i(x_{j_1},\dots,x_{j_{\alpha_i}})$, that is, each atom of $\psi$ applies some relation symbol~$R_i$ to a tuple of $\alpha_i$ variables.

\paragraph{Semantics}
So far, the formulas $\Phi$ and $\varphi$ only consist of syntax.
In order to be able to say that a formula is \emph{satisfiable}, we need to define a \emph{universe} $U$ as well as instantiate the relation symbols with actual relations over~$U$ of matching arities.
In our example, we might set the universe to be $U_\Phi=\{a,b,c,d\}$. Then the variables of $\Phi$ take values from~$U_\Phi$, and all quantification is over~$U_\Phi$.
We might then instantiate the relations via $\calR_1=\{(a,b),(b,c)\}$ and $\calR_2=\{(a,d,c)\}$.
We say that $\calS_\Phi=(U,\calR_1,\calR_2)$ is a \emph{structure} for $\nu_\Phi$.
Now we can see that $\Phi(a,b,c)$ is true and $\Phi(a,c,d)$ is false in the structure $\calS_\Phi$, and that $\Phi$ is \emph{satisfiable} in $\calS_\Phi$ because it has at least the satisfying assignment $(a,b,c)$.

In general, a \emph{structure} $\calS=(U,\calR_1,\dots,\calR_r)$ for the vocabulary $\nu$ consists of a finite \emph{universe} $U$ together with relations $\calR_1,\dots,\calR_r$ over $U$ such that the arity of each $\calR_i$ is equal to $\alpha_i$.
An \emph{assignment} $(y_1,\dots,y_\ell)\in U^\ell$ to the free variables of $\varphi$ is called \emph{satisfying} in $\calS$ if $\varphi(y_1,\dots,y_\ell)$ is true in the structure~$\calS$.
For all positive integers $k$, we define the following problem.

\begin{framed}
	\noindent
	\textbf{\klogic}\\
	\textit{Input:}
	\begin{enumerate}[(i)]
	    \item A vocabulary $\nu = (R_1, \dots, R_r,\alpha_1,\dots,\alpha_r)$ with $\alpha_i \le k$ for all $i$,
	    \item A structure $\calS = (U,\calR_1,\dots,\calR_r)$ on $\nu$, where all tuples in all $\calR_i$'s are explicitly given as a list and $\emptyset\ne\calR_i\subseteq U^{\alpha_i}$ holds for all $i$,
	    \item A first-order formula $\varphi$ in prenex normal form and with vocabulary $\nu$, free variables $x_1, \dots, x_\ell$, and at most $k-\ell$ quantifiers.
	\end{enumerate}
	\textit{Question:} Does $\varphi$ have a satisfying assignment in $\calS$?
\end{framed}

Property testing is an important problem in logic and in database theory (cf.~\cite{logicbook}).
The fine-grained complexity of property testing has recently been studied to some extent~\cite{DBLP:conf/icalp/DellRW19,DBLP:journals/talg/GaoIKW19,Williams14}. In particular, Gao et al.~\cite{DBLP:journals/talg/GaoIKW19} devise an algorithm for the property testing problem that runs in time $m^{k-1}/2^{\Theta(\sqrt{\log m})}$ for any fixed $\FOqr$ formula, where $m$ is the number of distinct tuples in the input relations.
This improves upon an already non-trivial $\tildeOO(m^{k-1})$ time algorithm.

We prove the following reduction from approximate counting to decision.
\begin{corollary}\label{cor:FO}
Fix $k\in\Z_{\ge 0}$, and suppose \klogic has an algorithm that runs in time $T(|\varphi|,n,m)$, where $|\varphi|$ is the size of the formula, $n$ is the size of the universe, and $m$ is the total number of tuples in the structure. Let $\calW$ be the set of satisfying assignments. Then there is a randomised algorithm to $\eps$-approximate~$|\calW|$, or draw an $\eps$-approximate sample from~$\calW$, in time $\varepsilon^{-2}\cdot\tildeOO(T(|\varphi|+\ell,n,m+n))$.
\end{corollary}
This corollary follows from \cref{thm:meta-informal} by using the assumed algorithm for \klogic to simulate the uncoloured independence oracle of an appropriate hypergraph.
\begin{proof}
In order to apply \cref{thm:meta-informal} to property testing, we need to show how \klogic{} is a uniform witness problem according to \cref{def:uwp}.
	Let $x=(\nu,\calS,\varphi)$ be an instance of \klogic, let $\calS = (U,\calR_1,\dots,\calR_r)$, and let $\ell$ with $\ell\le k$ be the number of free variables in $\varphi$.
	We define the hypergraph $G=G_x$ as follows:
	For all $i \in \{1,\dots,\ell\}$, let $U_i = U \times \{i\}$. Define an $\ell$-hypergraph $G$ with vertex set $U_1 \cup \dots \cup U_\ell$ whose edges are the sets $\{(y_1,1),\dots,(y_\ell,\ell)\}$ such that $(y_1,\dots,y_\ell)$ is a satisfying assignment of~$\varphi$ in~$\calS$.
	
	It is clear that this function~$x\mapsto G_x$ satisfies the conditions (i) and (ii) of \cref{def:uwp}.
	To prove that \klogic{} is a uniform witness problem, it remains to prove (iii).
	Indeed, let $X \subseteq V(G)$ be given as an input in addition to~$x$. Then we prepare in time $\tildeOO(|x|)$ an instance $I_x(X)=(\nu',\calS',\varphi')$ of \klogic such that $G_{I_x(X)}=G_x[X]$ holds:
	We set $X_i = X \cap U_i$ for all $i \in \{1,\dots,\ell\}$. We form the new instance $(\nu',\calS',\varphi')$ of \klogic as follows: form $\nu'$ from $\nu$ by adding $\ell$ additional relation symbols $\in_{X_1},\dots,\in_{X_\ell}$, each with arity 1; form~$\calS'$ from $\calS$ by instantiating each $\in_{X_i}$ with the set $\{y \colon (y,i)\in X_i\}$; and form~$\varphi'$ from $\varphi$ by conjoining it with the formula $({\in_{X_1}}(x_1) \wedge \dots \wedge {\in_{X_\ell}}(x_\ell))$. Then $G_{I_x(X)}=G_x[X]$ indeed holds.
	Thus, (iii) holds and \klogic{} is a uniform witness problem. In fact, since all hypergraphs involved are $k$-partite, it is a colourful uniform witness problem.

    A technical precondition of \cref{thm:meta-informal} is that any algorithm for the problem requires at least time~$\widetilde{\Omega}(|x|)$; this is true for \klogic, because any algorithm for it must read a constant proportion of the input.
    Since $\klogic$ is a colourful uniform witness problem, we obtain an algorithm for \pp{Colourful-$\Pi$} with an overhead of only $k^{\OO(k)}=\OO(1)$ time.
    Thus, we can apply the theorem and obtain an algorithm for $\eps$-approximate counting and sampling that runs in time \[
    \eps^{-2} (k\log n)^{\OO(k)} k^{\OO(1)}
    \cdot
    \max_{X\subseteq V(G)} T(I_x(X))\,.
    \]
    Since all instances $I_x(X)$ have a universe of size $n$, at most $n$ additional tuples, and a formula $\varphi'$ that is $\varphi$ conjoined with $\ell$ additional terms, we have $T(I_x(X))\le T(|\varphi|+\ell,n,m+n)$.
    The claimed running time then follows from $k=\OO(1)$.
\end{proof}
By applying \cref{cor:FO}, we are able to lift the algorithm of Gao et al.~\cite{DBLP:journals/talg/GaoIKW19}, to approximate counting and sampling.
For example, we obtain an algorithm to $\varepsilon$-approximately sample uniformly random satisfying assignments.
For any fixed $\FOqr{}$ formula $\varphi$, this algorithm runs in time~$\varepsilon^{-2} m^{k-1}/2^{\Theta(\sqrt{\log m})}$.
We remark the following subtleties:
Gao et al.~\cite{DBLP:journals/talg/GaoIKW19} state their algorithm for any fixed formula~$\varphi$, but actually their algorithm is uniform in~$\varphi$ with a running time that can be expressed as a product $f(|\varphi|)\cdot T(n,m)$. This means that increasing the size of $\varphi$ by a constant does not change the asymptotic running time for any fixed $\varphi$, since $f(|\varphi|+\ell)$ is constant as well.
Finally, for certain quantifier structures of $\varphi$, they obtain even faster decision algorithms; since our reduction does not change the quantifier structure, these running times transfer to approximate counting and sampling as well.

\subsection{Application to subgraph problems}

Recall from the discussion following the statement of \cref{thm:main-counting} that the theorem can be applied to the problem \#\pp{$k$-Clique} for every constant~$k$. Thus, every $T(n)$-time algorithm to decide the existence of a $k$-clique can be turned into a $\tildeOO(\eps^{-2} T(n))$-time algorithm to $\eps$-approximate the number of $k$-cliques.
We now generalise this result to the problems of finding colourful or zero-weight copies of an arbitrary subgraph, and we obtain similar consequences for approximately uniform sampling via \cref{thm:main-combined-colourful}.

\subsubsection{Colourful subgraphs}

In this section, we use \cref{thm:main-combined-colourful} to transform any decision algorithm for \pp{Colourful-$H$} into an approximate counting or sampling algorithm with roughly the same running time. For all graphs~$H$ with $k$ vertices, the \pp{Colourful-$H$} problem is stated formally as follows.
\begin{framed}
	\noindent
	\textbf{\pp{Colourful-$H$}}\\
	\textit{Input:} A graph $G$ in adjacency list format and a vertex-colouring $c:V(G)\to [k]$ (not necessarily proper).\\
	\textit{Question:} Does $G$ contain a (not necessarily induced) subgraph~$S$ such that $S$ is isomorphic to~$H$ and, for all $i\in [k]$, there is some $v\in V(S)$ with $c(v)=i$?
\end{framed}
D\'{i}az, Serna, and Thilikos~\cite{DBLP:journals/tcs/DiazST02} use dynamic programming to solve \pp{Colourful-$H$} in time $\tildeOO(n^{t+1})$, where $t$ is the treewidth of~$H$ --- their algorithm works for the exact counting version of the problem, too.
Marx~\cite{DBLP:journals/toc/Marx10} asks, loosely speaking, whether the decision problem can be solved in time~$n^{o(t)}$, and proves that $n^{o(t/\log t)}$ is impossible under the exponential-time hypothesis (ETH).
For all constant~$\gamma>0$, we show that a factor $n^\gamma$ improvement to D\'iaz et al.'s algorithm for the decision problem would immediately yield a corresponding improvement to the approximate counting or sampling problems.

\begin{corollary}\label{cor:colourful}
	Let~$H$ be any fixed graph. Suppose $n$-vertex $m$-edge instances of \pp{Colourful-$H$} can be solved by a randomised algorithm in time $T(m,n)$, and write $\mathcal{H}$ for the set of colourful $H$-subgraphs. Then there is a randomised algorithm to $\eps$-approximate $|\mathcal{H}|$, or draw an $\eps$-approximate sample from $\mathcal{H}$, in time $\eps^{-2}\cdot\tildeOO(T(m,n))$.
\end{corollary}
In the proof, we use the notion of a graph homomorphism (see~\cite{hombook}). A function $f:V(H)\to V(G)$ is called a \emph{homomorphism} if, for all $\{u,v\}\in E(H)$, we have $\{f(u), f(v)\}\in E(G)$. Note that injective homomorphisms correspond to subgraph embeddings.
\begin{proof}
	Let $(G,c)$ be an instance of \pp{Colourful-$H$}.
To motivate our proof, we first discuss a natural approach which fails. First, remove all edges $\{u,v\}\in E(G)$ with $c(u)=c(v)$ from the graph, and note that this makes $G$ $k$-partite with vertex classes $c^{-1}(1),\dots,c^{-1}(k)$ but does not change the answer.	
	It would be natural to apply \cref{thm:meta-informal} to the $k$-partite $k$-hypergraph $\calG$ on $V(G)$ in which $S \subseteq V(G)$ is an edge of $\calG$ if $G[S]$ contains a subgraph isomorphic to $H$, and indeed when~$H$ is a $k$-clique this approach works. However, in general each edge~$S$ of~$\calG$ may correspond to more than one copy of $H$ contained in $G[S]$. Instead, we will apply \cref{thm:main-combined-colourful} directly to count the edges in multiple $k$-hypergraphs corresponding to the $k!$ possible ways $H$ could be embedded as a subgraph in $G[S]$.

	For each bijective function $d:V(H)\to [k]$, we let $\mathcal G_d$ be a $k$-hypergraph with vertex set~$V(G)$ and edges given as follows. For each size-$k$ subset $S\subseteq V(G)$ with the property that $c|_S:S\to [k]$ is bijective (i.e.\ $S$ is colourful under $c$), we add $S$ to $E(\mathcal G_d)$ if the function $(c|_S^{-1}\circ d):V(H)\to S$ is an injective homomorphism from~$H$ to~$G[S]$ (i.e.\ $G[S]$ contains $H$ as a subgraph with vertex colours matching $d$).
	We now claim
	\begin{equation}\label{eq:colorful-aut}
	    |\mathcal{H}| = \frac{1}{\mathrm{Aut}(H)}\sum_d e(\calG_d),
	\end{equation}
	where $\mathrm{Aut}(H)$ is the number of automorphisms of $H$. Observe that since $H$ is fixed, we have access to $\mathrm{Aut}(H)$; thus given the claim, the problem of $\eps$-approximating $|\mathcal{H}|$ reduces to that of $\eps$-approximating each term $e(\calG_d)$.
	
	\textbf{Proof of \eqref{eq:colorful-aut}:} We proceed by double-counting the number $N$ of injective homomorphisms $h\colon V(H)\to V(G)$ such that the image $h(H)$ is colourful in~$G$ with respect to~$c$, i.e.\ the number of ways $H$ can be embedded in $G$ as a colourful subgraph. On the one hand, each such $h$ has a unique image $S=h(V(H))\subseteq V(G)$ and a unique labelling $d\colon V(H)\to [k]$ such that $h=c|_S^{-1}\circ d$; thus it corresponds bijectively to the edge spanning $S$ in $\calG_d$, and $\sum_d e(\calG_d) = N$. On the other hand, each copy of $H$ in $G$ corresponds to $\mathrm{Aut}(H)$ embeddings of $H$ into $G$, so we have $|\mathcal{H}|\mathrm{Aut}(H) = N$. Combining these two equations yields~\eqref{eq:colorful-aut} as required.
	
	\textbf{Approximating the terms:} We will now $\eps$-approximate each term $e(\calG_d)$ of the right-hand side of~\eqref{eq:colorful-aut} using \colourfulsamplecount{} from \cref{thm:main-combined-colourful} with $\delta=1/(3k!)$.
	By a union bound, the probability that at least one of the $k!$ invocations of \colourfulsamplecount{} fails is at most $1/3$.
	To approximate $e(\calG_d)$ using  \colourfulsamplecount{}, we first recall that $\calG_d$ is a $k$-partite $k$-hypergraph with explicitly given vertex classes, and then implement the uncoloured independence oracle for~$\calG_d$ using the assumed randomised decision algorithm for \pp{Colourful-$H$}.
	To this end, we define the following algorithm~$A(X)$:
	\begin{enumerate}
	    \item On input $X \subseteq V(G)$, the algorithm constructs a graph~$G'$ on the vertex set $V(G')=X$. Start with $G[X]$ and delete all edges internal to colour classes and all edges
	    between colour classes whose corresponding vertices are not joined in $H$.
	    (Conversely, we only add $\{u,v\}\in E(G[X])$ to $E(G')$ if  $\{d^{-1}(c(u)),d^{-1}(c(v))\} \in E(H)$ holds.)
	    \item Run the assumed algorithm for \pp{Colourful-$H$} on input $(G',c')$ where $c'=c|_{V(G')}$ and return its output.
	\end{enumerate}
	Note that $A$ runs in time $\tildeOO(n+m+T(m,n))$. Since any algorithm for \pp{Colourful-$H$} must read a constant proportion of the input vertices and edges, we have $T(n,m) = \Omega(n+m)$, so $A$ in fact runs in time $\tildeOO(T(n,m))$.
	If~$A$ implements the uncoloured independence oracle for~$\calG_d$ with error probability at most $1/3$, we can pass the description of this algorithm along with the vertex classes $X_i=X\cap c^{-1}(i)$ for all $i\in[k]$ and parameters $k$, $\eps$, and $\delta=1/(3k!)$ to \colourfulsamplecount{}.
    Our running time bounds therefore follow, and it remains to prove correctness.

	\textbf{Correctness:}
	First assume that $A(X)$ accepts, so that the goal is to show $e(\calG_d[X])>0$.
	Because $A$ accepts, there is an $H$-isomorphic colourful subgraph~$F$ of~$(G',c)$.
	Let $S=V(F)$. 
    By the construction of $G'$, all edges in $F$ must be between colour classes whose corresponding vertices are joined in $H$, so the function $c|_S^{-1}\circ d$ is an injective homomorphism from $H$ to~$G'[S]$. 
	Thus $S\in E(\mathcal G_d[X])$, so $e(\mathcal G_d[X]) > 0$ as required.

	For the reverse direction, suppose ${e(\mathcal{G}_d[X]) > 0}$.
	We need to show that $A(X)$ accepts with probability at least~$2/3$.
	Let $S \in E(\mathcal G_d[X])$.
	Then by the definition of $\mathcal G_d$, the function $c|_S^{-1}\circ d$ is an injective homomorphism from~$H$ to~$G[S]$. By the construction of~$G'$, it is also an injective homomorphism from~$H$ to~$G'[S]$, so $G'$ contains a colourful $H$-subgraph and so $(G',c)$ is a \texttt{Yes} instance.
	In this case, our assumed algorithm for \pp{Colourful-$H$}, and hence also $A(X)$, accepts with probability at least~$2/3$.
	We have proved correctness as required.
	
	\textbf{Sampling:} Our sampling algorithm is very similar to our approximate counting algorithm, again using the fact that each subgraph corresponds to $\mathrm{Aut}(H)$ embeddings, so we omit the details. The only differences are the following:
	
	\begin{enumerate}
	    \item We require that each invocation of our approximate counting algorithm has failure probability at most $\eps/(10k^{2k}\ln(5/\eps))$ and set $\delta$ accordingly.
	    \item Rather than outputting a weighted sum of $\eps$-approximations of each $e(\mathcal{G}_d)$, we use $(\eps/10)$-approximations of each $e(\mathcal{G}_d)$ to apply rejection sampling as in e.g.~Florescu~\cite[Proposition 3.3]{florescu}. Thus writing $N$ for our $(\eps/10)$-approximation of $|\mathcal H|$ and $N_d$ for our $(\eps/10)$-approximation of $e(\mathcal{G}_d)$, we choose a bijective function $d\colon [k]\to[k]$ uniformly at random, take an $(\eps/10)$-approximate sample from $E(\mathcal{G}_d)$, and accept and output this sample with probability $N_d/N$; otherwise, we reject it and resample. If we require more than $k^{2k}\ln(5/\eps)$ iterations, we return an arbitrary output.
	\end{enumerate}
	
	Note that the acceptance probability at each step is at least $1/(\mbox{Aut}(H)^2) \ge 1/k^{2k}$, and acceptance is independent at each step, so the probability we require more than $k^{2k}\ln(5/\eps)$ invocations is at most $(1-1/k^{2k})^{k^{2k}\ln(5/\eps)} \le \eps/5$. Combining this with the error in the rejection sampling from our $(\eps/10)$-approximations, and with the probability that our approximate counting algorithm fails to return a correct value in the first $k^{2k}\ln(5/\eps)$ invocations (which is at most $\eps/10$ by a union bound), we see that our algorithm returns an $\eps$-approximate sample as required. Since $\eps \ge n^{-k}$, the running time bound is immediate.
\end{proof}

\subsubsection{Weighted subgraphs}
In an edge-weighted graph, the graph~$G$ is augmented with a function $w:E(G)\to\Z$.
The weight~$w(F)$ of a subgraph~$F$ of~$G$ is the sum $\sum_{e\in E(F)} w(e)$ of all edge-weights in~$F$.
We now consider the following computational problem for any fixed unweighted graph~$H$.

\begin{framed}
	\noindent
	\textbf{\pp{Exact-Weight-$H$}}\\
	\textit{Input:} An edge-weighted graph $G$ with (perhaps negative) integer weights.\\
	\textit{Question:} Does $G$ have a subgraph isomorphic to $H$ with total weight zero?
\end{framed}

The special case where~$H$ is a $k$-clique has been studied in fine-grained complexity under the name \pp{Exact-Weight $k$-Clique}.
It has been conjectured~\cite{DBLP:conf/stoc/AbboudBDN18} that there does not exist any real $\varepsilon>0$ and integer $k\ge 3$ such that the \pp{Exact-Weight $k$-Clique} problem on $n$-vertex graphs and with edge-weights in $\{-M,\ldots,M\}$ can be solved in time $n^{(1-\varepsilon)k} \polylog(M)$.
For the closely related \pp{Min-Weight $k$-Clique} problem, only subpolynomial-time improvement over the exhaustive search algorithm is known \cite{DBLP:conf/stoc/AbboudBDN18,DBLP:journals/siamcomp/Williams18,DBLP:conf/soda/ChanW16}, with a running time of $n^{k}/\exp(\Omega(\sqrt{\log n}))$.
As we now show, \cref{thm:meta-informal} implies that any sufficiently non-trivial improvement on the running time of an \pp{Exact-Weight $k$-Clique} algorithm will carry over to the approximate counting and sampling versions of the problem.
\begin{corollary}\label{cor:exact-clique}
	Fix $k \ge 3$, suppose an $n$-vertex $m$-edge instance of \pp{Exact-Weight $k$-Clique} with weights in $[-M,M]$ can be solved in time $T(n,m,M)$, and write $\mathcal{C}$ for the set of zero-weight $k$-cliques. Then there is a randomised algorithm to $\eps$-approximate $|\mathcal{C}|$, or draw an $\eps$-approximate sample from $\mathcal{C}$, in time $\eps^{-2}\cdot\tildeOO(T(n,m,M))$.
\end{corollary}
\begin{proof}
    First observe that \pp{Exact-Weight $k$-Clique} is a uniform witness problem by \cref{def:uwp}: Given an instance $(G,w,k)$ of \pp{Exact-Weight $k$-Clique}, we have $V(G_x) = V(G)$ and the edges of $G_x$ are precisely the $k$-cliques of $G$, with $I_{(G,w,k)}(S) = (G[S], w|_S, k)$ for all $S \subseteq V(G)$. Moreover, any randomised algorithm for \pp{Colourful-Exact-Weight $k$-Clique} (see \cref{def:colourful-problems}) must read at least a constant proportion of the input bits, and so the lower bound on its running time required by \cref{thm:meta-informal} is satisfied. Hence, by \cref{thm:meta-informal}, it suffices to give an algorithm for \pp{Colourful-Exact-Weight $k$-Clique} with running time $\tildeOO(T(n,m,M))$.
    
    Let $(G,w,k)$ be an instance of \pp{Exact-Weight $k$-Clique}, and let $X_1, \dots, X_k$ be a partition of $V(G)$. Then we form a graph $G'$ from $G$ in linear time by removing all edges internal to each vertex set $X_i$, and apply our \pp{Exact-Weight $k$-Clique} decision algorithm to $G'$ in time at most $T(n,m,M)$. The cliques of $G'$ are precisely the colourful cliques of $G$ with respect to $V_1,\dots,V_k$, so this solves \pp{Colourful-Exact-Weight $k$-Clique} in $\tilde{\OO}(m+n+T(n,m,M))$ time as required.
\end{proof}
There is a generalisation of \pp{Exact-Weight $k$-Clique} to edge-weighted $d$-hypergraphs, for which a fine-grained complexity conjecture exists~\cite{DBLP:conf/stoc/AbboudBDN18}.
A result analogous to \cref{cor:exact-clique} holds for this problem as well, but we do not state it formally.

Instead, we now state a more interesting corollary for \pp{Exact-Weight-$H$}. When $H$ is a $k$-clique, it was sufficient in the proof of \cref{cor:exact-clique} to delete some edges to make sure that the colourful independence oracle only counts \emph{colourful} copies of~$H$. For general graphs~$H$, we control this by relying on basic bit tricks that are commonly used in subset sum or exact-weight type problems.
Unfortunately, we do not know how to do so inside the colourful independence oracle, at least not in the way in which we have formalised it.
Instead, we perform an additional colour-coding step before running the algorithm of \cref{thm:main-combined-colourful} as a black box, which leads to an additional $\eps^{-2}$ factor overhead in the running time. However, we believe that this is merely an artefact that can be avoided by doing some surgery on the proofs of our main results. Since doing so would be lengthy, and does not add any insight, we accept the loss in the running time to get a cleaner proof.

\begin{corollary}
\label{cor:exact-weight-H}
	Let $H$ be any fixed graph with~$k$ vertices.
	Suppose that $n$-vertex $m$-edge instances~$G$ of \pp{Exact-Weight-$H$} with weights in $[-M,M]$ can be solved in time $T(m,n,M)$, and write $\calS$ for the set of subgraphs of~$G$ that are isomorphic to~$H$. Then there is a randomised algorithm to $\eps$-approximate $|\calS|$, or draw an $\eps$-approximate sample from $\calS$, in time $\eps^{-4}\cdot\tildeOO(T(m,n,k^2M))$. 
\end{corollary}

\begin{proof}
	We first set out the approximate counting algorithm, and we suppose for simplicity that our decision algorithm is deterministic.
	Let $G$ be an instance of \pp{Exact-Weight-$H$}, and let $k = |V(H)|$.
	If $n<k$, we return the correct answer~$0$.
	If~$H$ has exactly $i$ isolated vertices, we remove them from~$H$, produce an estimate for the reduced graph, and multiply this estimate by~$\binom{n-k+i}{i}$ to obtain the estimate for~$H$.
	Thus, we can assume without loss of generality that $n \ge k$ and that $H$ has no isolated vertices.
	The algorithm now proceeds as follows:
	\begin{enumerate}
	\item Let $t=\eps^{-2}\cdot 100e^{2k}$. For all $r$ from $1$ to $t$:
	\begin{enumerate}
	    \item Sample a partition $V_1\cup\dots\cup V_k=V(G)$ uniformly at random.
	    \item For all bijective functions $\pi:[k]\to V(H)$:
	    \begin{enumerate}
	        \item Define the graph~$G_{r,\pi}$ by deleting from~$G$ all edges $E(V_i,V_j)$ between parts $V_i$ and $V_j$ for which $\pi(i)\pi(j)$ is not an edge of~$H$.
	        \item Define the $k$-partite $k$-hypergraph~$\calG_{r,\pi}$ with vertex set $V(G)$ such that each set~$S\subseteq V(G)$ is an edge of $\calG_{r,\pi}$ if and only if $G_{r,\pi}[S]$ is isomorphic to~$H$, contains one vertex from each~$V_i$, and is zero-weight with respect to~$w$.
	        \item Call the algorithm \colourfulsamplecount from \cref{thm:main-combined-colourful} on~$\calG_{r,\pi}$ with $\delta=1/(100tk!)$ and with error parameter $\eps/3$ to obtain an estimate~$N_{r,\pi}$ for the number of edges in $\calG_{r,\pi}$.
	    \end{enumerate}
	\end{enumerate}
	\item Output the estimate $\sum_{r,\pi} N_{r,\pi}\cdot k^k/(tk!\cdot\operatorname{Aut}(H))$.
	\end{enumerate}
	Here the scaling factor $k^k/(tk!\cdot\operatorname{Aut}(H))$ arises from the fact that a random $k$-colouring of a set of size $k$ is colourful with probability $k! / k^k$, each edge in our hypergraph corresponds to $\operatorname{Aut}(H)$ injective homomorphisms from $H$ to $G$, and we repeat the whole process $t$ times.
	
	We will explain how to simulate the uncoloured independence oracles of the hypergraphs $\calG_{r,\pi}$ (as required by \colourfulsamplecount) shortly. First, we argue for the correctness of the algorithm.
	
    \textbf{Proving correctness given uncoloured independence oracles:} \aau is called $tk!$ times, and each invocation fails to produce an $(\eps/3)$-approximation with probability at most~$\delta$. By a union bound, this implies that, with probability at least~$0.99$, \emph{all} estimates $N_{r,\pi}$ are $(\eps/3)$-approximations to the respective numbers $e(\calG_{r,\pi})$ of edges in the hypergraphs~$\calG_{r,\pi}$.
	Conditioned on this event, the sum that is produced as output is thus an $(\eps/3)$-approximation to $\sum_{r,\pi} e(\calG_{r,\pi})\cdot k^k/(tk!\cdot\operatorname{Aut}(H))$. Let $s_r$ be the number of subgraphs of~$G$ that are isomorphic to~$H$, are zero-weight with respect to~$w$, and are colourful with respect to $V_1,\dots,V_k$. As in the proof of \cref{cor:colourful}, each colourful zero-weight copy of $H$ corresponds to exactly $\mathrm{Aut}(H)$ terms in $\sum_{\pi} e(\calG_{r,\pi})$, and so the algorithm produces an $\eps/3$-approximation to $\sum_r s_rk^k/(tk!)$.
	
	We now prove that $\sum_r s_r k^k/(tk!)$ is likely to be an $(\eps/3)$-approximation to the total number $|\calS|$ of zero-weight subgraphs of~$G$ that are isomorphic to~$H$. To do so, we first show that it is equal to $|\calS|$ in expectation and then apply Hoeffding's inequality to prove concentration. For any zero-weight subgraph~$F$ of~$G$ that is isomorphic to~$H$, let $X_{r,F} \in \{0,1\}$ be the indicator random variable for the event that $F$ is a subgraph of $G[V_1,\dots,V_k]$ in the $r$'th iteration (that is, the event that there is some~$\pi$ with $V(F) \in E(\calG_{r,\pi})$). Then $\E(X_{r,F}) = k!/k^k$, so we have
	\[
	    \E\bigg(\sum_{r=1}^t s_r\frac{k^k}{tk!}\bigg) = \E\bigg(\sum_{r=1}^t \sum_{F\in\calS} X_{r,F} \frac{k^k}{tk!}\bigg) = \sum_{r=1}^t \sum_{F\in\calS} \frac{1}{t} = |\calS|.
	\]
    Since each $s_r$ lies in $[0,|\calS|]$, it follows by Hoeffding's inequality (\cref{lem:hoeffding}) that
    \begin{align*}
      \Pr\Big(\Big|\frac{k^k}{tk!}\sum_r s_r - |\calS| \Big| > \frac{\eps}{3} |\calS| \Big) &= \Pr\Big(\Big|\sum_r s_r - \frac{tk!\cdot |\calS|}{k^k} \Big| > \eps\frac{tk!}{3k^k} |\calS| \Big)\\
      &\le 2\exp\bigg({-}2\eps^{2}\Big(\frac{tk!}{3k^k}|\calS|\Big)^2/t|\calS|^2 \bigg)
      \\
      &= 2\exp\big({-}2\eps^{2}t(k!/3k^k)^2\big).
    \end{align*}
    By Stirling's formula and our definition of $t$, it follows that
    \[
	    \Pr\Big(\Big|\frac{k^k}{tk!}\sum_r s_r - |\calS| \Big| > \frac{\eps}{3} |\calS| \Big) \le 2\exp\big({-}t\eps^{2}e^{-2k}/9\big) \le 1/10.
    \]
	Thus, with overall probability at least $4/5$, we have $(1-\tfrac{\eps}{3})|\calS| \le \tfrac{k^k}{tk!}\sum_r s_r \le (1+\tfrac{\eps}{3})|\calS|$.
	
	We now note that for all $0 < \eps < 1$, $(1-\eps/3)^2 > 1-\eps$ and $(1+\eps/3)^2 < 1+\eps$; hence an $(\eps/3)$-approximation of an $(\eps/3)$-approximation is an $\eps$-approximation. It therefore follows by a union bound that with probability at least $2/3$, the output of our algorithm is an $\eps$-approximation to $|\calS|$.
	This completes our correctness analysis.

    \textbf{Implementing the uncoloured independence oracles:} For each hypergraph $\calG_{r,\pi}$, we will implement the uncoloured independence oracle using our assumed decision algorithm for \pp{Exact-Weight-$H$}. Let $X\subseteq V(G)$; then we must determine whether $\calG_{r,\pi}[X]$ contains an edge, i.e. whether $G_{r,\pi}[X]$ contains a subgraph which is isomorphic to~$H$, is zero-weight with respect to~$w$, and is colourful with respect to the partition~$V_1,\dots,V_k$. We cannot simply apply the decision algorithm to $G_{r,\pi}[X]$ with the same weight function $w$, because this algorithm may find zero-weight isomorphic copies of~$H$ that are not colourful.
	To deal with this issue, we give $G_{r,\pi}$ a new weight function $w_{r,\pi}$ which will enforce colourfulness using standard bit tricks.
	
	\newcommand{\dd}{\iota}
	For each $i \in [k]$, let $X_i = X \cap V_i$. Let $h$ be the number of edges of~$H$.
    We define the function $\dd:E(G)\to[h]$ as follows: We set~$\dd(\{u,v\})=\dd$ for all $\{u,v\}\in E(G)$ with $u\in V_i$ and $v\in V_j$ such that $\{\pi(i),\pi(j)\}\in E(H)$ is the $\dd$'th edge of~$H$ in lexicographic order.
    We define the edge weights of~$G_{r,\pi}$ by $w_{r,\pi}(e)=w(e)+\delta_{r,\pi}(e)$ for all~$e$, where
	\begin{equation*}
	\delta_{r,\pi}(e)=
	\begin{cases}
	  (hM+1)\cdot (2h)^{\dd(e)} & \text{if $\dd(e)>1$;}\\
	  -(hM+1)\cdot \sum_{\dd=2}^h (2h)^{\dd}& \text{if $\dd(e)=1$.}
	\end{cases}
	\end{equation*}
	Let $\phi$ be the canonical homomorphism from $G_{r,\pi}$ to $H$, that is, the function that maps vertices from~$V_i$ to the vertex~$\pi(i)$ in $H$.
	
	We claim that for any subgraph $F$ of $G_{r,\pi}$ isomorphic to $H$, we have $w_{r,\pi}(F) = 0$ if and only $w(F)=0$ and $F$ is colourful with respect to the partition $V_1,\dots,V_k$.  If this is true, then we can simulate the uncoloured independence oracle of $\calG_{r,\pi}$ applied to a set $X$ by applying our decision algorithm to $G_{r,\pi}[X]$.
	
	To prove the claim, first note that
	\begin{equation}\label{eq:hom-bit-tricks}
	    w_{r,\pi}(F)=\sum_{f\in E(F)} \big(w(f) + \delta_{r,\pi}(f)\big)=w(F)+\sum_{f\in E(F)}\delta_{r,\pi}(f).
    \end{equation}
    This latter sum has exactly $h$ summands, and each summand is either $(hM+1)(2h)^\dd$ for some $\dd\ge 2$, or it is $-(hM+1)\sum_{\dd=2}^h (2h)^\dd$.
	If $F$ is colourful, then $\phi$ maps every edge of $F$ to a distinct edge of $H$, so the summands that occur are all different and trivially cancel; we therefore have $w_{r,\pi}(F)=w(F)$ in this case.
	
	Suppose instead that $F$ is not colourful, so that $\phi$ maps two distinct vertices $x$ and $y$ of~$F$ to the same vertex $z$ of~$H$. Since $\phi$ is a homomorphism, $x$ and $y$ are each joined to $d_H(z)$ colour classes out of $\{V_1,\dots,V_k\}$ in $F$; moreover, by the construction of $G_{r,\pi}$, at most $d_H(z)$ of these classes are distinct. Since $H$ does not contain isolated vertices, we have $d_H(z)>0$ and so some colour class must appear in both $x$'s neighbourhood and $y$'s neighbourhood. In other words, $\phi$ must map two different edges $f,f'$ of $F$ to the same edge $e$ of~$H$; this implies there is some $\iota \in [h]$ such that $\phi$ maps no edges to the $\iota$'th edge of $H$. Thus, the summands on the right-hand side of~\eqref{eq:hom-bit-tricks} do not cancel, since for all $\iota \in [h]$ we have $h\cdot (2h)^\iota < (2h)^{\iota+1}$ and $(2h)^{\iota+1}-h\cdot(2h)^\iota > (2h)^\iota$, and so $w_{r,\pi} \ne 0$ as required. We have therefore successfully implemented the uncoloured independence oracle for $\calG_{r,\pi}$, the missing ingredient in our algorithm.
	
	\textbf{Bounding the running time:} Our algorithm creates $tk! = O(\eps^{-2})$ weighted graphs $G_{r,\pi}$ and calls \colourfulsamplecount once for each. By \cref{thm:main-combined-colourful}, this takes time $\tildeOO(\eps^{-4}(m+n))$ and makes $\tildeOO(\eps^{-4})$ queries to the independence oracle.
	Each query to the independence oracle takes time $T(n,m,k^2M)=\Omega(n+m)$.
	Combining these facts, we obtain an overall running time of $\tildeOO(\eps^{-4} T(n,m,k^2M))$ as claimed.

	\textbf{Sampling algorithm:}
	Our sampling algorithm is very similar to our approximate counting algorithm, so we omit the details. The only differences are the following.
	
	\begin{enumerate}
	    \item We take only a single uniformly random partition $V_1 \cup \dots \cup V_k$ with associated graphs $G_\pi$ for every $\pi\colon [k]\to V(H)$.
	    \item We require that each invocation of our approximate counting algorithm has failure probability at most $\eps/(\ln(5/\eps)10k^{2k})$ and set $\delta$ accordingly.
	    \item Rather than outputting a weighted sum of $\eps$-approximations of each $e(G_\pi)$, we use $(\eps/10)$-approximations of each $e(G_\pi)$ to apply rejection sampling as in e.g.~Florescu~\cite[Proposition 3.3]{florescu}. Thus writing $N$ for our $(\eps/10)$-approximation of $e(G)$ and $N_\pi$ for our $(\eps/10)$-approximation of $e(G_\pi)$, we choose a bijective function $\pi\colon [k]\to[k]$ uniformly at random, take an $(\eps/10)$-approximate sample from $E(G_\pi)$ using the \colourfulsamplecount algorithm from \cref{thm:main-combined-colourful}, and accept and output this sample with probability $N_\pi/N$; otherwise, we reject it and resample. If we require more than $k^{2k}\ln(5/\eps)$ iterations, we return an arbitrary output.
	\end{enumerate}
	
	Note that the acceptance probability at each step is at least $1/k^{2k}$, and acceptance is independent at each step, so the probability we require more than $k^{2k}\ln(5/\eps)$ invocations is at most $(1-1/k^{2k})^{k^{2k}\ln(5/\eps)} \le \eps/5$. Combining this with the error in the rejection sampling from our $(\eps/10)$-approximations, and with the probability that our approximate counting algorithm fails to return a correct value in the first $k^{2k}\ln(5/\eps)$ invocations (which is at most $\eps/10$ by a union bound), we see that our algorithm returns an $\eps$-approximate sample as required. Since $\eps \ge n^{-k}$, the running time bound is immediate.
\end{proof}

\cref{cor:exact-weight-H} can be combined with known, non-trivial decision algorithms for \pp{Exact-Weight-$H$}. For example, Abboud and Lewi~\cite[Corollary~5]{DBLP:conf/icalp/AbboudL13} prove that \pp{Exact-Weight-$H$} can be solved in time $\tildeOO(n^{\gamma(H)})$, where $\gamma(H)\ge 1$ is a graph parameter that is small whenever~$H$ has a balanced separator.
We obtain the following.
\begin{corollary}\label{cor:gamma}
	Let $H$ be a fixed graph. When $\calS$ denotes the set of zero-weight $H$-subgraphs of a given $n$-vertex graph $G$, there is a randomised algorithm to $\eps$-approximate $|\calS|$, or draw an $\eps$-approximate sample from $\calS$, in time~$\tildeOO(\varepsilon^{-4}n^{\gamma(H)})$.
\end{corollary}

\bibliographystyle{siamplain}
\bibliography{references}
\end{document}